\documentclass[twoside, 11pt]{article}

\usepackage{graphicx, subcaption}
\usepackage[margin=1in, footnotesep=0.5in]{geometry}
\usepackage[sort&compress, numbers]{natbib} 
\usepackage{amsfonts, amsmath, amssymb, amsthm}
\usepackage[shortlabels]{enumitem}
\usepackage{booktabs}
\usepackage{mathtools}
\mathtoolsset{showonlyrefs}

\usepackage{color}

\usepackage{hyperref}
\definecolor{customdarkred}{RGB}{150,0,0}
\definecolor{customdarkgreen}{RGB}{0,150,0}
\definecolor{customdarkblue}{RGB}{0,0,150}
\hypersetup{colorlinks=true, linkcolor=customdarkred, citecolor=customdarkgreen, urlcolor=customdarkblue}
\usepackage{url}
\usepackage[linesnumbered, ruled]{algorithm2e}
\usepackage{authblk}

\DeclareMathOperator{\card}{card}

\def\d{\mathsf{d}}

\theoremstyle{definition}

\theoremstyle{remark}

\numberwithin{equation}{section}
\numberwithin{figure}{section}

\def\beq{\begin{equation}} % \setcounter{equation}{1}}
\def\eeq{\end{equation}}
\def\beqn{\begin{eqnarray*}}
\def\eeqn{\end{eqnarray*}}
\def\Bitem{\begin{itemize}\setlength{\itemsep}{.2in}}
\def\bitem{\begin{itemize}\setlength{\itemsep}{.05in}}
\def\eitem{\end{itemize}}
\def\Benum{\begin{enumerate}\setlength{\itemsep}{.2in}}
\def\benum{\begin{enumerate}\setlength{\itemsep}{.05in}}
\def\eenum{\end{enumerate}}
\def\bmult{\begin{multline*}}
\def\emult{\end{multline*}}
\def\bcenter{\begin{center}}
\def\ecenter{\end{center}}
\def\bframe{\begin{frame}}
\def\eframe{\end{frame}}

\newcommand{\secref}[1]{Section~\ref{sec:#1}}
\newcommand{\figref}[1]{Figure~\ref{fig:#1}}

\newcommand{\algref}[1]{Algorithm~\ref{alg:#1}}

\DeclareMathOperator*{\argmax}{arg\, max}

\def\cE{\mathcal{E}}

\def\cG{\mathcal{G}}

\def\cV{\mathcal{V}}

\def\bbR{\mathbb{R}}

\def\1{\mathbbm{1}}

\title{On the Selection of Tuning Parameters\\ for Patch-Stitching Embedding Methods}
\author[1]{Phong Alain Chau} 
\author[1,2]{Ery Arias-Castro}
\affil[1]{\small Department of Mathematics, University of California, San Diego} 
\affil[2]{\small Halıcıoğlu Data Science Institute, University of California, San Diego}
\date{}

\begin{document}
\maketitle
\thispagestyle{empty}

\begin{abstract}
While classical scaling, just like principal component analysis, is parameter-free, other methods for embedding multivariate data require the selection of one or several tuning parameters. This tuning can be difficult due to the unsupervised nature of the situation. We propose a simple, almost obvious, approach to supervise the choice of tuning parameter(s): minimize a notion of stress.
We apply this approach to the selection of the patch size in a prototypical patch-stitching embedding method, both in the multidimensional scaling (aka network localization) setting and in the dimensionality reduction (aka manifold learning) setting. 
In our study, we uncover a new bias--variance tradeoff phenomenon.
\end{abstract}

\section{Introduction}

In the general problem known as {\em multidimensional scaling (MDS)}, the primary objective is to represent a set of items as points within a Euclidean space of a specified dimension. This representation should ideally preserve the given pairwise dissimilarities as accurately as possible, by ensuring that the Euclidean distances between these points mirror the original dissimilarities.
MDS is a extensively researched problem found in diverse fields such as psychometrics \cite{borg2005modern}, mathematics, and computer science \cite{blumenthal1953theory, laurent2001matrix, battista1998graph}, engineering (where it is also known as {\em network localization}) \cite{mao2007wireless}, as well as statistics \cite{seber2009multivariate, MVA} and machine learning \cite[Ch 14]{hastie2009elements}.

{\em Dimensionality reduction (DR)} aims at embedding data points in a Euclidean space into a lower-dimensional Euclidean space while preserving, as much as possible, the geometry of the point cloud \cite{lee2007nonlinear, ghojogh2023elements}. When the data points are assumed to be on or near a smooth submanifold, a variant of DR known as {\em manifold learning}, this typically means preserving the pairwise intrinsic distances to the greatest extent. As is well-known, the two problems, MDS and DR, are closely related. 
%We will mostly consider the MDS problem, and later derive implications for DR. 

While classical scaling, and its equivalent in DR, principal component analysis, do not require the choice of a tuning parameter (other than the embedding dimension, whose choice we only discuss in \secref{dimension}), other methods for embedding data necessitate the specification of one or several parameters, and being the situation unsupervised in that the items (in MDS) or the points (in DR) are not labeled, it is not obvious how to tune these parameters. In fact, we are not aware of any data-driven procedures for choosing such parameters that are currently in use. There is no equivalent to cross-validation --- a widely used method for parameter tuning in the context of supervised learning such as regression or classification --- that we know of. 
We propose a simple approach: to use a notion of stress --- a measure of quality of fit --- to supervise the choice of tuning parameters. 
We apply this approach to the selection of the patch size in a prototypical patch-stitching embedding method, both in MDS and in DR. 

Although the proposed approach this is rather natural, bordering on the obvious, the specter of overfitting might have dissuaded its use. We argue in this paper that there is no danger of overfitting when choosing tuning parameters other than the embedding dimension. 

In our investigation, we uncover a new form of bias--variance tradeoff. While such a tradeoff is well-known in regression \cite[Ch 7]{hastie2009elements} --- where it is foundational in the textbook understanding of statistical complexity and the need to appropriately select method parameters --- the discussion of such a phenomenon appears to be absent from the MDS literature in particular. For patch-stitching methods, the choice of patch size needs to balance out how much noise is present in the dissimilarities with how complex the configuration domain is: a high level of noise requires a larger patch size, while a domain that is far from convex requires a smaller patch size. 
Our numerical experiments show that choosing a patch size that minimizes a notion of stress helps strike a seemingly good compromise when these two aspects need to be balanced out.

The remainder of the paper is organized as follows.
In \secref{MDS}, we consider patch-stitching methods in the context of MDS. 
This is where we discuss the bias--variance phenomenon mentioned above.
In \secref{DR}, we consider patch-stitching methods in the context of DR. The particular variant that we introduce can be seen as a local form of {\em Isomap}.
We conclude the paper with a brief discussion in \secref{discussion}.

\section{Multidimensional scaling}
\label{sec:MDS}

In this section we consider multidimensional scaling (MDS). We first formalize the setting in \secref{setting MDS}. We then discuss methods in \secref{methods MDS}. We use a popular class of methods known under the umbrella name of patch-stitching to showcase the data-driven choice of tuning parameter --- here the patch size --- by stress minimization, and also the bias--variance tradeoff at play. We present the result of some numerical experiments in \secref{experiments MDS} meant to illustrate the proposed approach. 

\subsection{Setting} 
\label{sec:setting MDS}
In MDS, the data consist of a weighted undirected graph $\cG = (\cV, \cE, d)$, with node set $\cV = [n] := \{1, \dots, n\}$ and edge set $\cE \subset \cV \times \cV$, together with non-negative weights on the edges. The weight on $(i, j) \in \cE$ is referred to as the dissimilarity between $i$ and $j$, and denoted $d_{ij}$. The matrix $D = (d_{ij})$, which is incomplete unless the graph is complete, gathers these dissimilarities. Given a dimension $p \ge 1$, we seek a configuration $y_1, \dots, y_n \in \bbR^p$ such that $\|y_i - y_j\| \approx d_{ij}$ for all or most $(i,j) \in \cE$. 
The problem is further formalized by translating it into an optimization problem that consists in minimization a notion of what is traditionally called {\em stress} in Psychometrics, for example,
\begin{align}
\label{stress}
\sum_{(i,j) \in \cE} \big(\|y_i - y_j\|^2 - d_{ij}^2\big)^2.
\end{align}
(This stress variant was proposed by \citet{takane1977nonmetric} and is called the s-stress in the psychometrics literature.)

We will focus on the noisy realizable situation in which 
\begin{align}
\label{setting mult}
d_{ij} = (1+\eta_{ij}) \|x_i - x_j\|, \quad (i,j) \in \cE,
\end{align}
where $\{x_1, \dots, x_n\} \in \bbR^p$ will be referred to as the {\em latent configuration} --- although it is only determined up to a rigid transformation, as no anchor is assumed available --- and $\{\eta_{ij}: (i,j) \in \cE\}$ stands for (multiplicative) measurement error. The latent positions will be assumed to be somewhat dense in a subset of $\bbR^p$ referred to as the latent domain. 

Throughout, the dimension $p$ will be assumed given (see \secref{dimension}) and $\|\cdot\|$ will denote the Euclidean norm in the appropriate space (which will be $\bbR^p$ in the entire section).

\subsection{Methods}
\label{sec:methods MDS}
Many approaches have been suggested in the literature, including classical scaling \cite{torgerson1958theory, torgerson1952multidimensional, gower1966some} and other spectral methods \cite{hall1970r}; first-order  
\cite{kruskal1964nonmetric, kruskal1964multidimensional}, second-order  \cite{kamada1989algorithm}, as well as other Newton and quasi-Newton approaches \cite{kearsley1998solution, glunt1993molecular}; augmentation and majorization \cite{de1975alternating, heiser1988multidimensional}, including the popular {\em SMACOF} \cite{de1977applications, de2009multidimensional, mair2022more}; incremental approaches \cite{cohen1997drawing, williams2004steerable, bronstein2006multigrid}; semidefinite programs (SDP) \cite{alfakih1999solving, biswas2006semidefinite, biswas2006semidefinite_based, javanmard2013localization, weinberger2006graph, so2007theory, drusvyatskiy2017noisy, cayton2006robust}; dissimilarity matrix completion by graph distances, including the original {\em MDS-D} of \citet{kruskal1980designing}, and its multiple incarnations \cite{priyantha2003anchor, shang2003localization, niculescu2003dv}; and sequential lateration \cite{kearsley1998solution, aspnes2006theory, fang2009sequential, eren2004rigidity, laurent2001polynomial, bakonyi1995euclidian, grone1984positive}.

\subsubsection{Patch-stitching methods}
We focus on {\em patch-stitching methods}. These are divide-and-conquer methods that consist in embedding appropriately selected subgraphs as `patches' in the target Euclidean space and then `stitching' these patches together by applying a form of Procrustes analysis to align patches that have a large enough intersection. This alignment (aka synchronization) can be done in a greedy manner, by sequentially aligning a new patch with a sufficient overlap with the current embedding; or by a more global approach that attempts to align all patches simultaneously based on all (multiway) intersections.  
A wide variety of patch-stitching approaches have been proposed \cite{yang2006fast, tzeng2008multidimensional, koren2005patchwork, moore2004robust, cucuringu2012sensor, singer2008remark, zhang2010rigid, drusvyatskiy2017noisy, krislock2010explicit, shang2004improved, hendrickson1995molecule, costa2006distributed}, motivated by two main reasons. 
The first reason is computational: the computation of patches can be done in parallel, and the overall procedure can have, depending on the variant, low run time. 
The second reason is that such methods can work well even when the underlying domain that the latent positions populate has a complex shape. By contrast, for example, methods that rely on shortest path distances like MDS-D, and also some SDP methods like semidefinite embedding of \citet{weinberger2006graph}, can have a substantial bias when the latent domain is far from non-convex. 

\begin{figure}[ht!]
\centering
\includegraphics[width=0.32\textwidth]{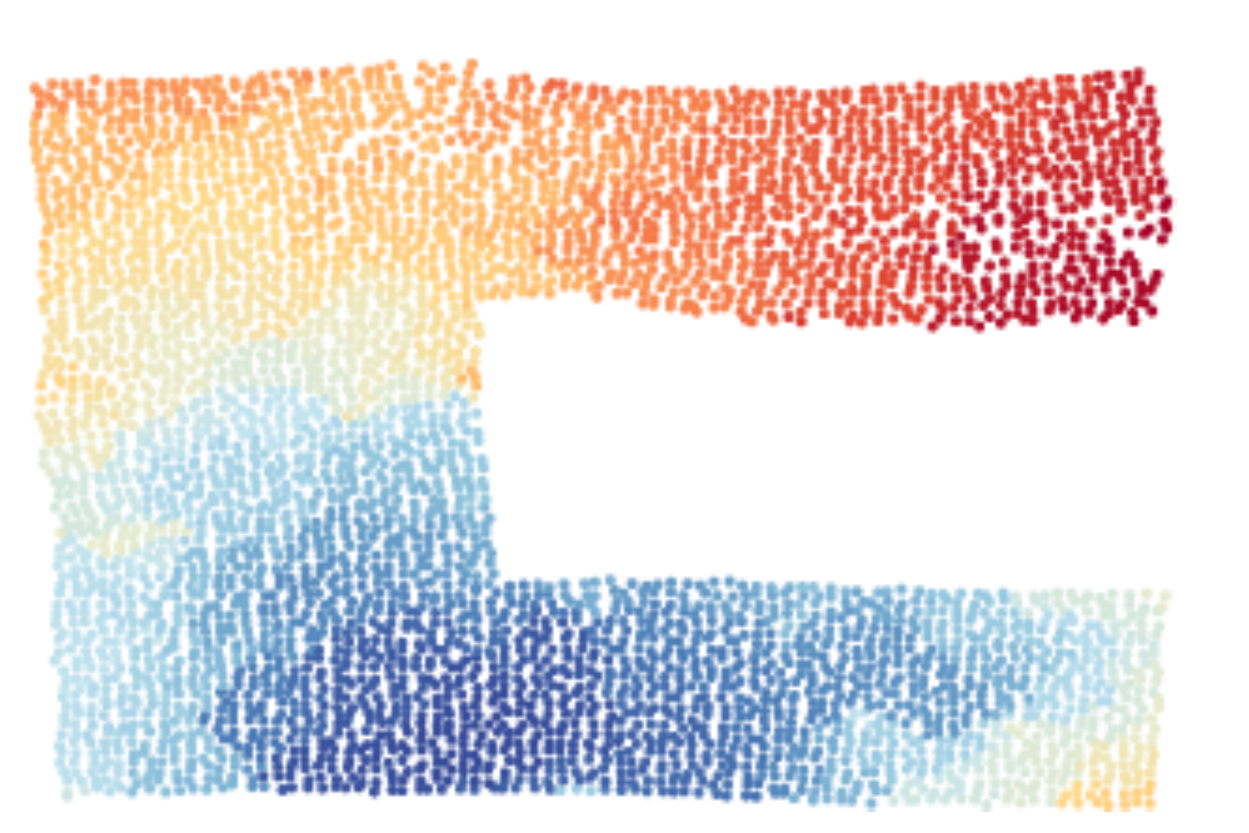}
\includegraphics[width=0.32\textwidth]{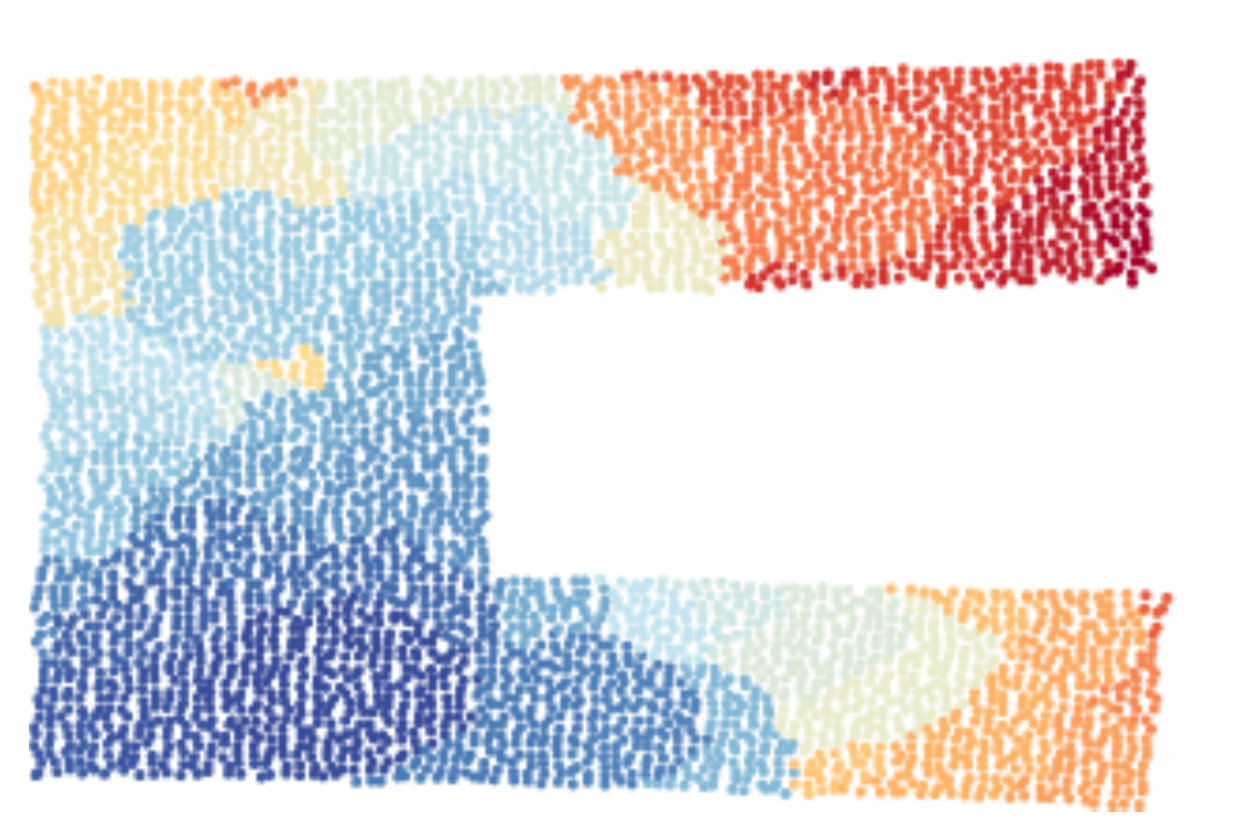}
\includegraphics[width=0.32\textwidth]{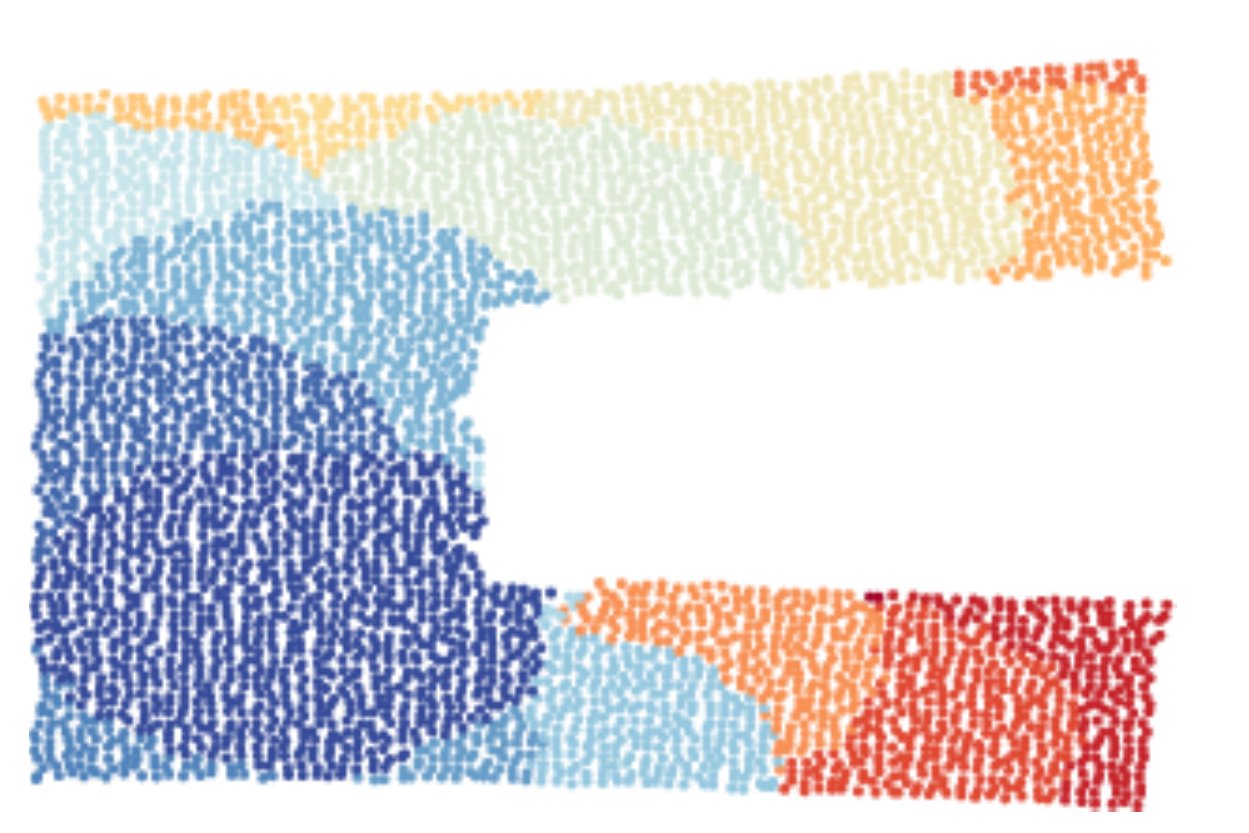}
\caption{In the MDS-MAP(P) algorithm, each patch is embedded separately and then merged sequentially. The points in the seed patch are in dark blue, while points added with subsequent patches are more and more red as the stitching progresses. Plotted above are embeddings with number of hops $h \in \{1,2,5\}$.}
\label{fig:patches}
\end{figure}

\citet{shang2004improved} cite both reasons as motivation for their patch-stitching method, {\em MDS-MAP(P)}, and present them as advantages, in particular as compared to a method they had previously suggested, MDS-MAP \cite{shang2003localization}, which was in fact a rediscovery of MDS-D. MDS-MAP(P) stitches the patches in a greedy fashion: see \figref{patches} for a visualization of the patches and their sequential stitching.
 
\subsubsection{Tuning by stress minimization}
MDS-MAP(P), like any other patch-stitching method, requires the choice of a parameter controlling the patch size. In this particular case, it's the number of hops. 
A detail the variant that we implemented in \algref{MDS-MAP(P)}, which is a somewhat simplified variant of the original, which, for example, weighs connections according to the number of hops and uses these weights in the refinement step --- while we do not do that.
Although we could have used almost any other patch-based method, we adopt MDS-MAP(P) simply to illustrate how to choose that tuning parameter in a data-driven manner. 

\begin{algorithm}[!t] \small
\KwData{weighted graph $\cG=(\cV, \cE, D)$, number of hops $h$, embedding dimension $p$}
\KwResult{configuration $Y = [y_1 \cdots y_n]^\top \in\bbR^{n \times p}$}
\For{$v\in \cV$}{
$N_v \gets$ $h$-hop neighborhood of $v$\;
%$W[v,w]\gets 1$ if $w\in N_v$\; \label{opW}
$D_v \gets D[N_v, N_v] = (d_{ij} : i,j \in N_v; (i,j) \in \cE)$\;
$\Lambda_v \gets$ MDS-D applied to $D_v$\;  \label{opSP}
$X_v \gets$ classical scaling applied to $\Lambda_v$\; \label{opcmds}
$X_v \gets$ SMACOF applied to $X_v$\; \label{opsmacof}
}
$v^* \gets \argmax_{v} \card (N_v)$\;
$Y \gets X_{v^*}$\;
$N_Y \gets N_{v^*}$\;
\While{there exist unmapped nodes}{
$v^* \gets \argmax_{v} \card(N_v \cap N_Y)$\;
$X_* \gets$ align $X_{v^*}$ to $Y$ by Procrustes\;
$Y \gets Y \cup X_{*}$\; \label{opmerge}
}
$Y \gets$ SMACOF applied to $Y$\; \label{opfsmacof}
return $Y$\;
\caption{A variant of MDS-MAP(P)}
\label{alg:MDS-MAP(P)}
\end{algorithm}

%In terms of computational complexity, the Floyd--Warshall algorithm~\cite{Floyd1962} runs in $O(k^3)$ time on a graph with $k$ nodes and is used in Step \ref{opSP} to calculate all pairs of shortest path distances for each neighborhood. This is the main computational bottleneck. 
%Some improvement may be possible using an algorithm that is able to exploit the sparsity of $D_v$~\cite{Pettie2002, Sao2020}.
%Classical scaling is used in Step~\ref{opcmds}. Using a naive eigendecomposition results in a complexity of $O(k^3)$ for $k\times k$ matrices, but there are approximate algorithms that run in linear time~\cite{Silva2004, Yang2006}. 
%The stitching in Step~\ref{opmerge} is done by Procrustes, while the refinement in Steps~\ref{opsmacof} and~\ref{opfsmacof} is done by SMACOF.
%(The final SMACOF refinement is an optional step that can be extremely costly for large datasets. It can be useful when the number of patches is relatively small.)

The approach we propose for choosing one or even several tuning parameters, such as the number of hops in MDS-MAP(P), consists in minimizing a notion of stress. We adopt the variant \eqref{stress} for no particular reason other than it is fairly popular. 
Although the embedding dimension may be seen as a tuning parameter, minimizing the stress is not an appropriate way to select it, simply because it will always lead to choosing the largest possible embedding dimension, which is $p = n-1$. See \secref{dimension} for further discussion.

Although this approach would seem rather natural, we have not seen it suggested in the literature, where the choice of tuning parameter is often ad hoc or just left to the user.  
There might be some hesitation to use the stress, as it stands for what is called empirical risk in statistical learning, and minimizing the risk is known to lead to overfitting unless the model complexity is under control. This is particularly true in nonparametric regression. The situation in MDS is not immediately translatable to regression, which is by now well-understood, but we can reason in similar terms. On the one hand, the parameter that we need to estimate is very high dimensional: it is the latent configuration $\{x_1, \dots, x_n\}$ in \eqref{setting mult} modulo an arbitrary rigid transformation. Thus, even if the embedding dimension is small, say $p = 2$, the parameter is of $O(n)$ dimension. This needs to be contrasted with the number of observations, which is $|\cE|$. It turns out that, as long as the latent configuration is in general position and the graph is connected enough that it is {\em generically globally rigid} \cite{connelly2005generic} or even a {\em lateration graph} \cite{aspnes2006theory}, minimizing the stress recovers the latent configuration in the noiseless setting ($\eta_{ij} = 0$ for all $(i,j) \in \cE$); and although the recovery is no longer exact in the presence of noise, it degrades graceful with the noise level as shown in \cite{anderson2010formal} and \cite{arias2023stability} in the same situations, respectively. Therefore, minimizing the stress is a reasonable target, and this can be done by all means necessary, as long as the embedding dimension is fixed, since being generically globally rigid or a lateration graph depends in a crucial manner on the dimension $p$.

\subsubsection{Bias--variance tradeoff}
\label{sec:tradeoff MDS}
In the standard textbook exposition of statistical complexity such as \cite[Ch 7]{hastie2009elements}, one is taught that, in the context of regression, as the model being fitted to the data increases in complexity, the bias decreases while the variance increases. Model complexity is often driven by one or several tuning parameters (e.g., the bandwidth in kernel regression), and a `good' selection of these parameters is one that results in a `good' compromise between (squared) bias and variance, often understood as being equivalent to minimizing the prediction error. In \cite[Fig~7.1]{hastie2009elements}, we see that, as the model complexity increases, the empirical error (as measured on training data) decreases, while the prediction error (as measured on the test data) decreases at first but then increases --- and a `good' selection of model complexity would be so that the prediction error is at its minimum.

The discussion of such a bias--variance tradeoff, or the choice of model complexity, seems absent from the MDS literature, except for the choice of embedding dimension (see \secref{dimension}). But it is clearly observed in our experiments involving a non-convex domain, in the case of a hollow rectangle (Figures~\ref{fig:rect_hole_a}--\ref{fig:rect_hole_b}), a C-shaped domain (Figures~\ref{fig:cshape_a}--\ref{fig:cshape_b}); and an H-shaped domain or `dumbbell' (Figures~\ref{fig:dumbbell_a}--\ref{fig:dumbbell_b}). 
Indeed, we can clearly see that, as the number of hops increases, the embedding error decreases and then increases. On the other hand, we observe that the stress does not function as the empirical error does in regression.

In the particular case of MDS-MAP(P), we may explain this as follows. When a domain is non-convex, using a large enough patch that covers the entire domain, MDS-MAP(P) coincides with MDS-D, and MDS-D is known to be biased unless the domain is convex. This is because the shortest path distances are consistent for the intrinsic distances \cite{bernstein2000graph, arias2019unconstrained}, and the intrinsic distances are not Euclidean unless the domain is convex.
The choice of a smaller patch size enables MDS-MAP(P) to better avoid that bias, as it only relies on being able to cover the domain with approximately convex balls the size of the patches.  
Thus the number of hops can be understood as controlling model complexity here: the smaller the number of hops, the smaller the patch size, and the more flexible and therefore complex the domain shape model being implicitly fitted. 
However, in the presence of noise, one also has to contend with the variance, as the smaller a patch is, the more difficult it may be to accurately embed it.

\subsection{Experiments}
\label{sec:experiments MDS}

\subsubsection{Synthetic data}
We start with some synthetic datasets that exemplify the setting of \secref{setting MDS}. All our experiments are in dimension $p = 2$. We first describe how the datasets used in the experiments are constructed. Recall that the latent configuration is denoted by $x_1, \dots, x_n \in \bbR^2$. In all cases, these points are chosen dense in a domain with varying shape. This is done by considering a fine square grid of points inside the domain to which we add some jitter. The added jitter is small and plays two roles: it makes the configuration generic and it also prevents some possible systematic bias from arising when computing shortest path distances in the graph (which is a building block of MDS-MAP(P)). The jittered grid inside the domain gives the configuration. The graph structure is given by connecting each point to its $k=15$ nearest neighbors. 
The pairwise Euclidean distances between configuration points that are connected in the graph are then corrupted by multiplicative noise as in \eqref{setting mult}, where the $\eta_{ij}$ are drawn iid from the uniform distribution on $[-\sigma, \sigma]$, where the noise amplitude $\sigma$ varies from experiment to experiment.

\begin{figure}[ht!]
\centering
\includegraphics[width=0.5\textwidth]{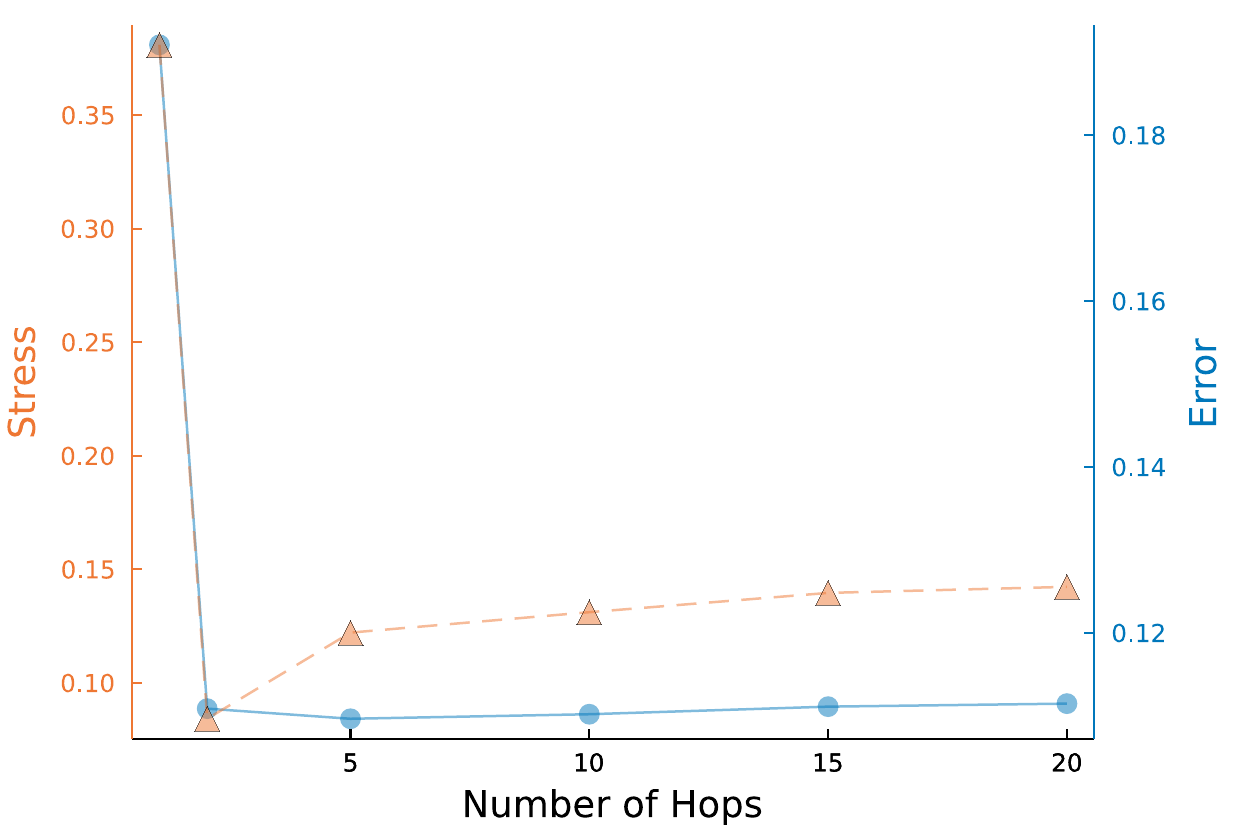}
\caption{Experiment with $n=4000$ points on a rectangle with $\sigma=0.15$.}
\label{fig:rect_grid_a}
\end{figure}

\begin{figure}[ht!]
\centering
\includegraphics[width=.197\textwidth, trim={0 0 2.2in 0}, clip]{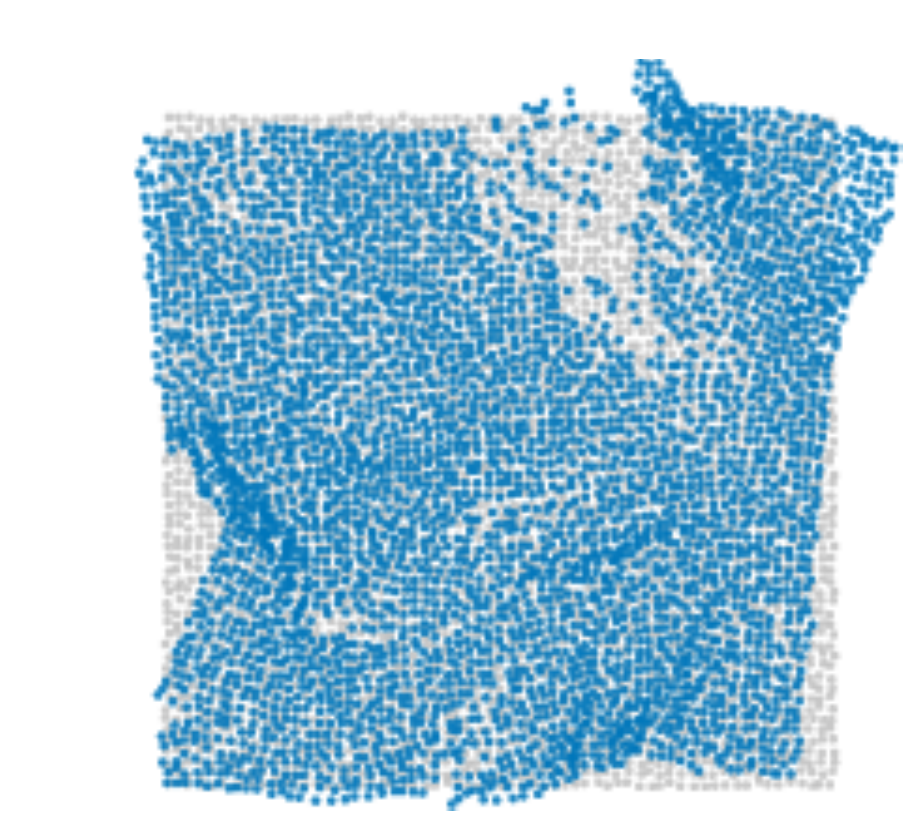}
\includegraphics[width=.18\textwidth, trim={0 0 2.2in 0}, clip]{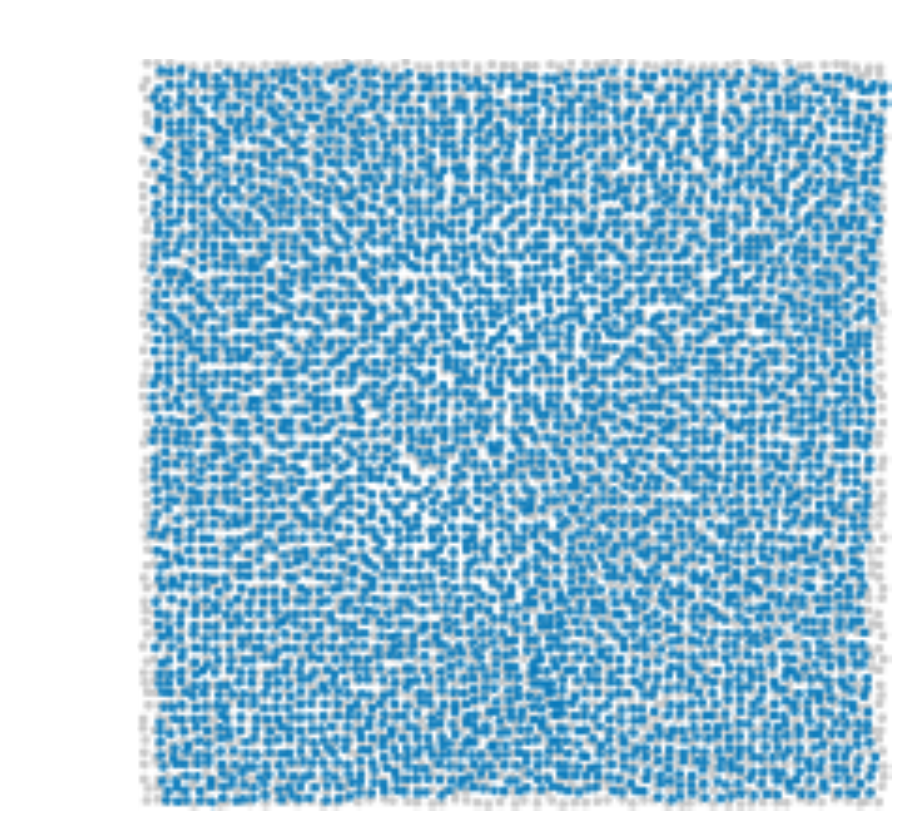} 
\includegraphics[width=.18\textwidth, trim={0 0 2.2in 0}, clip]{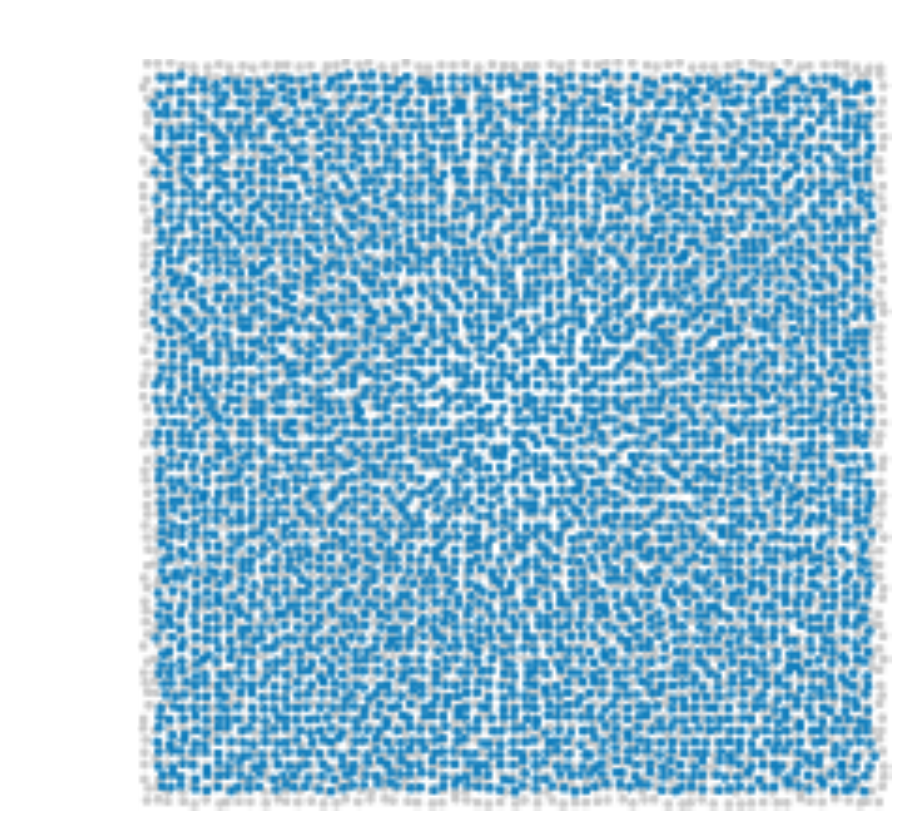}
\includegraphics[width=.18\textwidth, trim={0 0 2.2in 0}, clip]{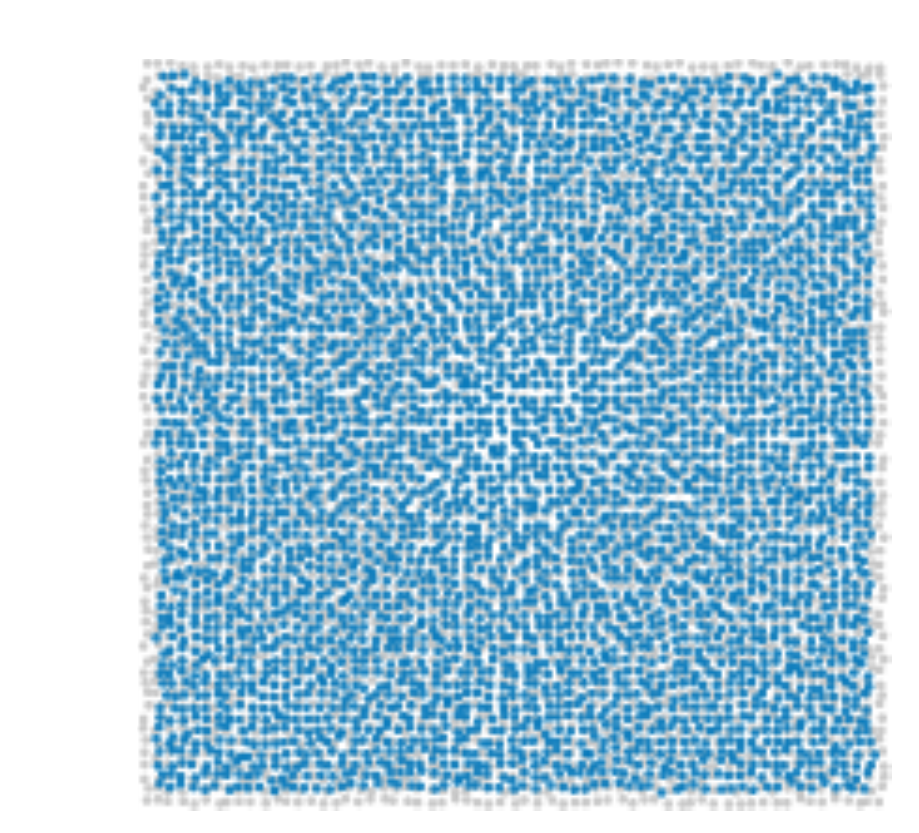}
\includegraphics[width=.18\textwidth, trim={0 0 2.2in 0}, clip]{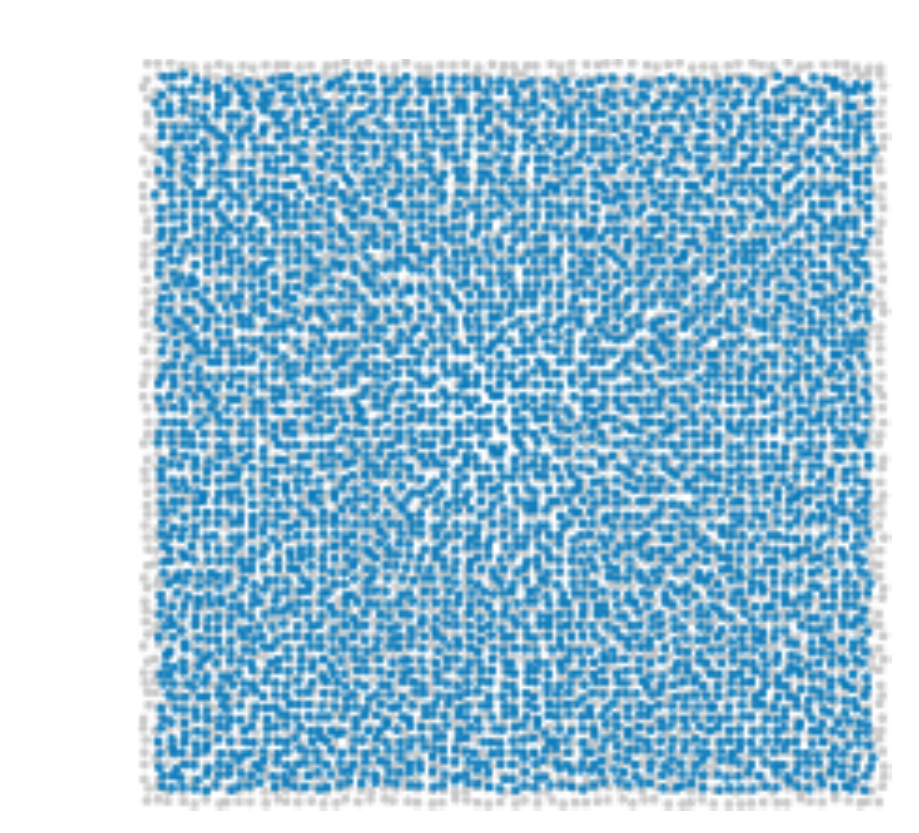}
\caption{Same setting as \figref{rect_grid_a}. Examples of embeddings with number of hops $h = 1, 2, 3, 5, 10$.}
\label{fig:rect_grid_b}
\end{figure}

\begin{figure}[ht!]
\centering
\includegraphics[width=.5\textwidth]{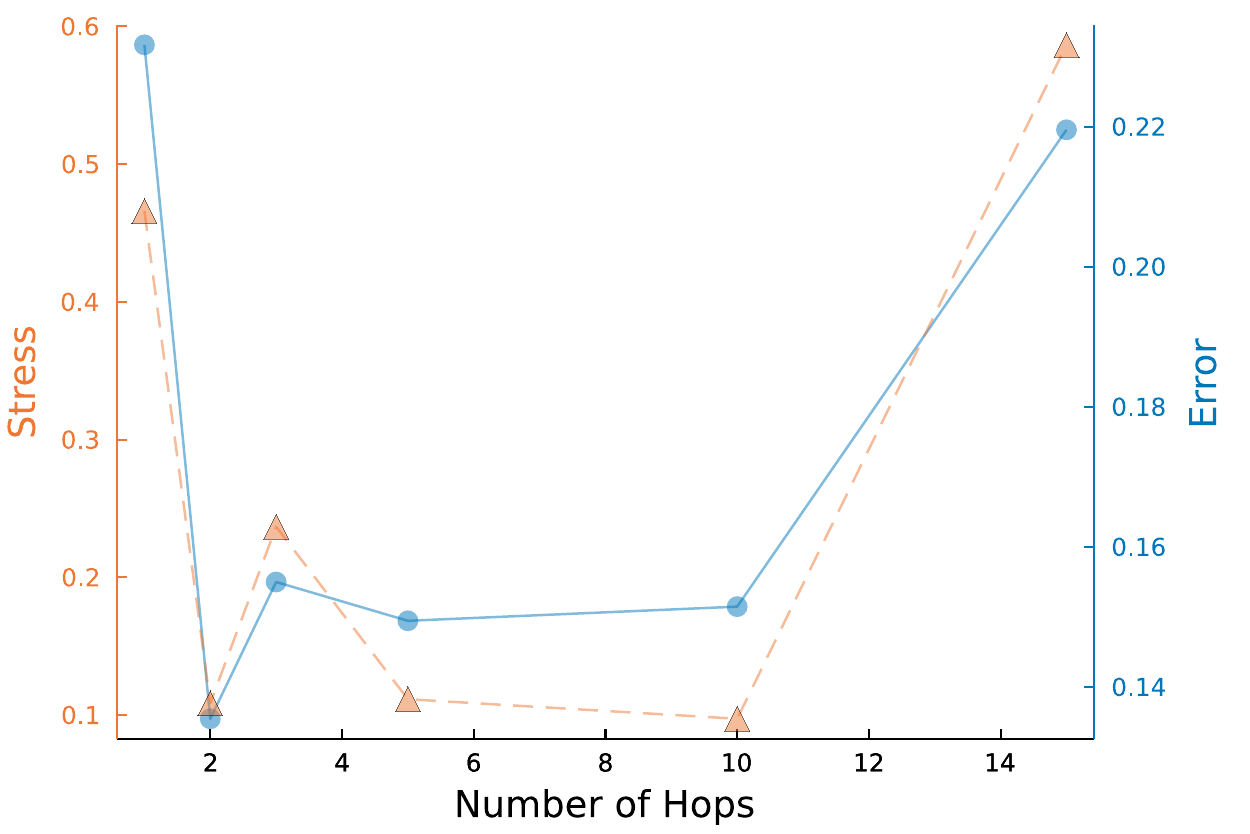}
\caption{Experiment with $n=4140$ points on a rectangle with a rectangular hole with $\sigma=0.15$.}
\label{fig:rect_hole_a}
\end{figure}

\begin{figure}[ht!]
\centering
\includegraphics[width=.197\textwidth, trim={0 0 2.2in 0}, clip]{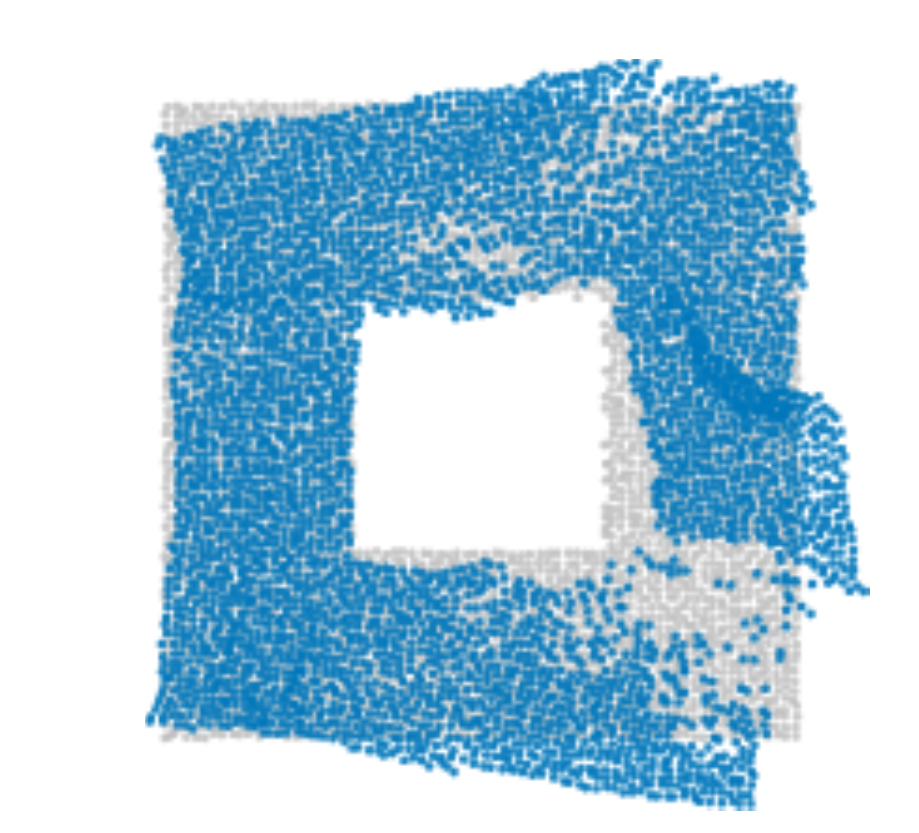}
\includegraphics[width=.18\textwidth, trim={0 0 2.2in 0}, clip]{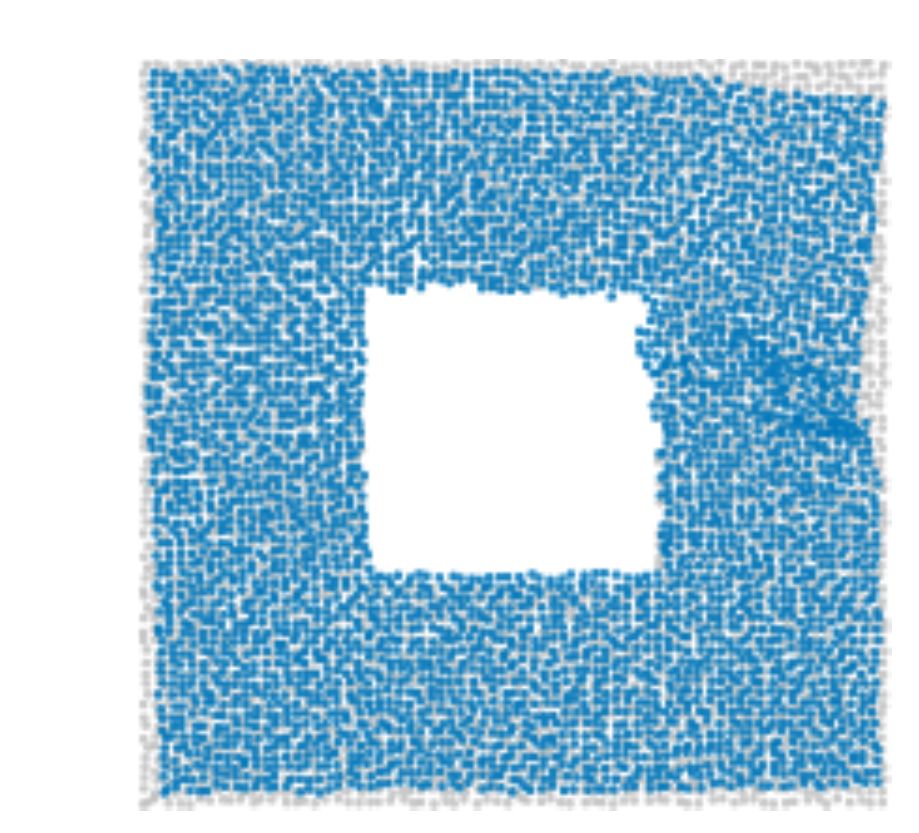}
\includegraphics[width=.18\textwidth, trim={0 0 2.2in 0}, clip]{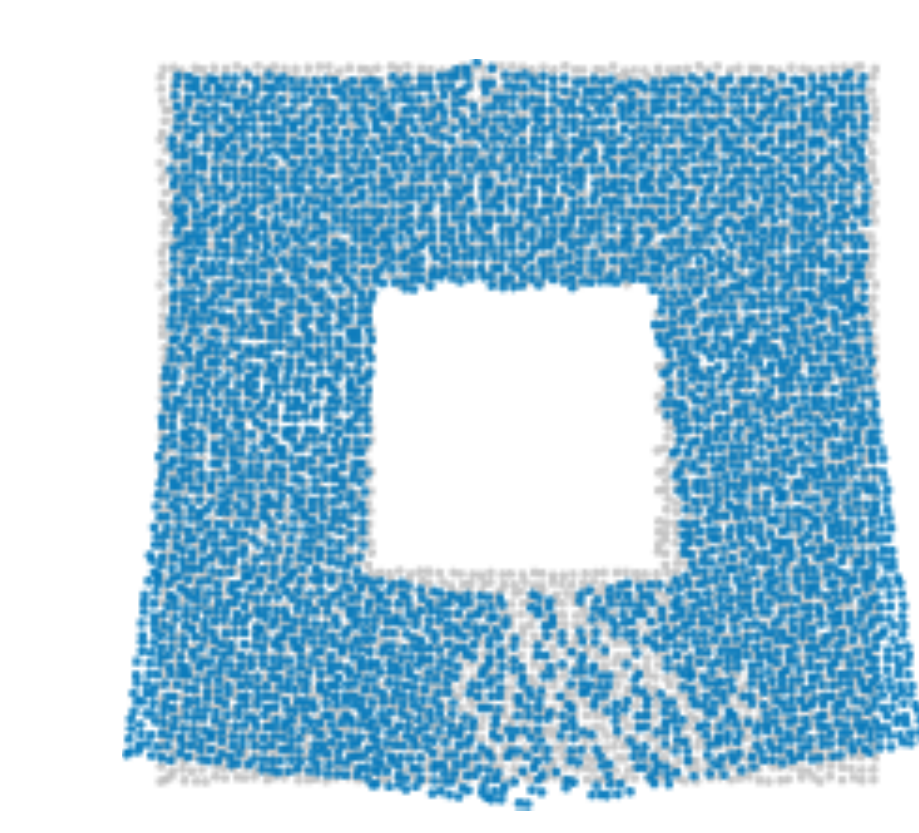}
\includegraphics[width=.18\textwidth, trim={0 0 2.2in 0}, clip]{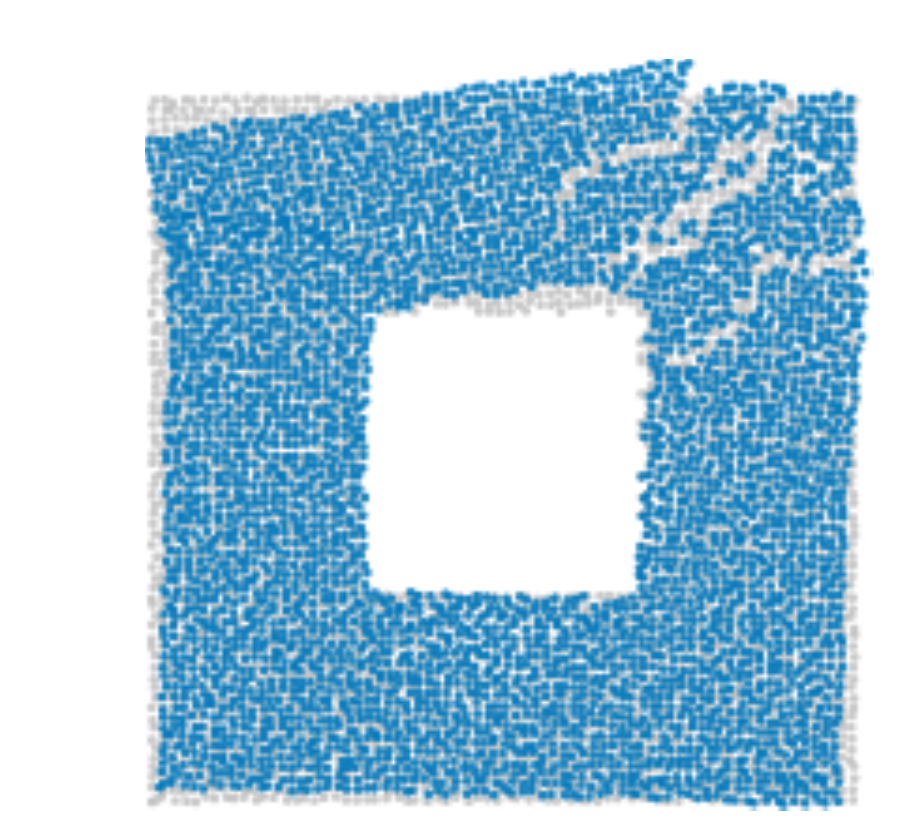}\includegraphics[width=.18\textwidth, trim={0 0 2.2in 0}, clip]{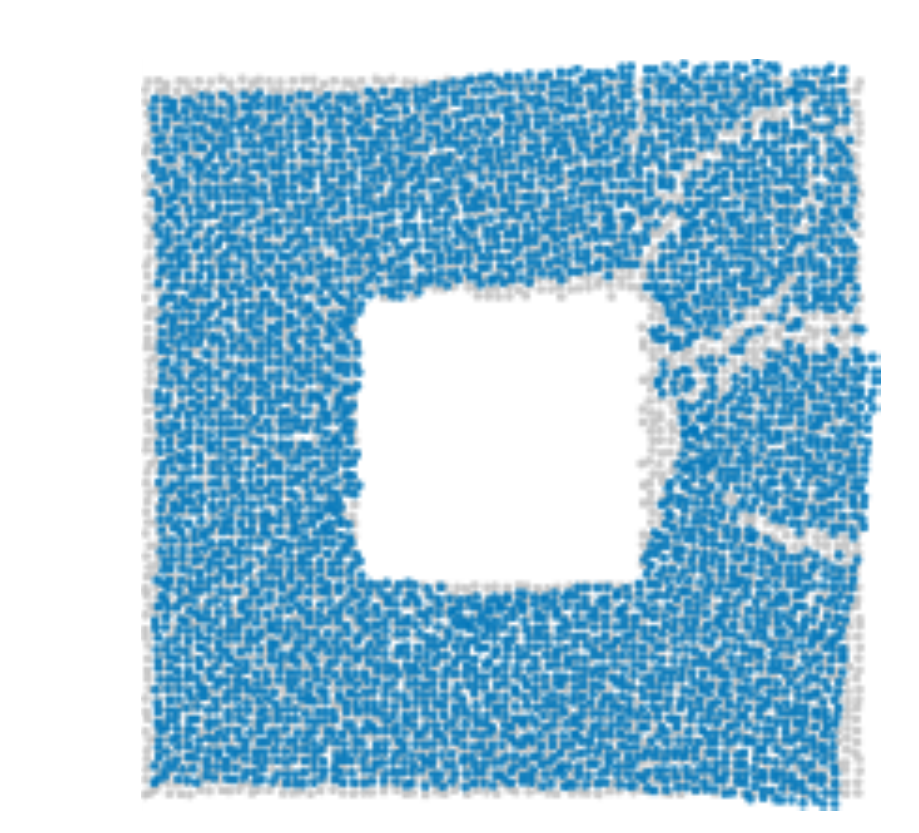}
\caption{Same setting as \figref{rect_hole_a}. Examples of embeddings with number of hops $h = 1, 2, 3, 5, 10$.}
\label{fig:rect_hole_b}
\end{figure}

\begin{figure}[ht!]
\centering
\includegraphics[width=.5\textwidth]{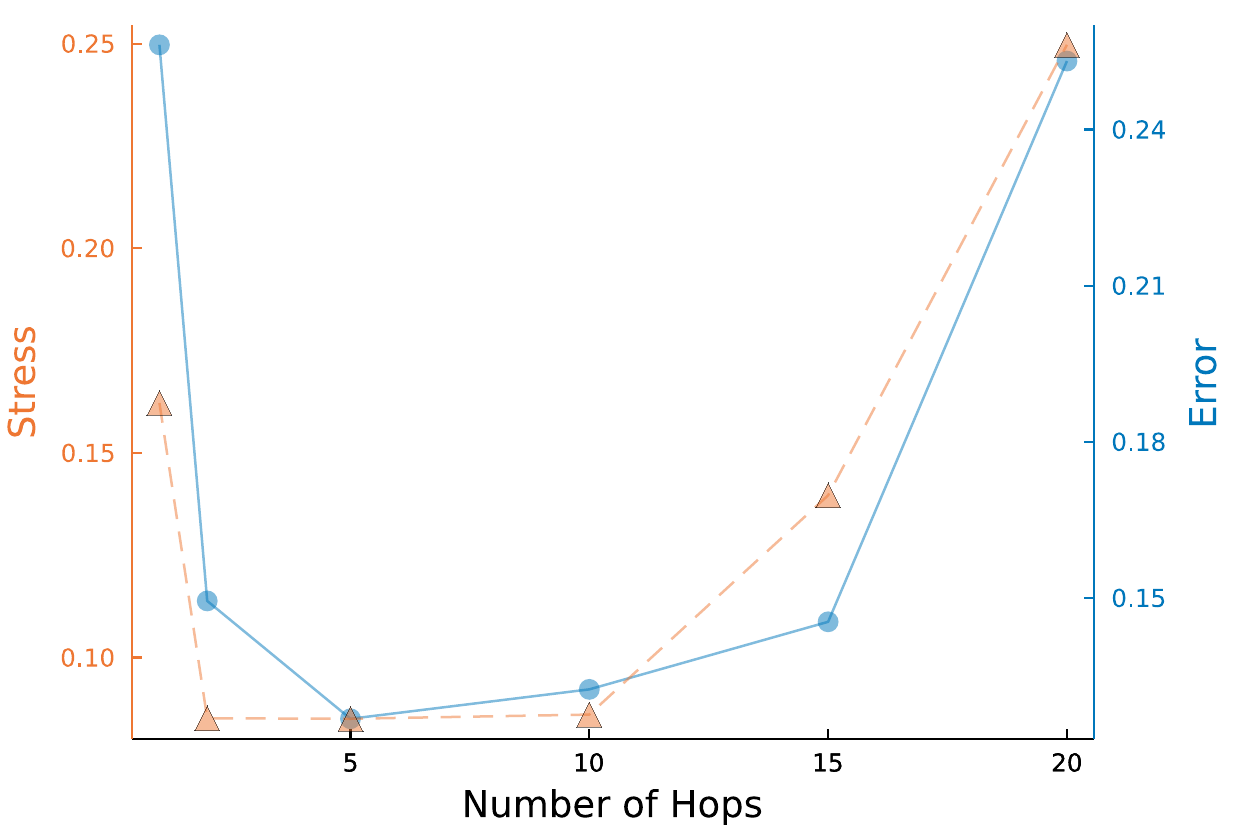}
\caption{Experiment with $n=4528$ points on a C-shaped domain with $\sigma=0.15$.}
\label{fig:cshape_a}
\end{figure}

\begin{figure}[ht!]
\centering
\includegraphics[width=.18\textwidth, trim={0 0 2.2in 0}, clip]{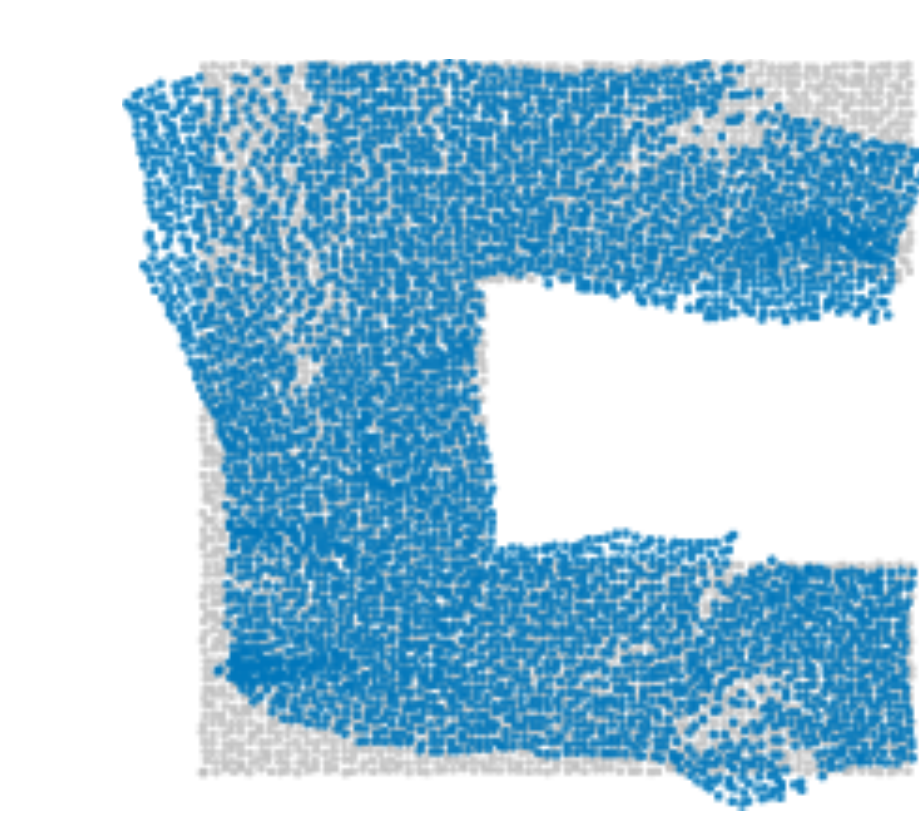}
\includegraphics[width=.18\textwidth, trim={0 0 2.2in 0}, clip]{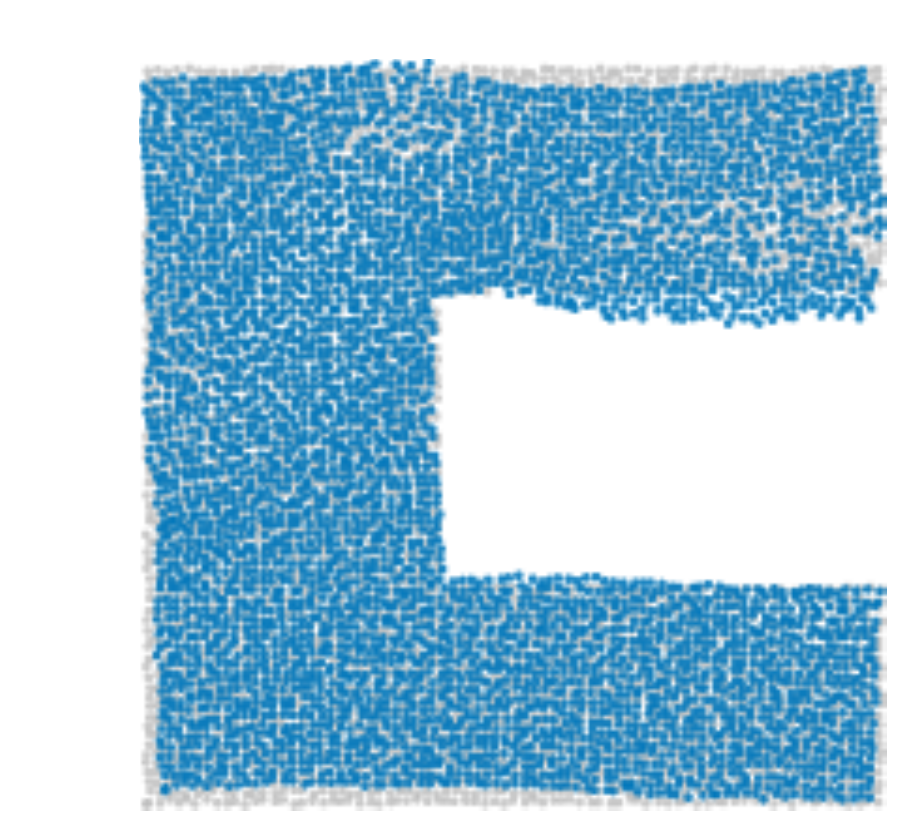}
\includegraphics[width=.18\textwidth, trim={0 0 2.2in 0}, clip]{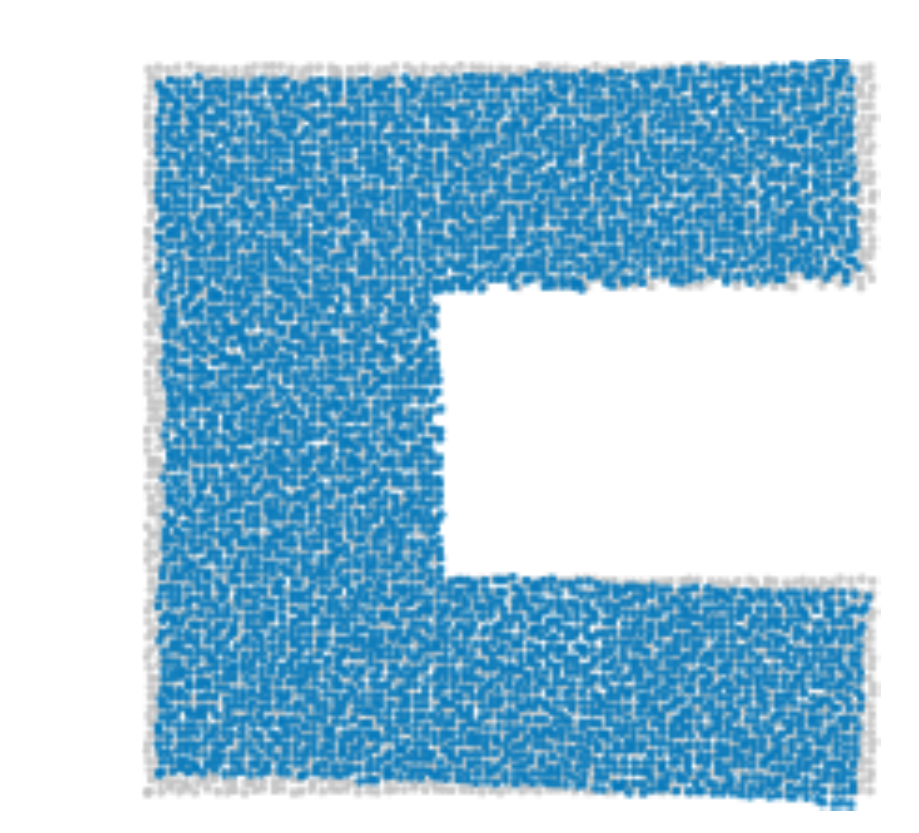}
\includegraphics[width=.18\textwidth, trim={0 0 2.2in 0}, clip]{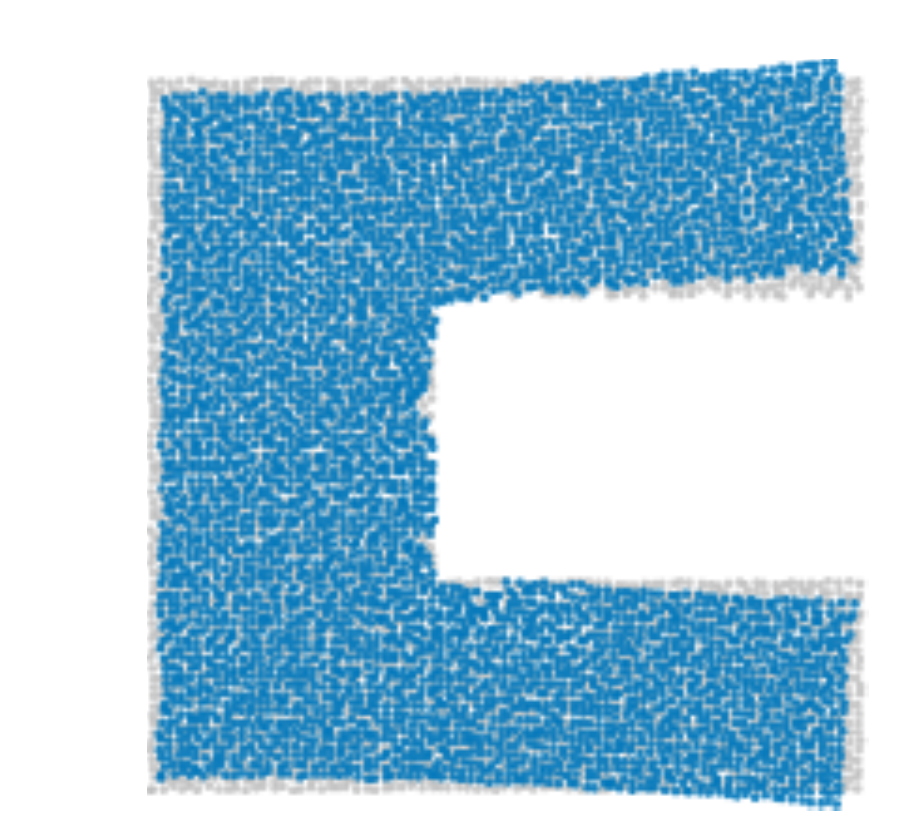}\includegraphics[width=.21\textwidth, trim={0 0 2.2in 0}, clip]{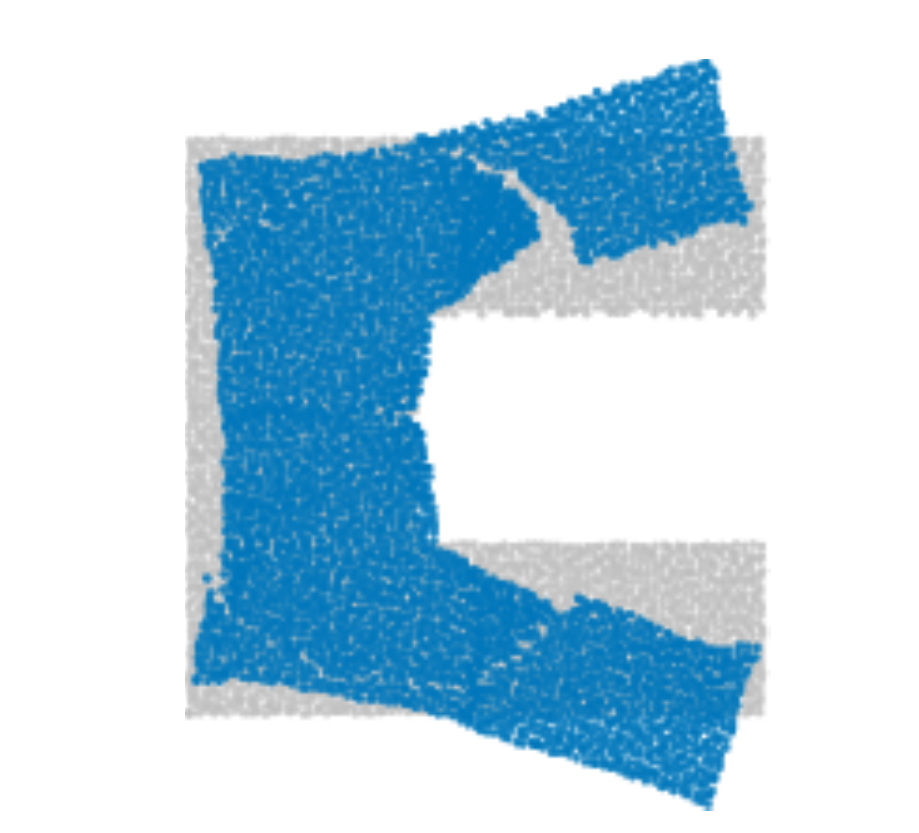}
\caption{Same setting as \figref{cshape_a}. Examples of embeddings with number of hops $h = 1, 2, 5, 10, 20$.}
\label{fig:cshape_b}
\end{figure}

\begin{figure}[ht!]
\centering
\includegraphics[width=.5\textwidth]{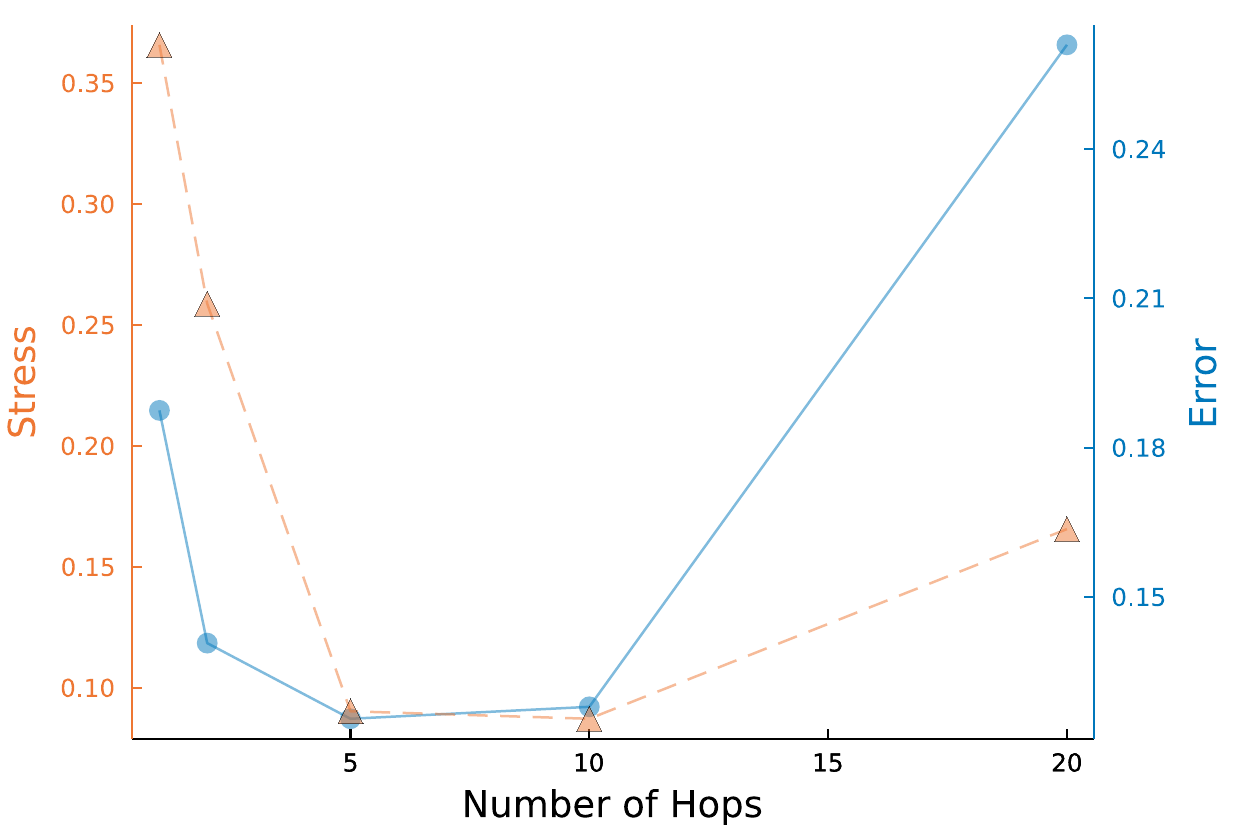}
\caption{Experiment with $n=4278$ points on an H-shaped domain or `dumbbell' with $\sigma=0.15$.}
\label{fig:dumbbell_a}
\end{figure}

\begin{figure}[ht!]
\centering
\includegraphics[width=.18\textwidth, trim={0 0 2.2in 0}, clip]{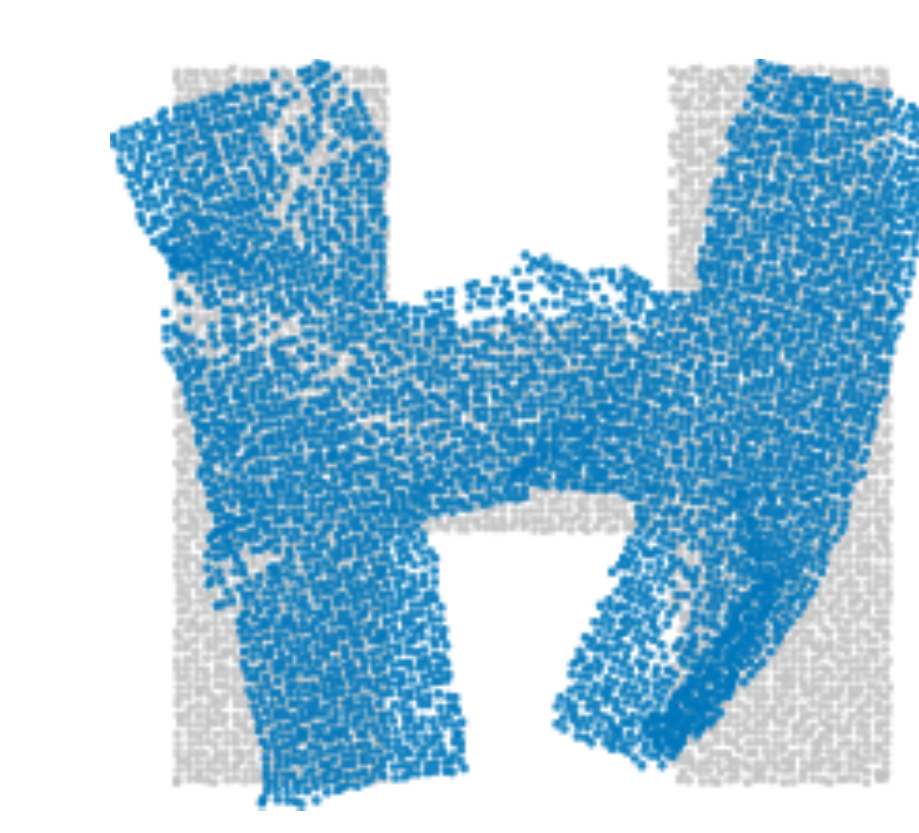}
\includegraphics[width=.18\textwidth, trim={0 0 2.2in 0}, clip]{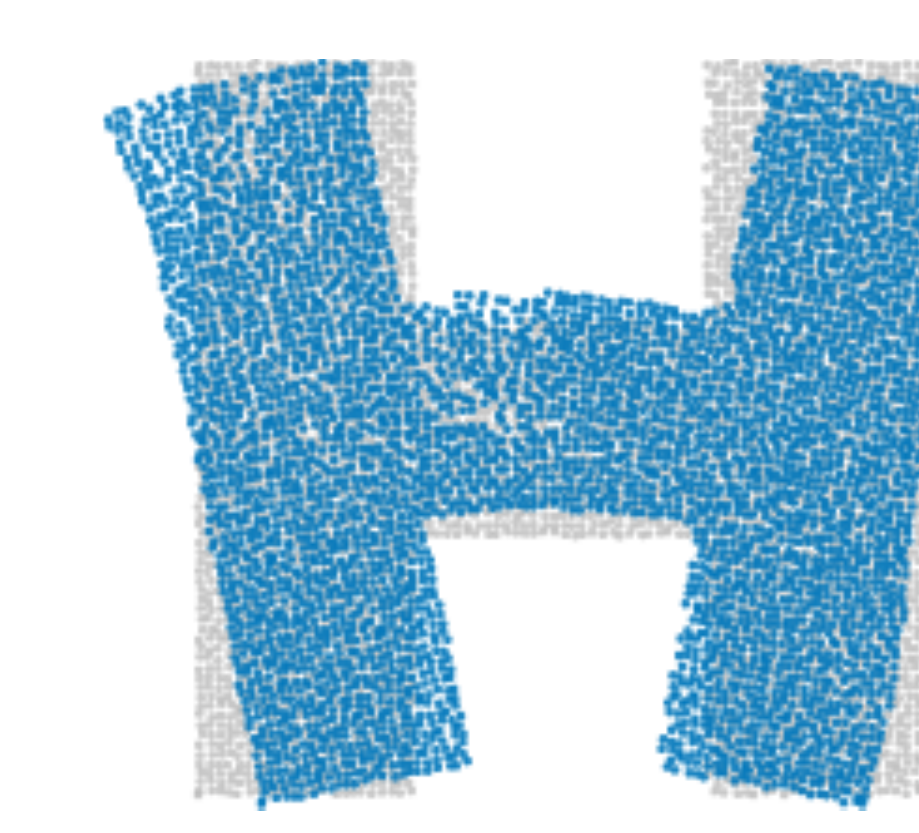}
\includegraphics[width=.18\textwidth, trim={0 0 2.2in 0}, clip]{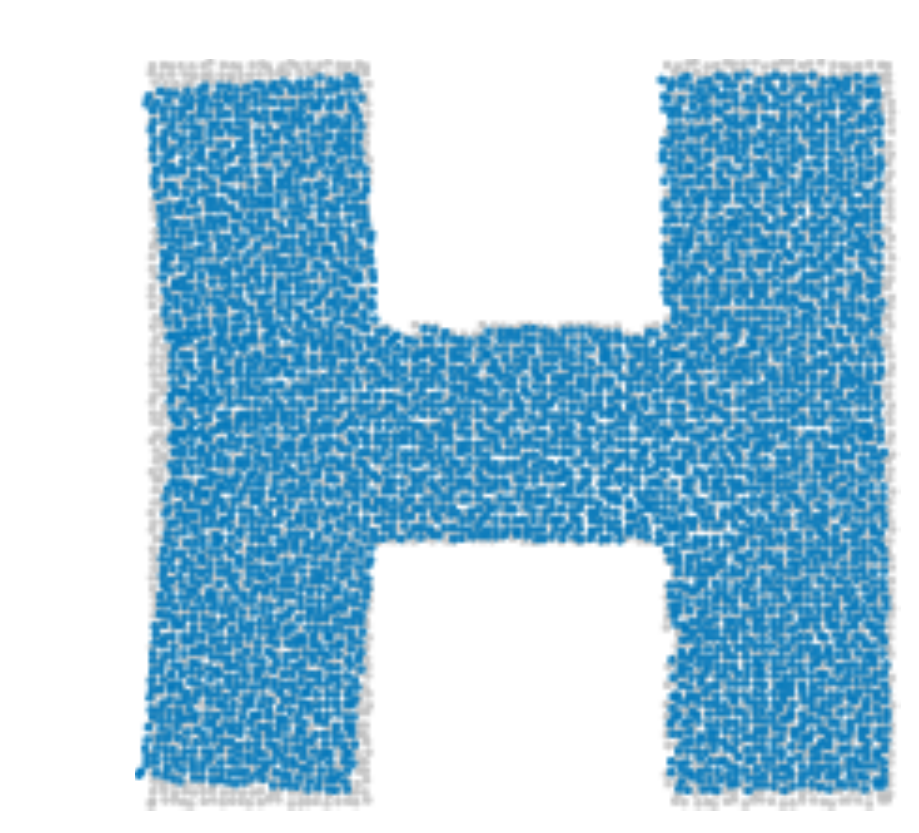}
\includegraphics[width=.18\textwidth, trim={0 0 2.2in 0}, clip]{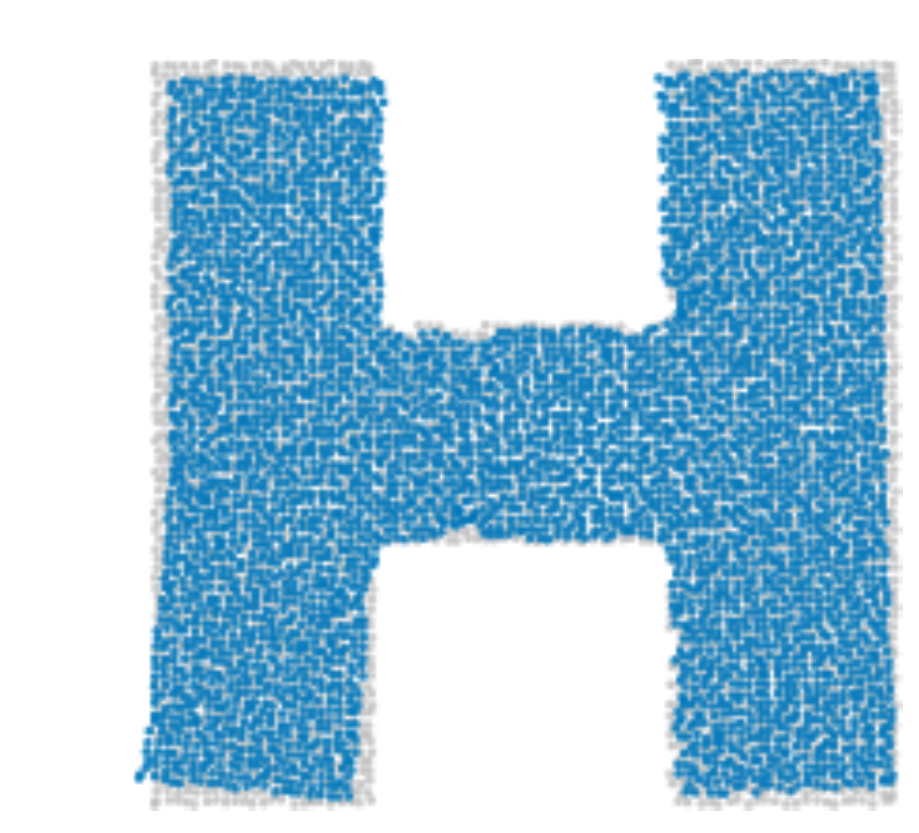}\includegraphics[width=.18\textwidth, trim={0 0 2.2in 0}, clip]{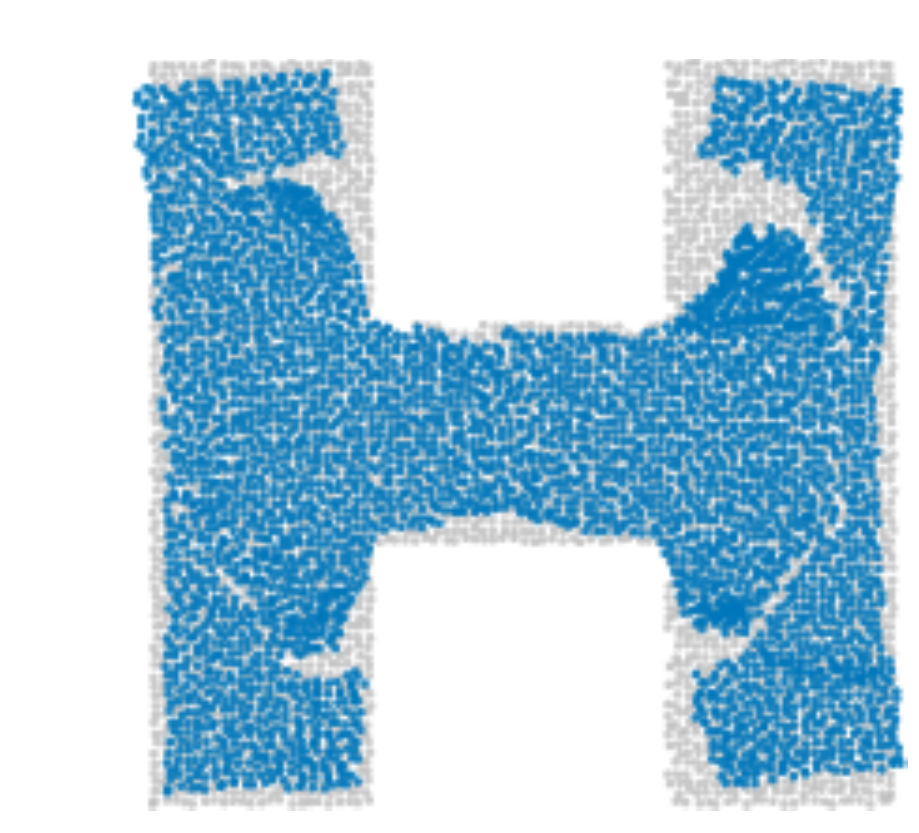}
\caption{Same setting as \figref{dumbbell_a}. Examples of embeddings with number of hops $h = 1, 2, 5, 10, 20$.}
\label{fig:dumbbell_b}
\end{figure}

\begin{figure}
\centering
\begin{tabular}{cccc}
\includegraphics[width=.15\linewidth]{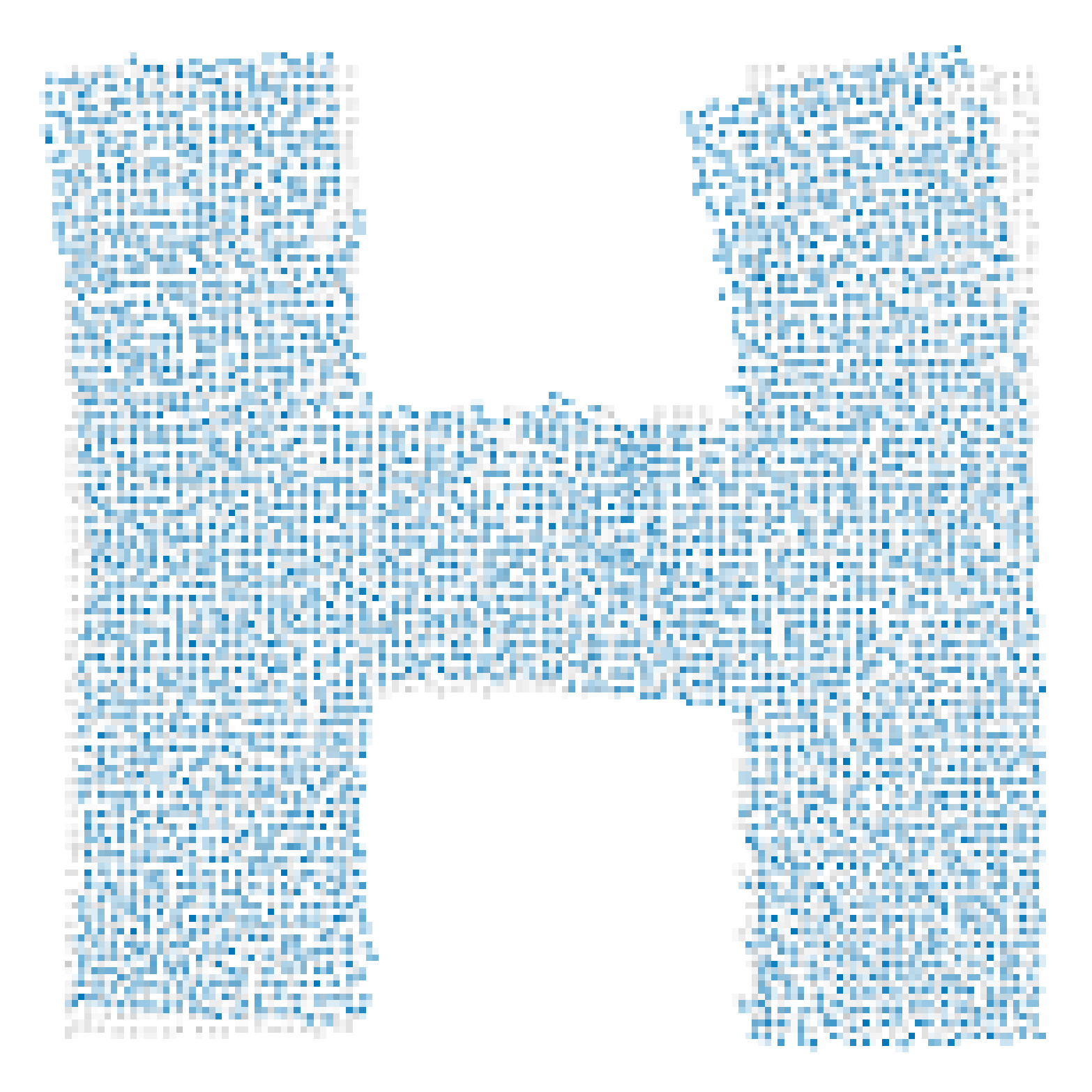} &
\includegraphics[width=.15\linewidth]{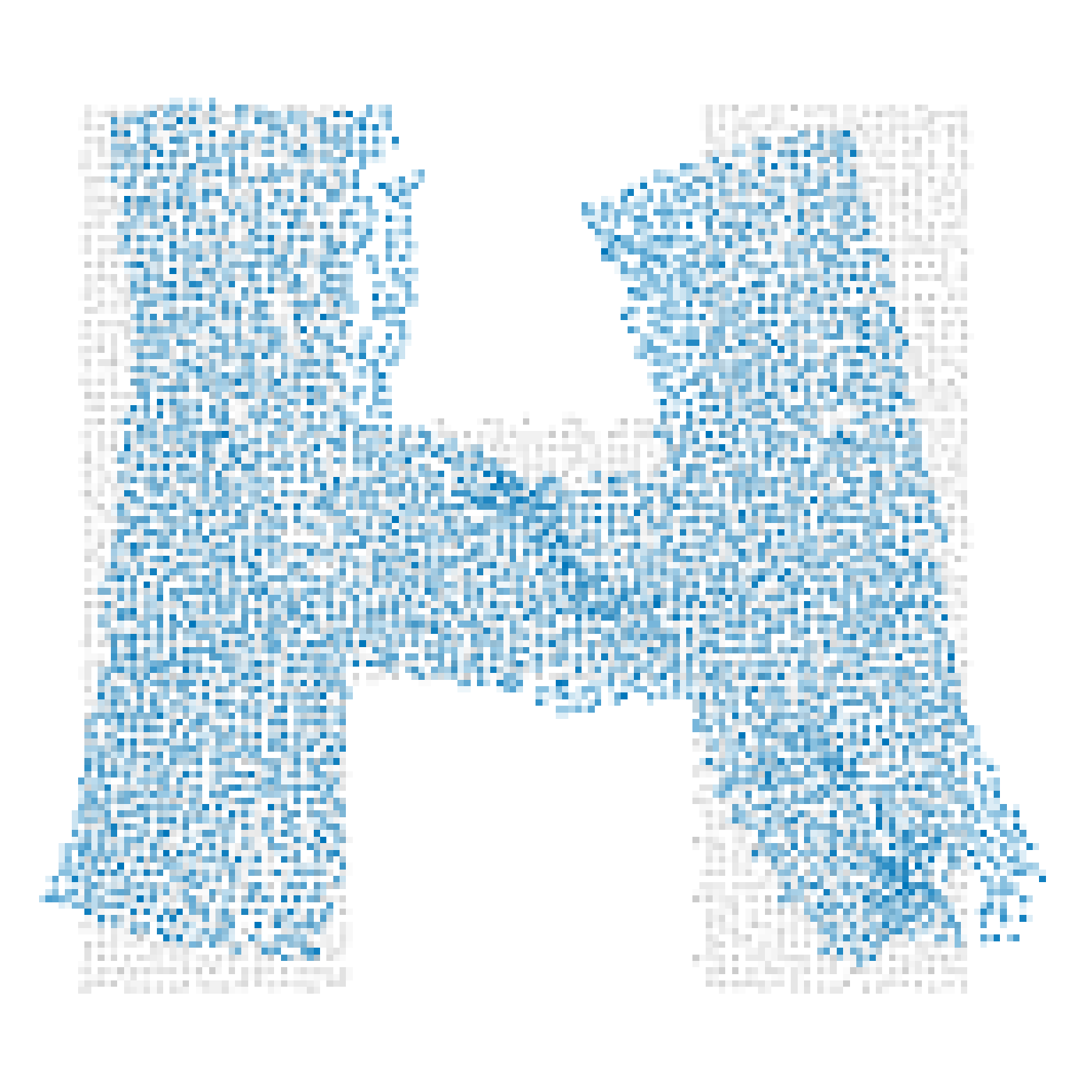} &
\includegraphics[width=.15\linewidth]{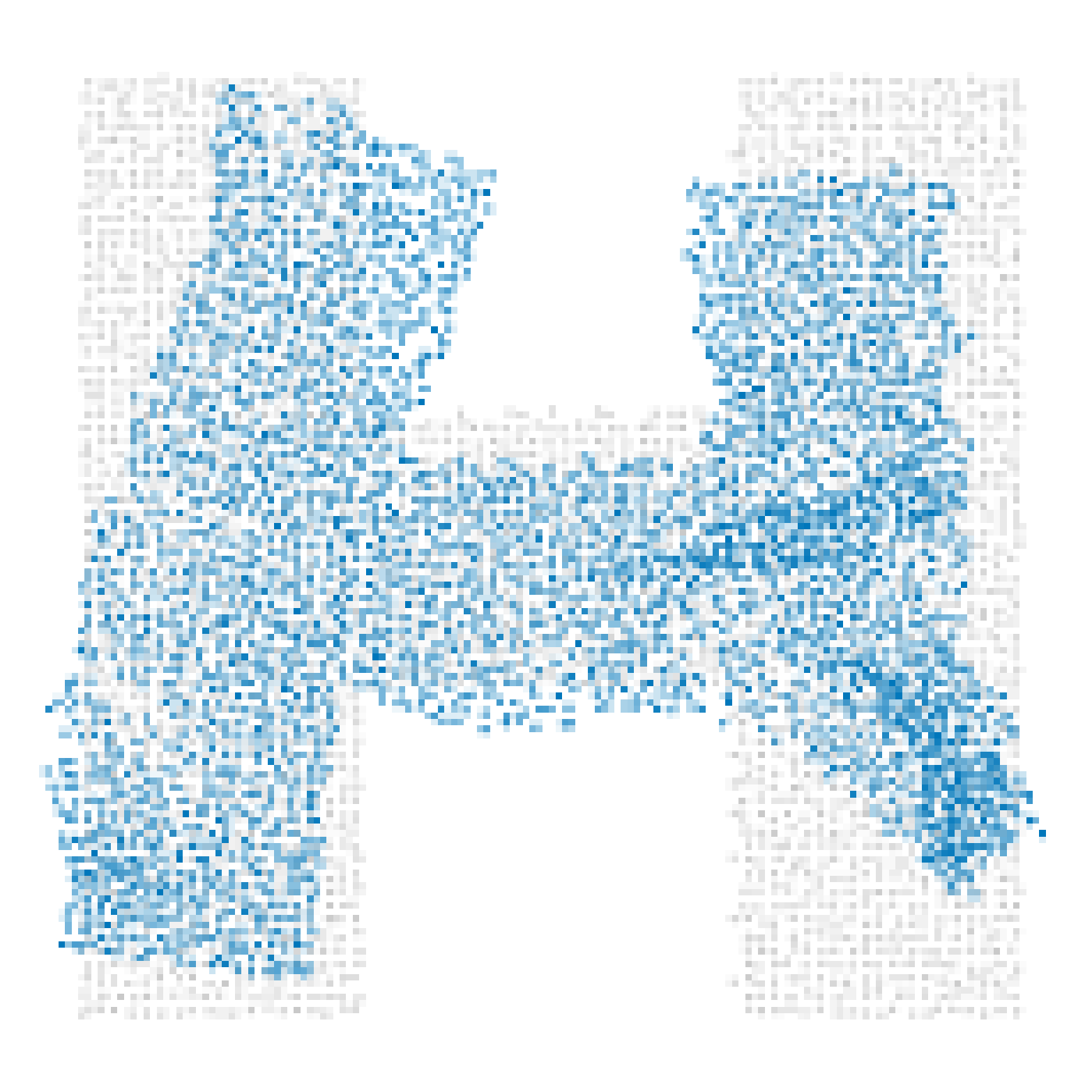} &
\includegraphics[width=.15\linewidth]{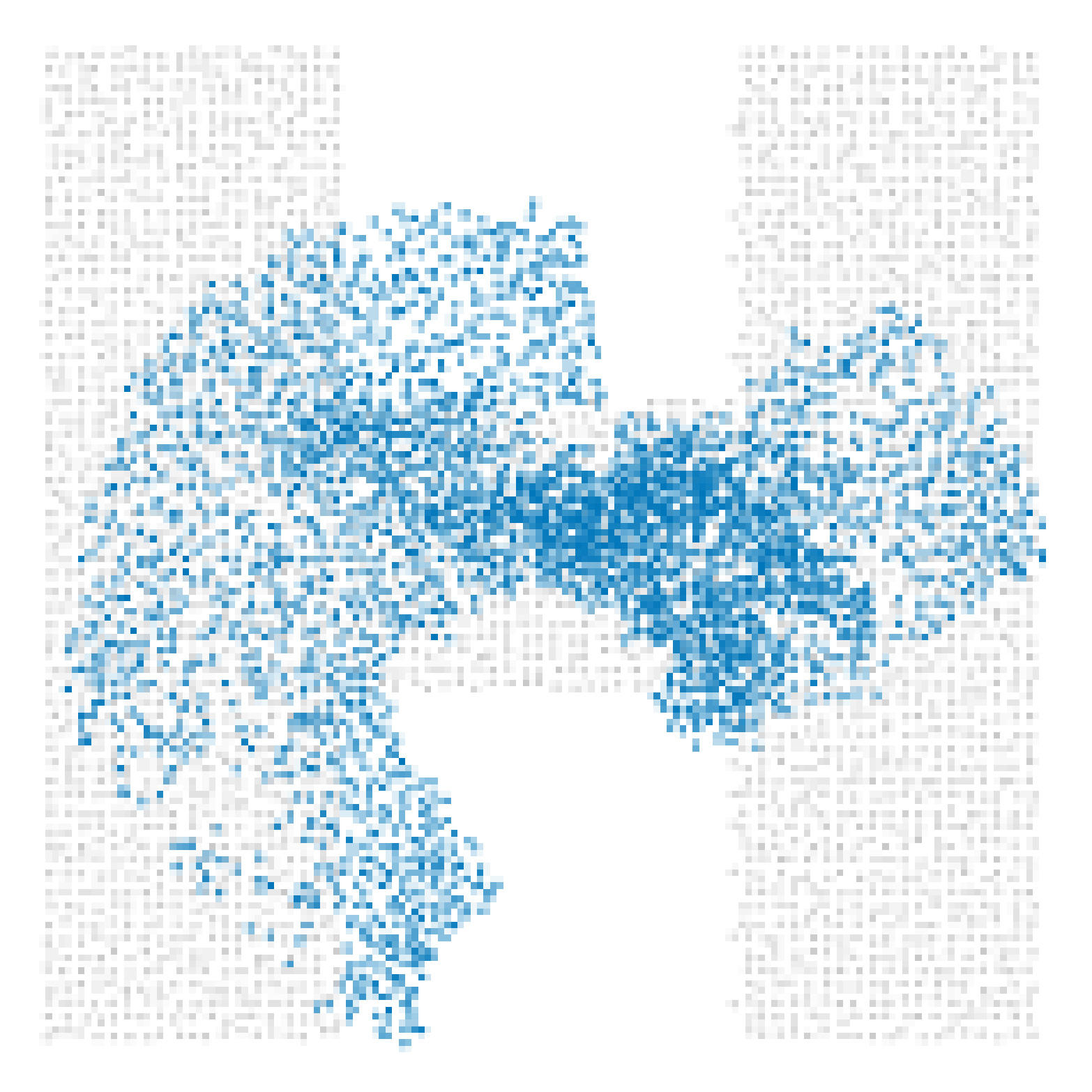} \\
\includegraphics[width=.15\linewidth]{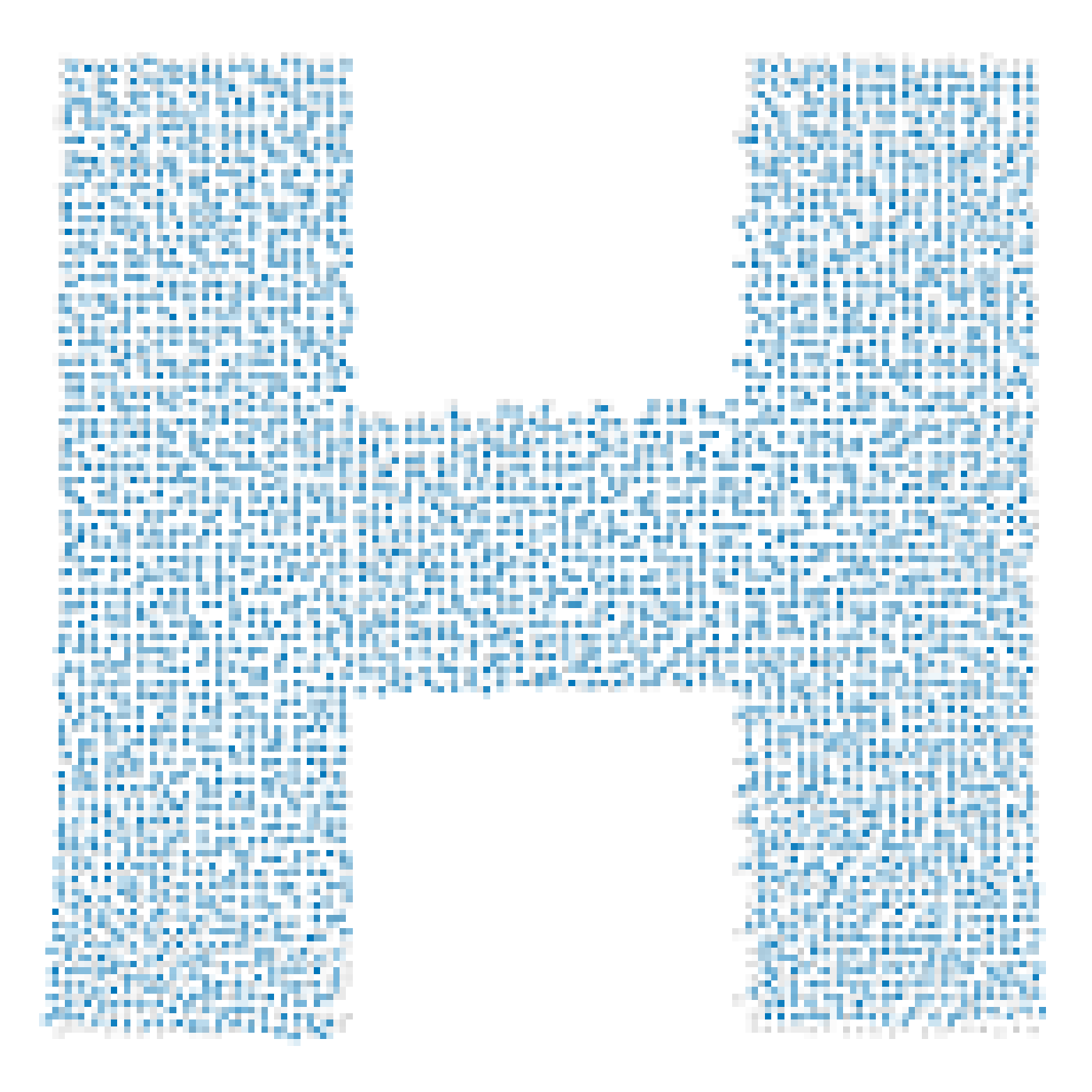} &
\includegraphics[width=.15\linewidth]{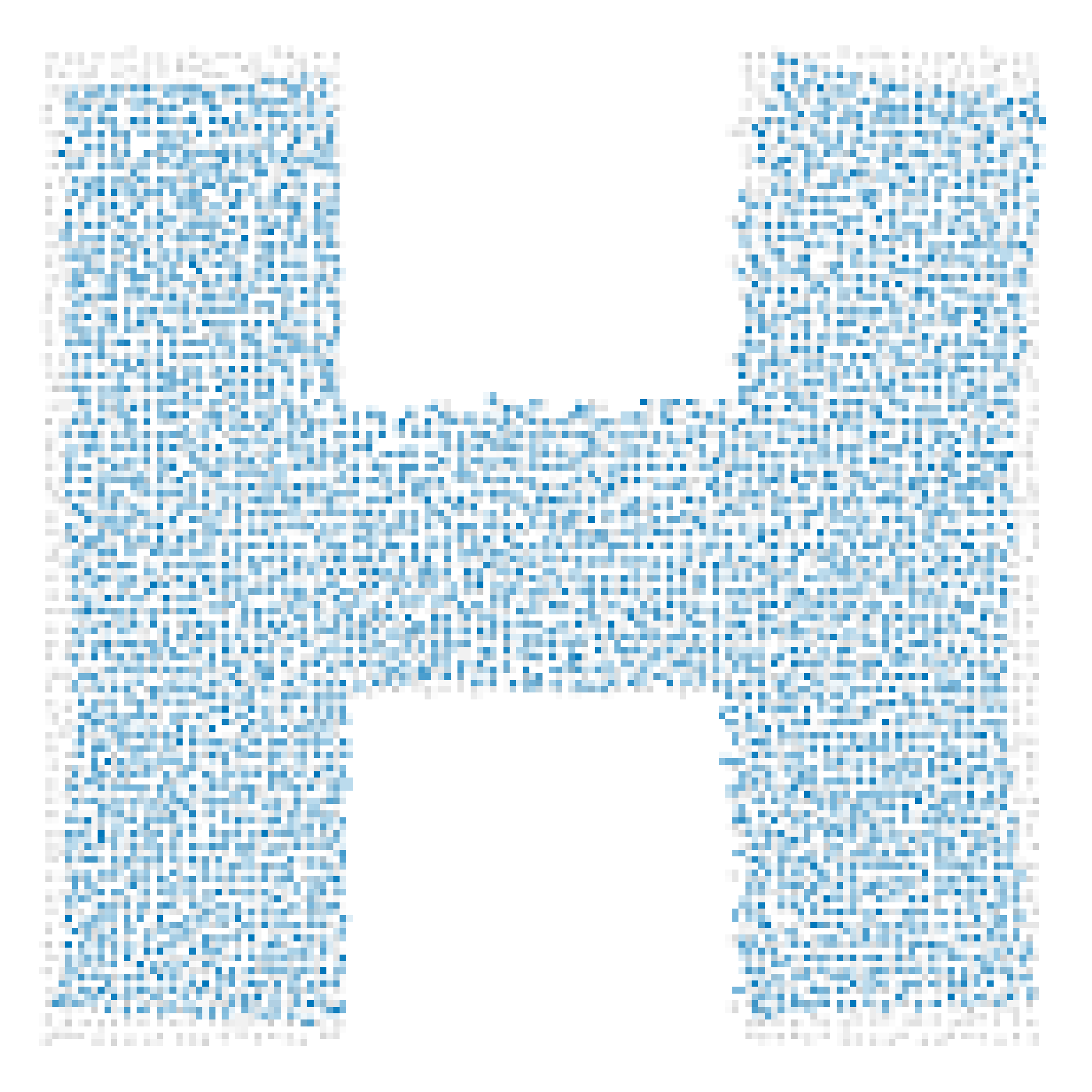} &
\includegraphics[width=.15\linewidth]{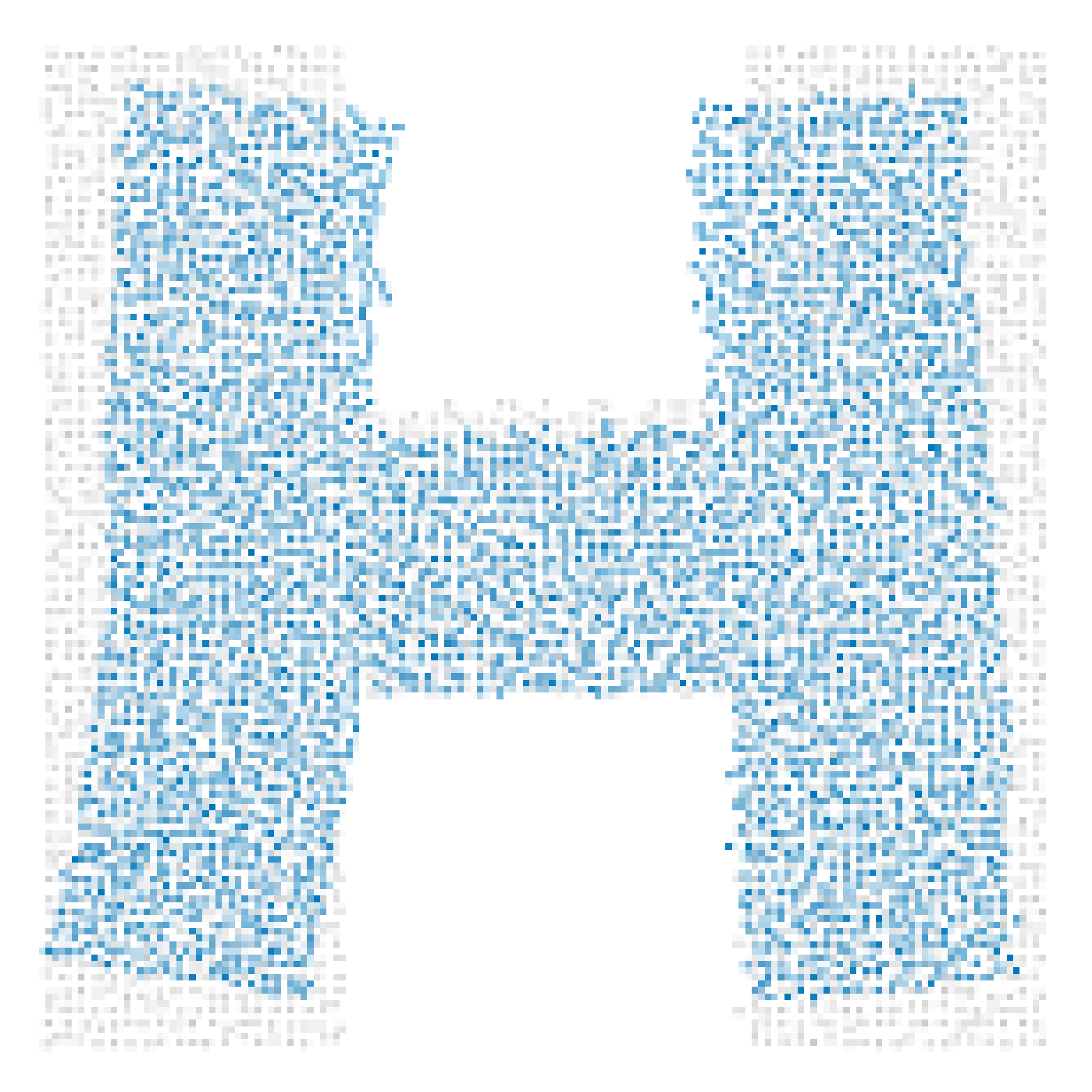} &
\includegraphics[width=.15\linewidth]{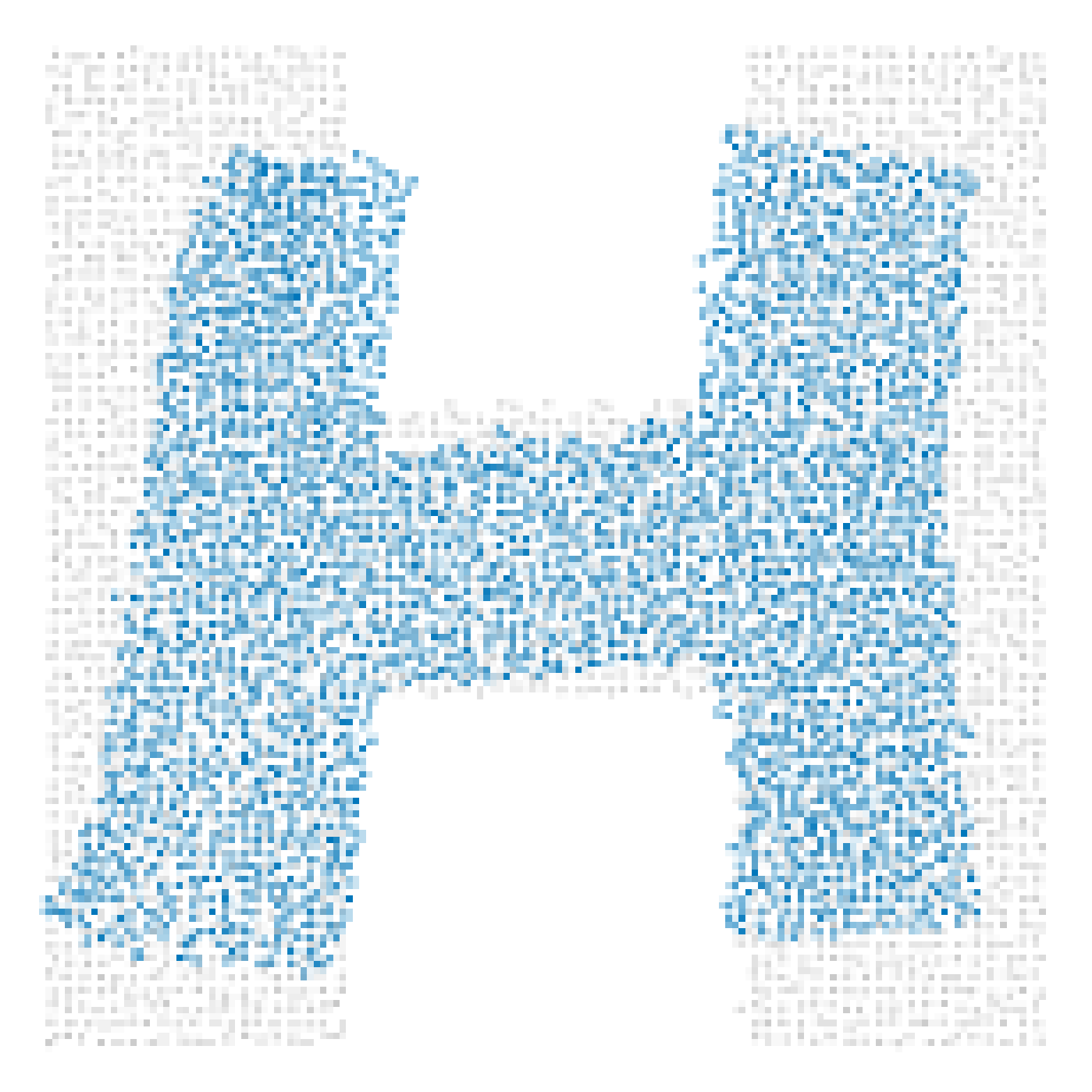} \\
\includegraphics[width=.15\linewidth]{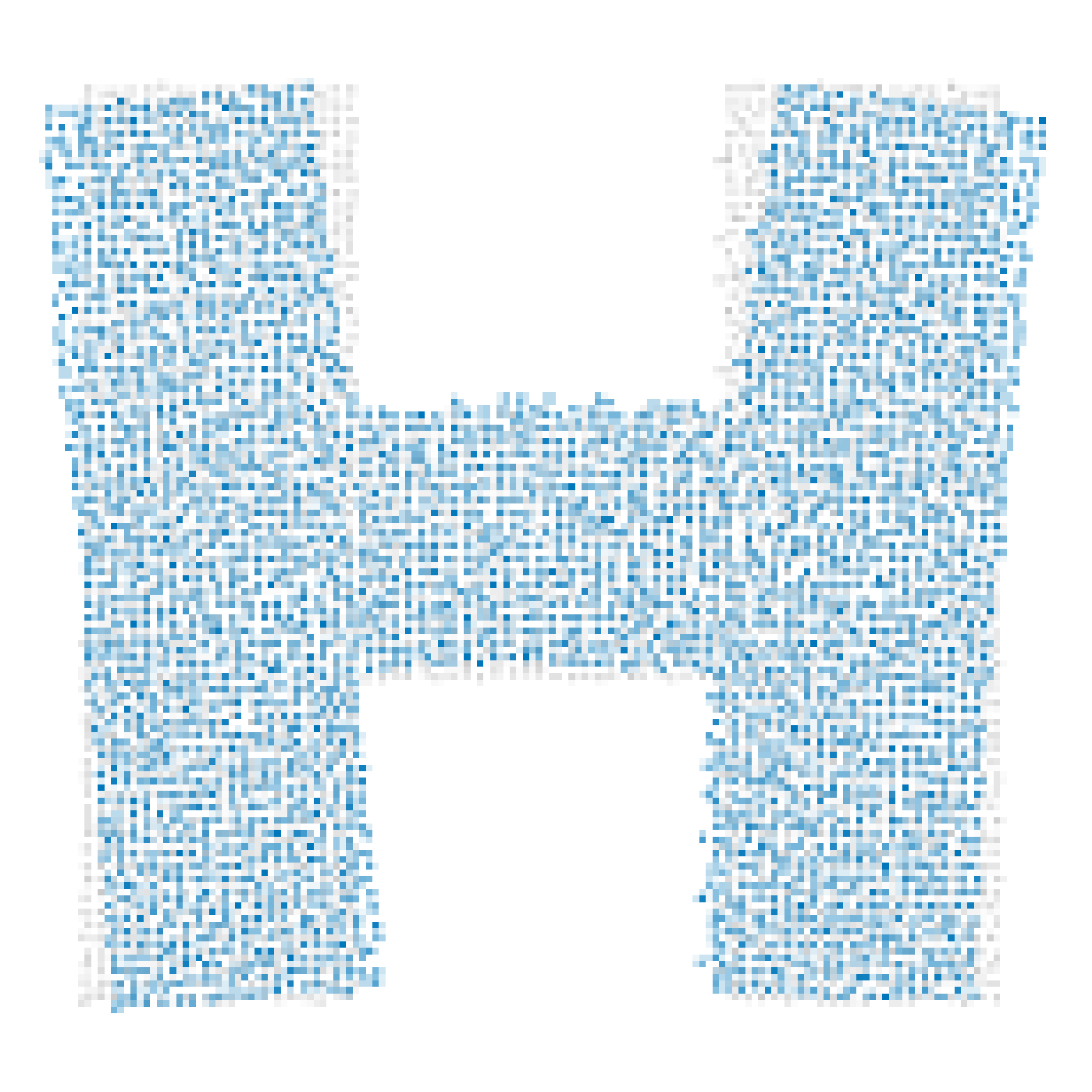} &
\includegraphics[width=.15\linewidth]{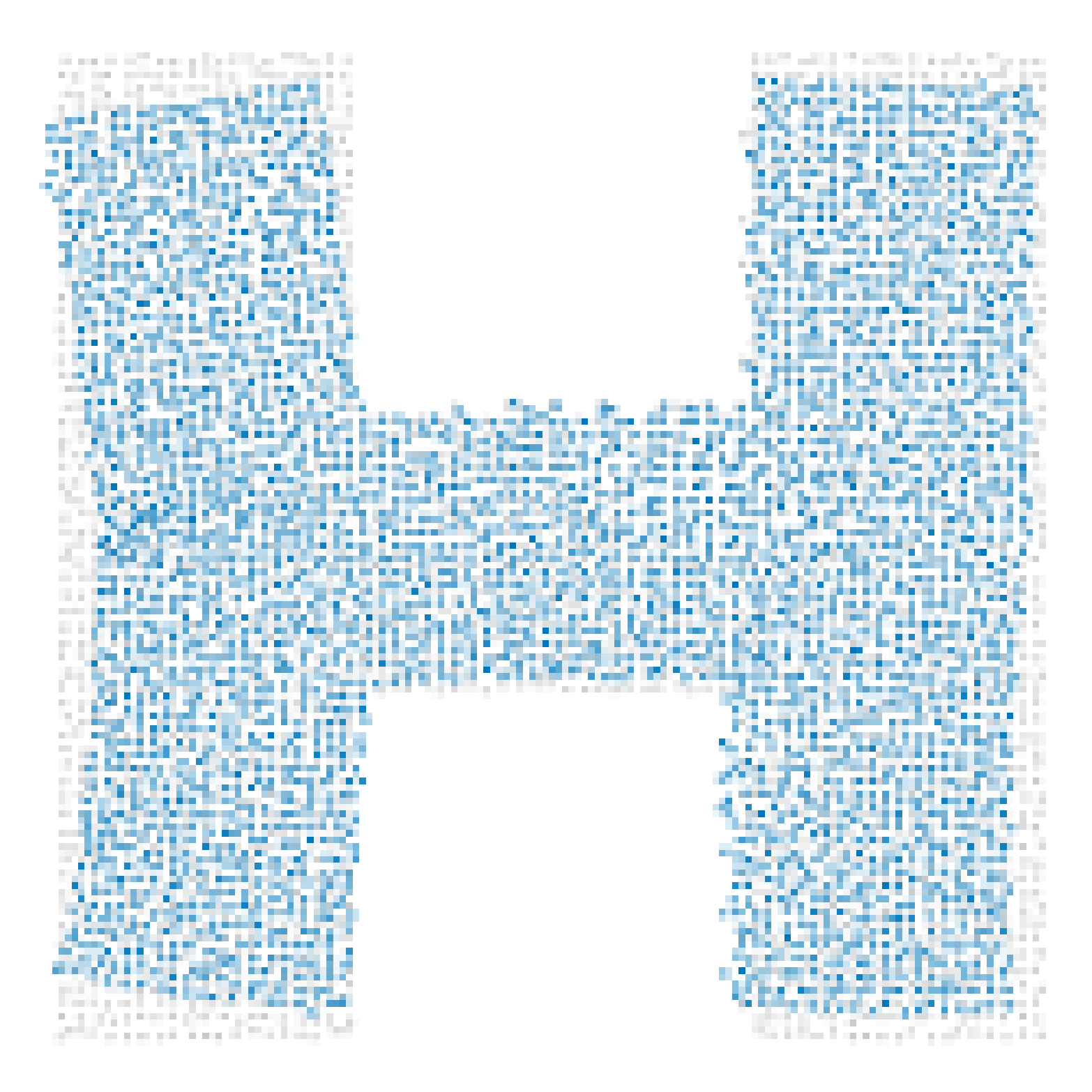} &
\includegraphics[width=.15\linewidth]{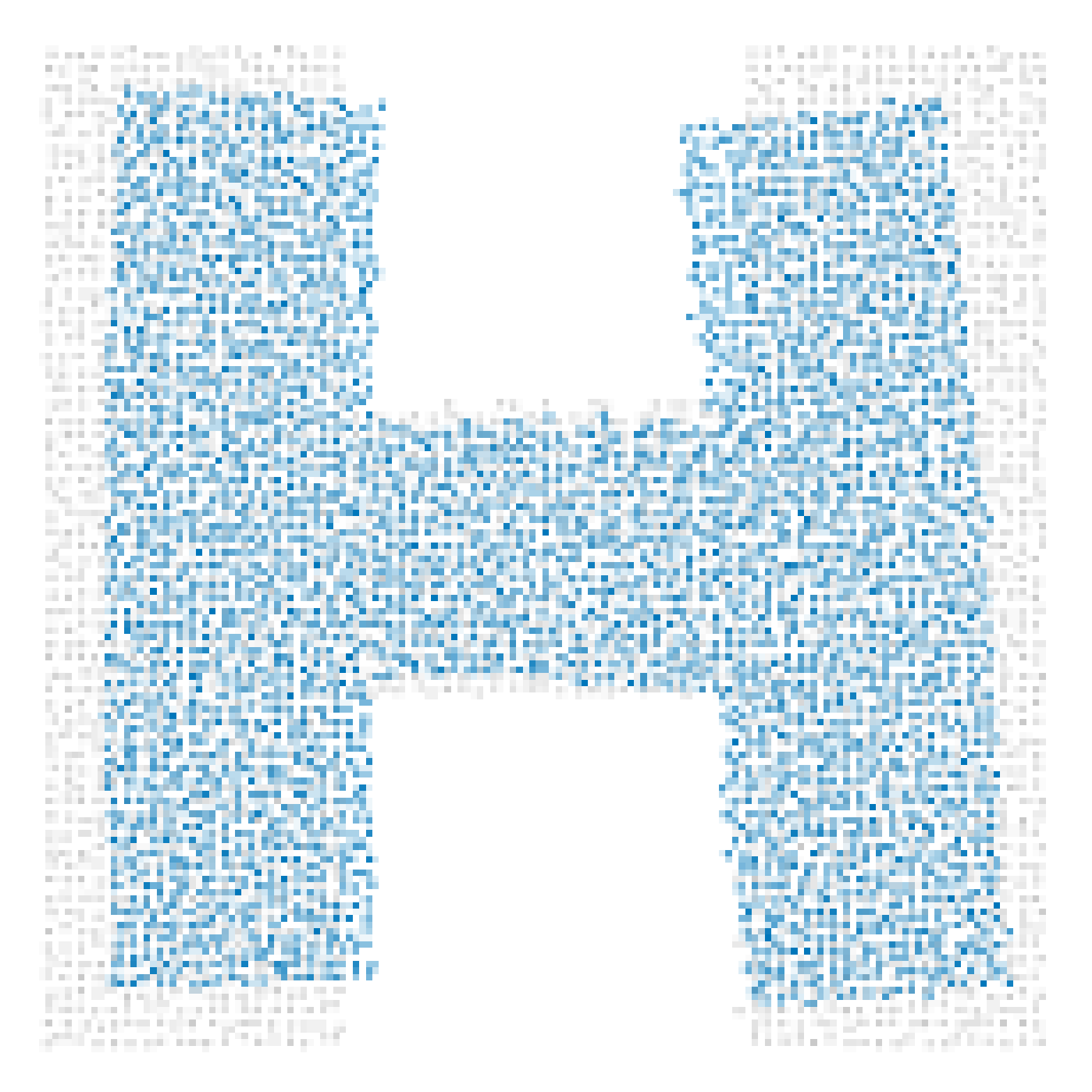} &
\includegraphics[width=.15\linewidth]{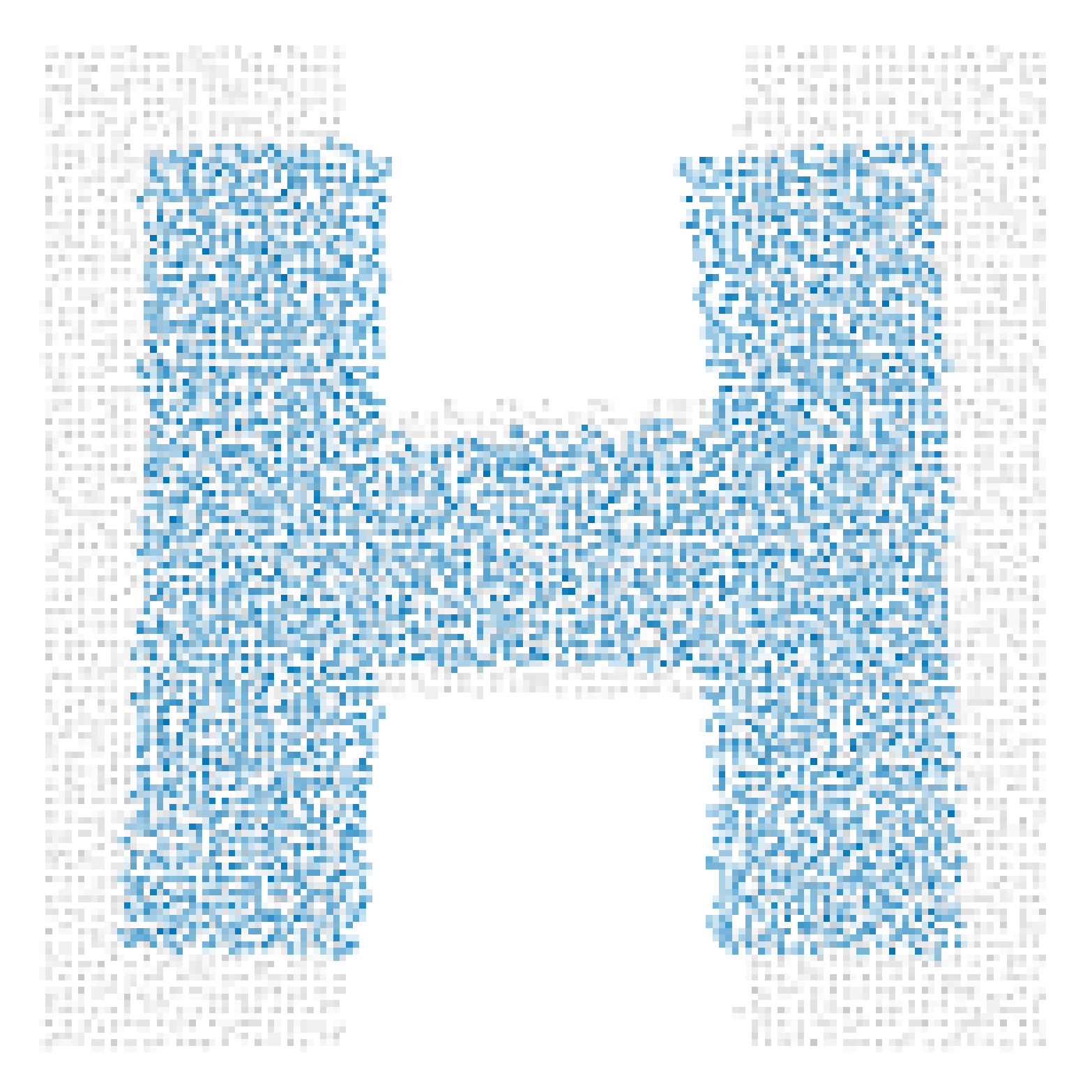} \\
\includegraphics[width=.15\linewidth]{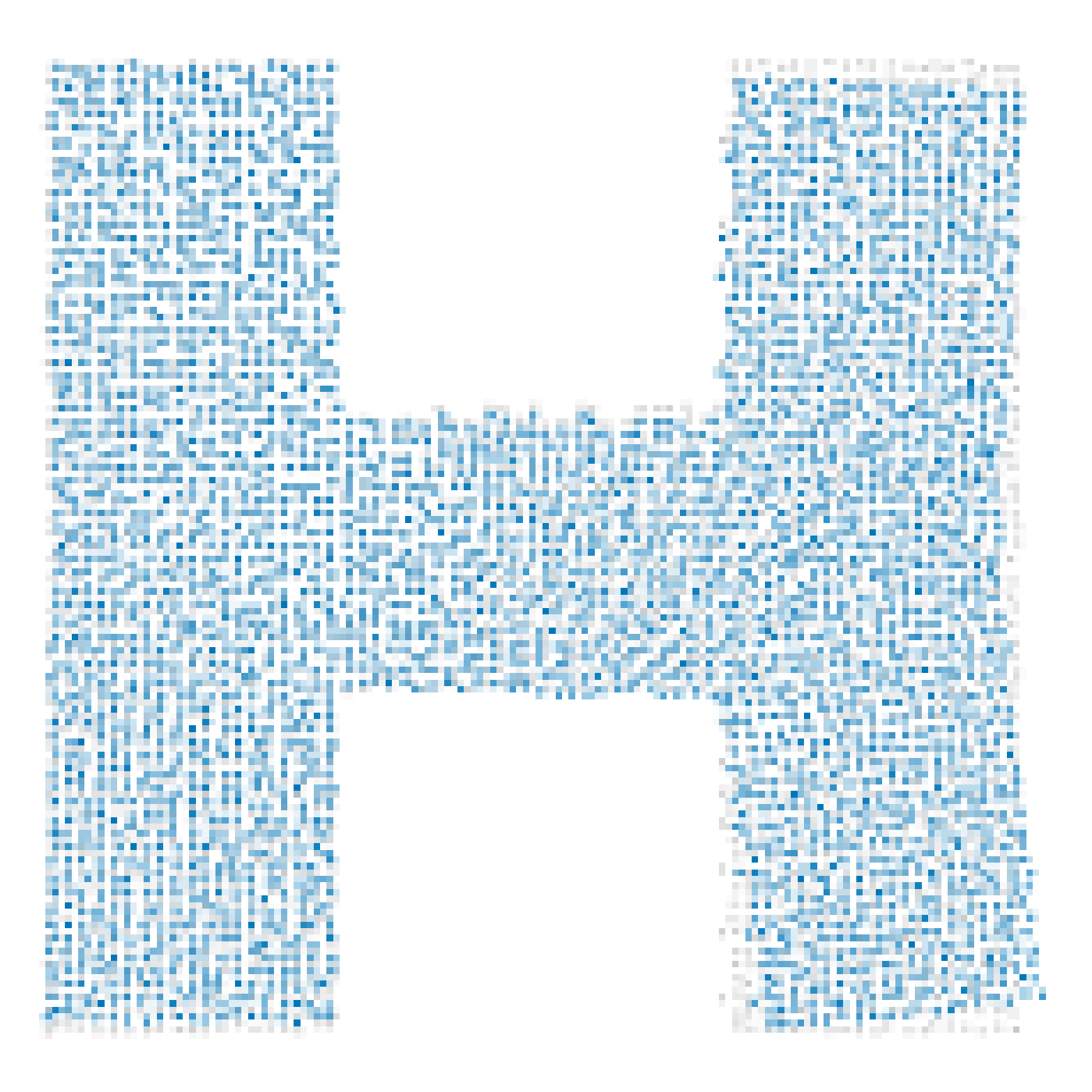} &
\includegraphics[width=.15\linewidth]{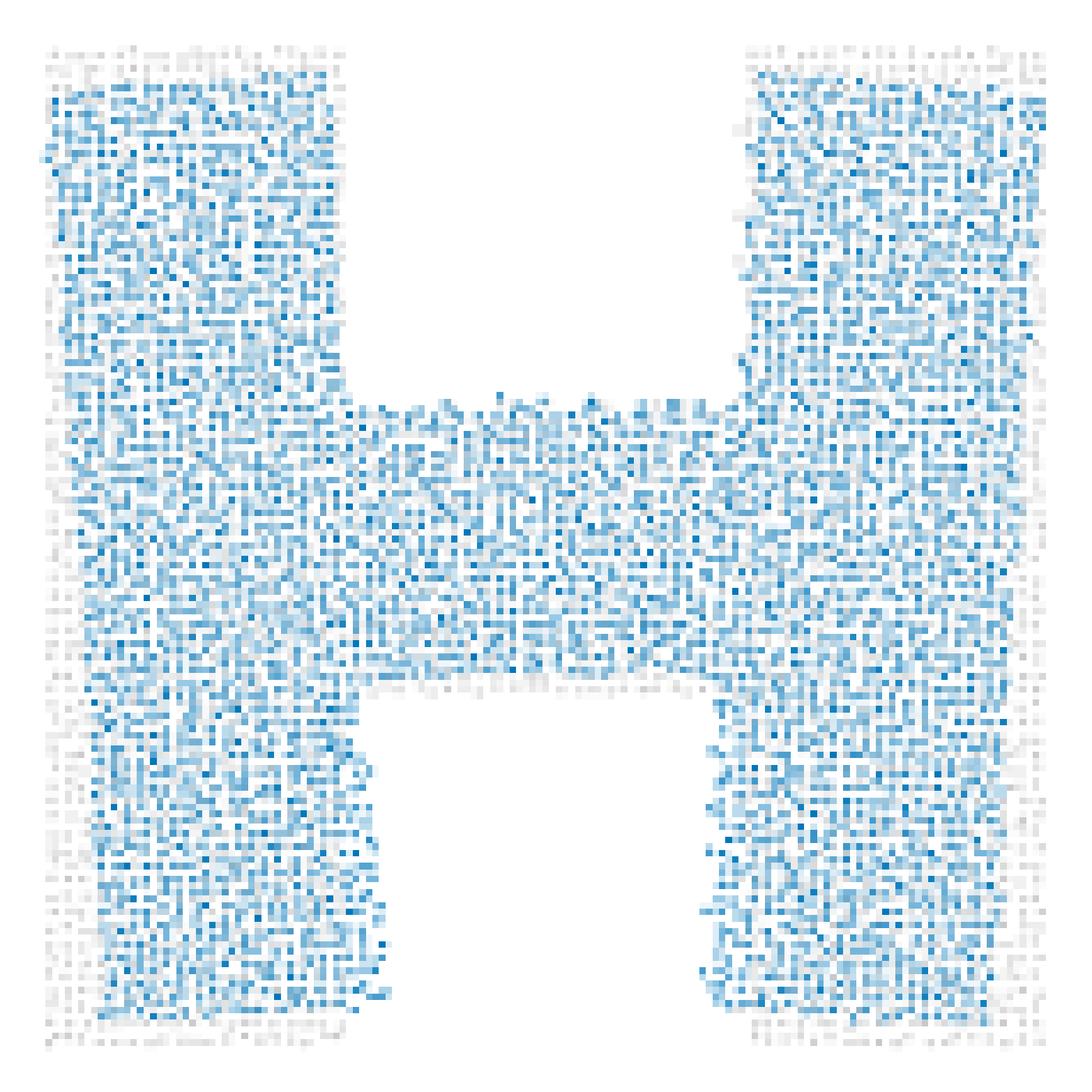} &
\includegraphics[width=.15\linewidth]{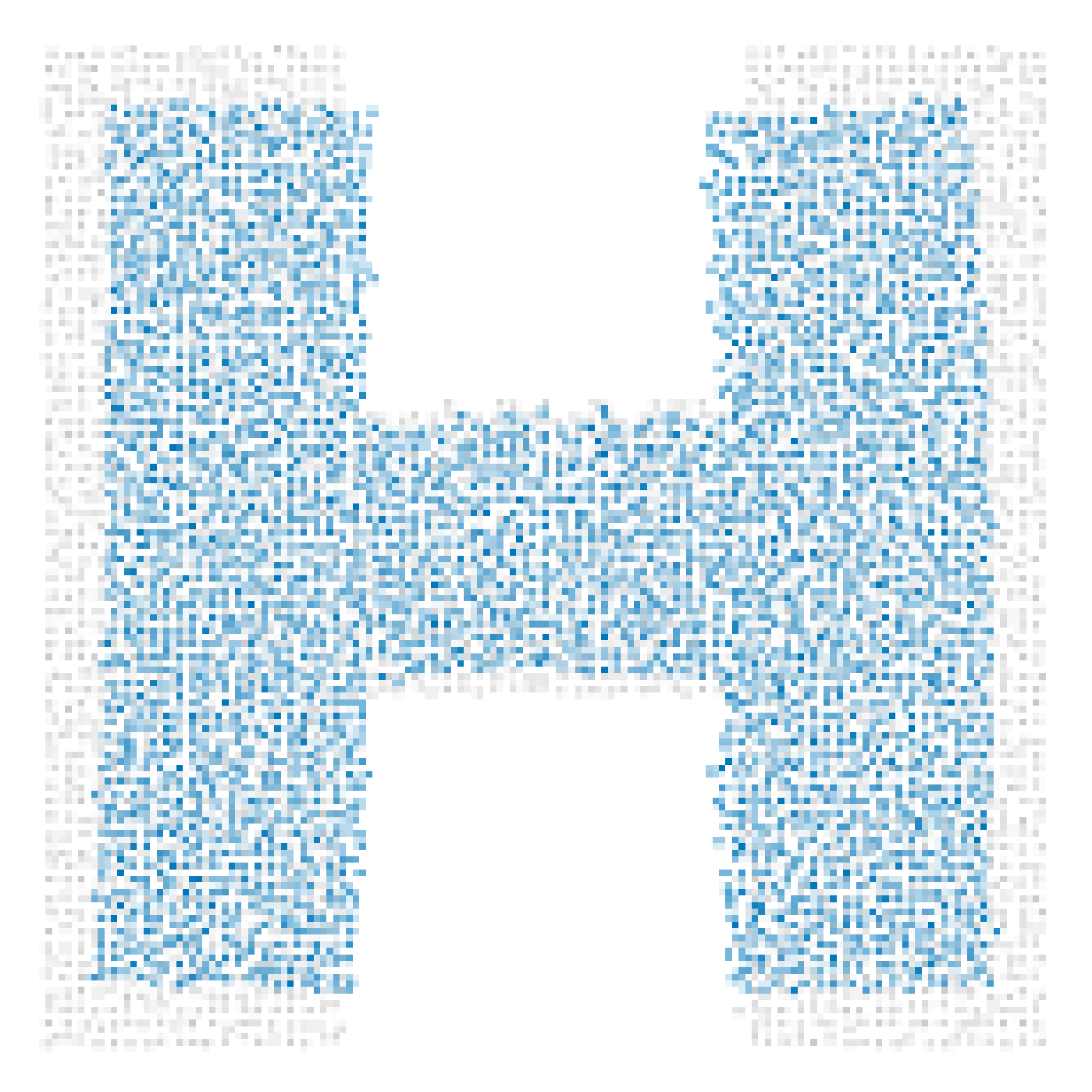} &
\includegraphics[width=.15\linewidth]{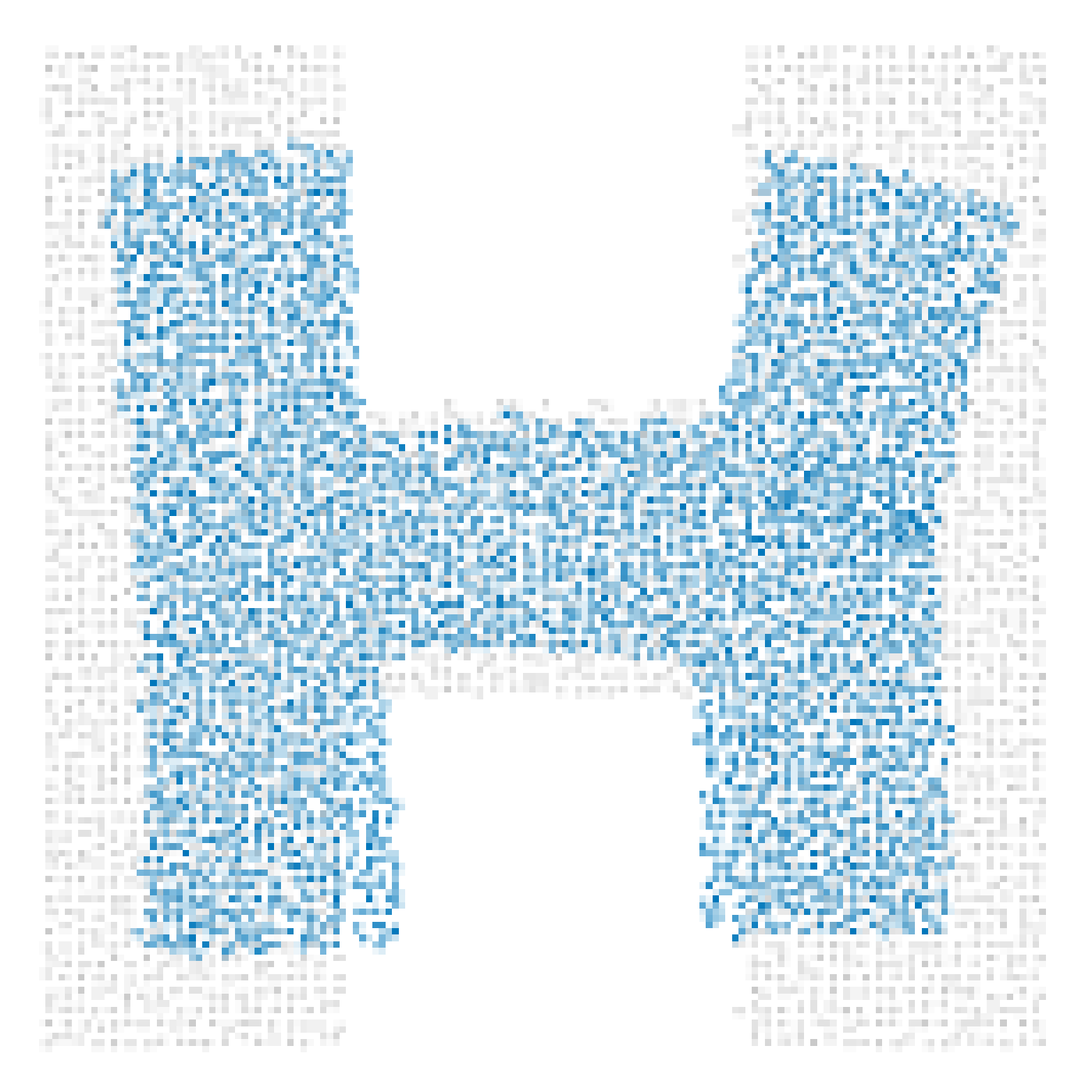} \\
\includegraphics[width=.15\linewidth]{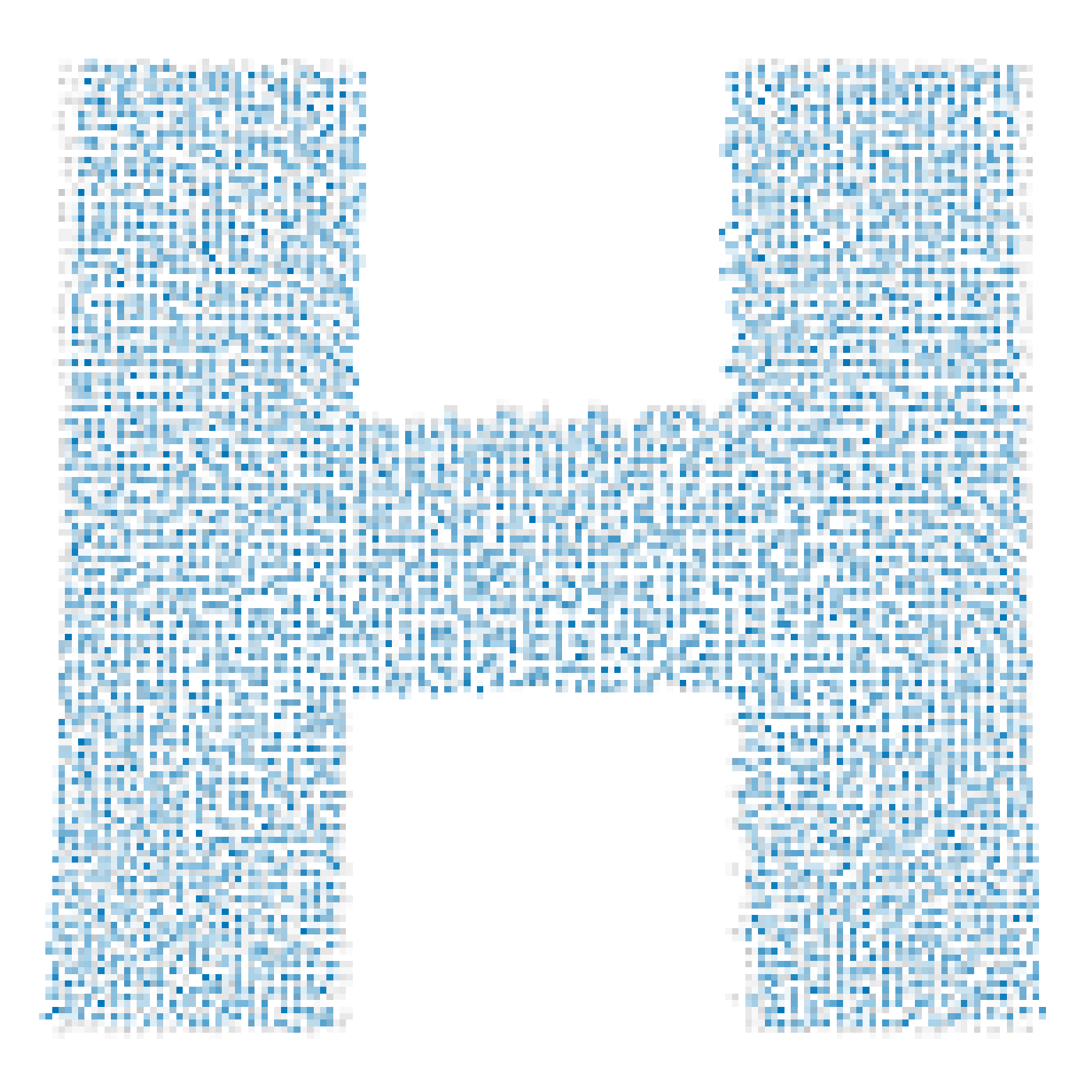} &
\includegraphics[width=.15\linewidth]{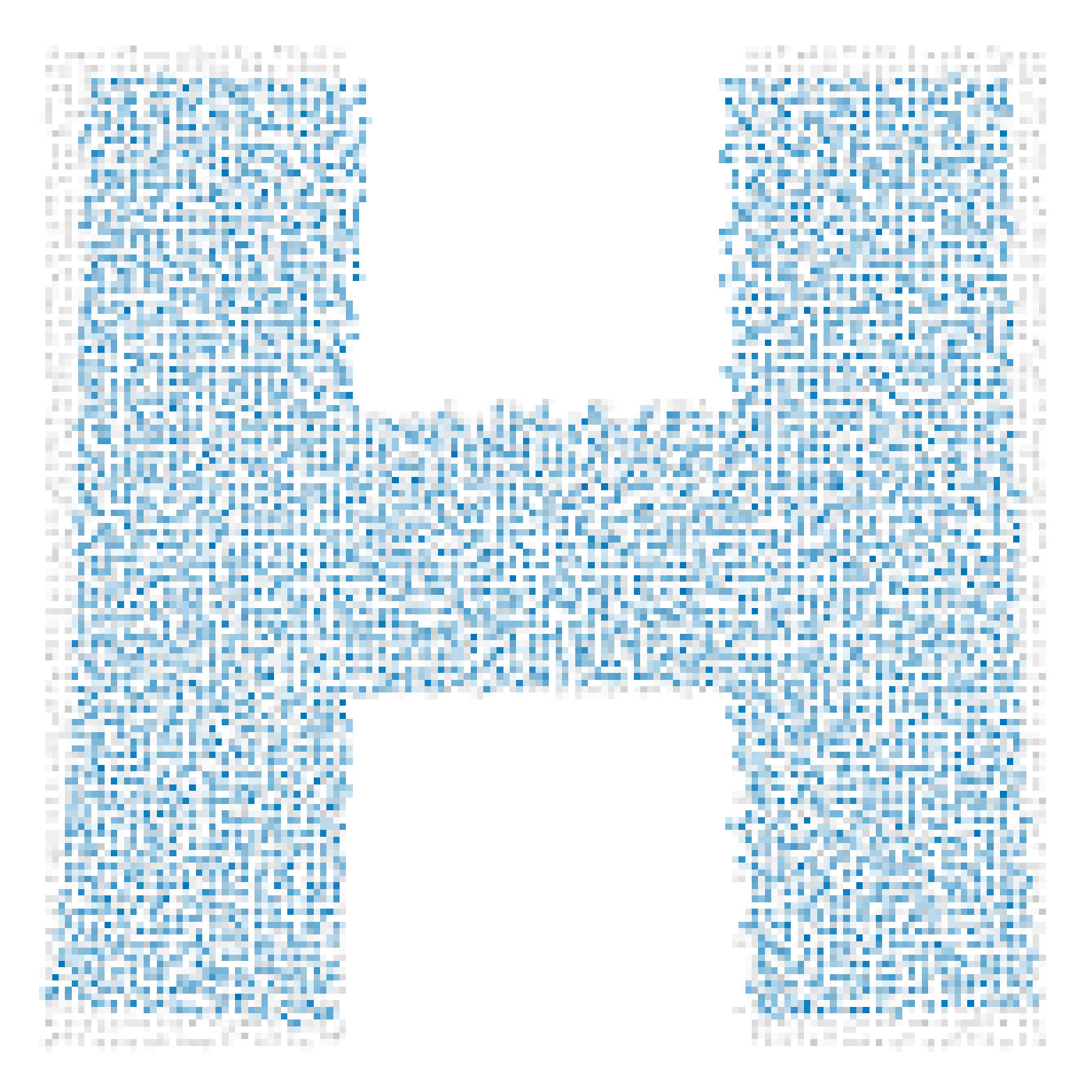} &
\includegraphics[width=.15\linewidth]{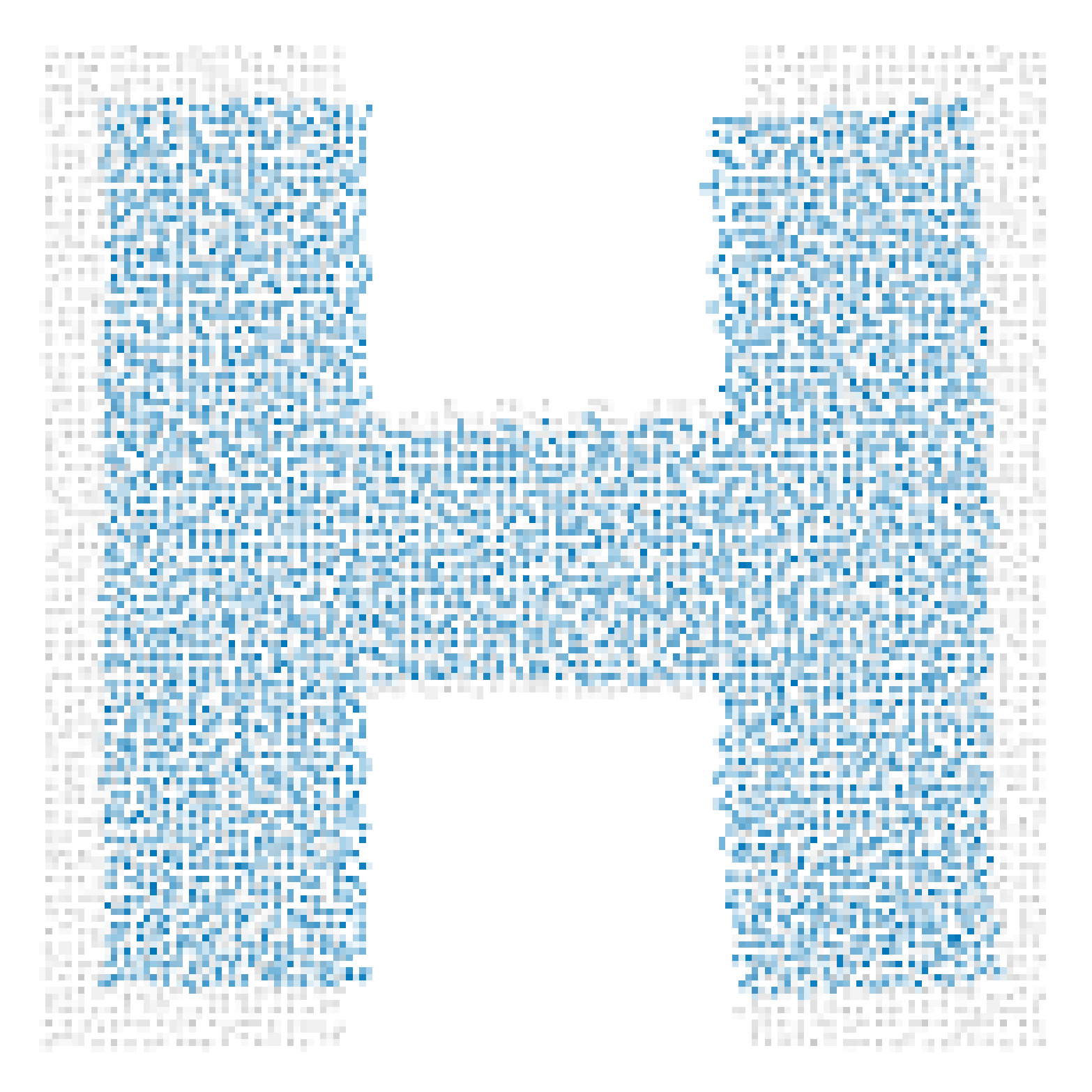} &
\includegraphics[width=.15\linewidth]{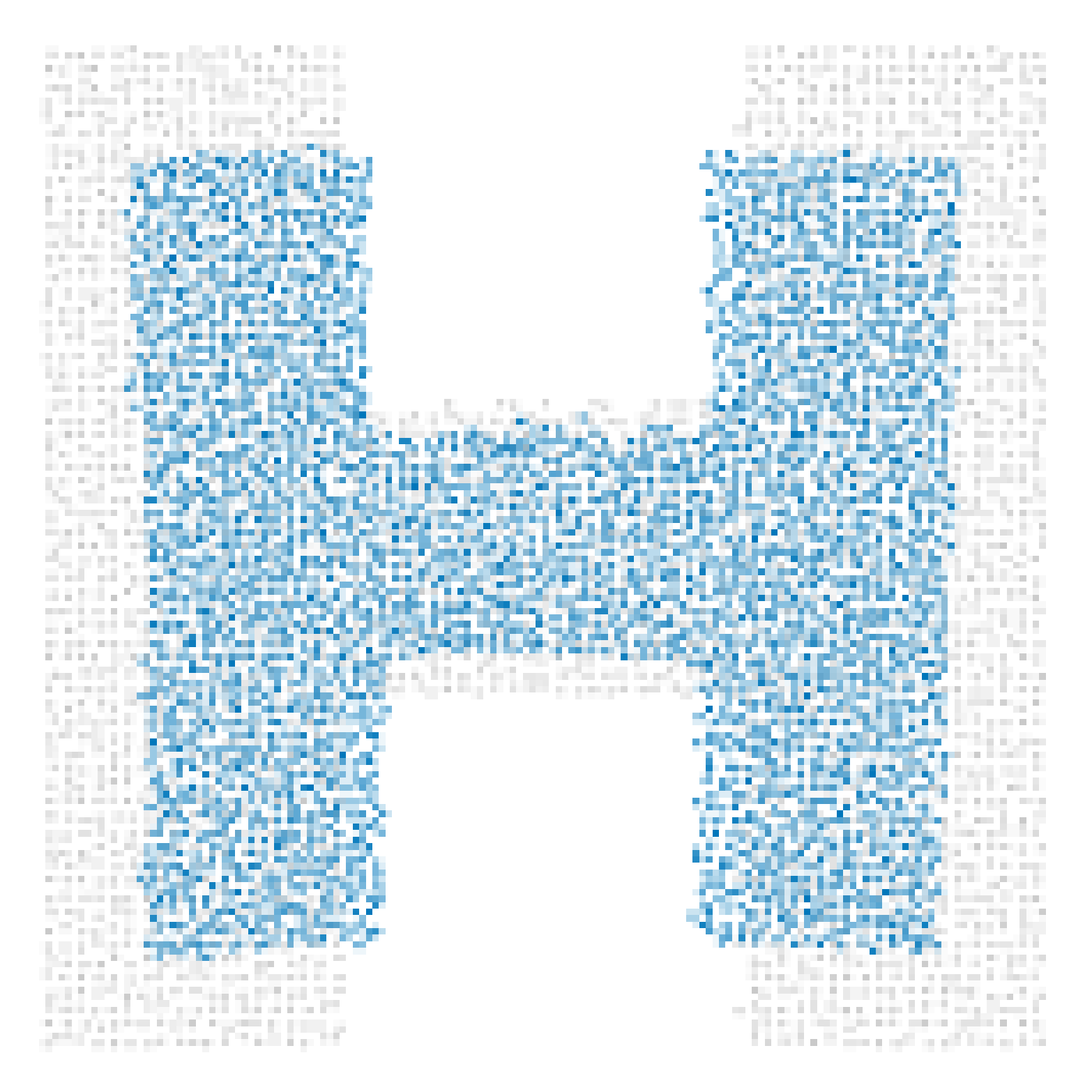} \\
\includegraphics[width=.15\linewidth]{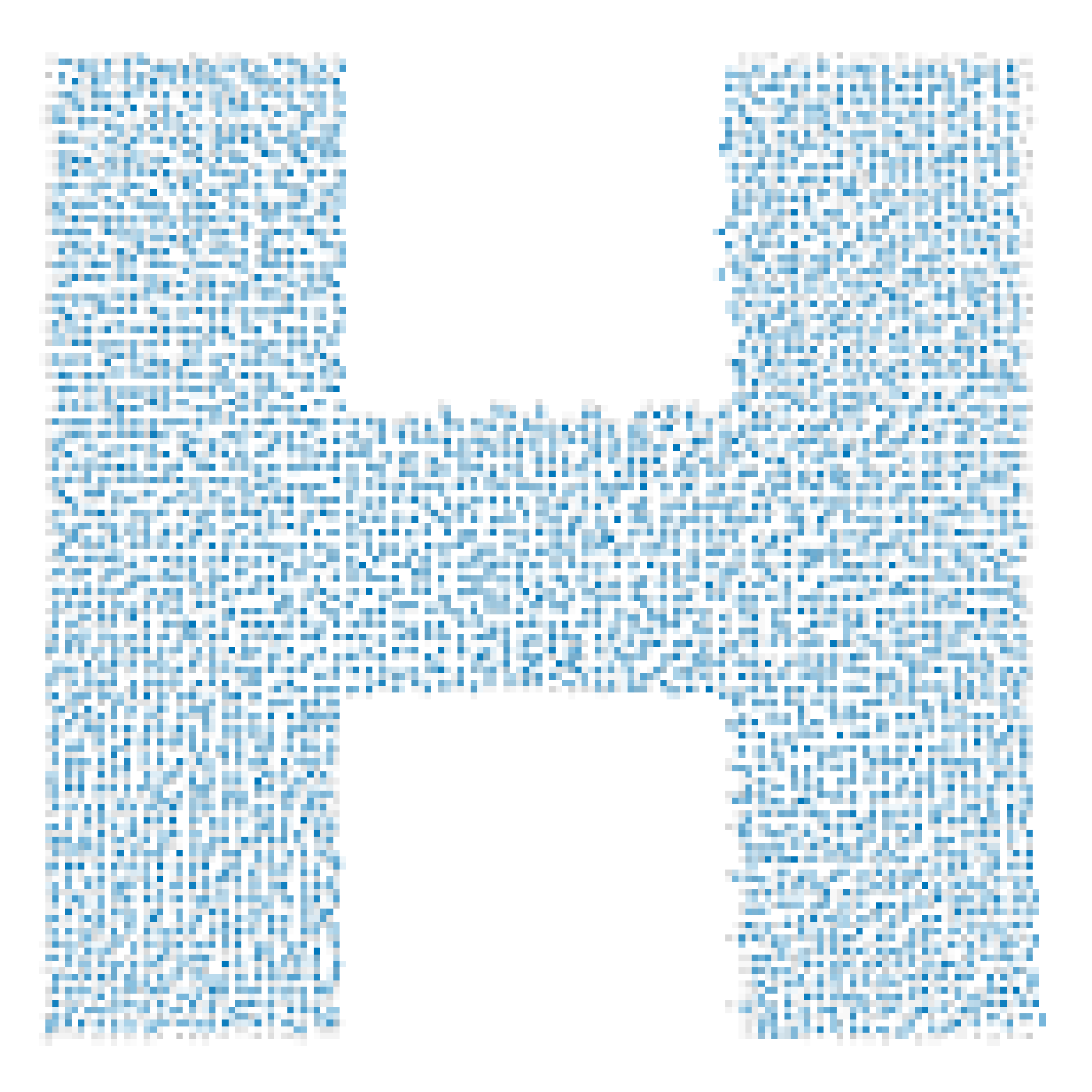} &
\includegraphics[width=.15\linewidth]{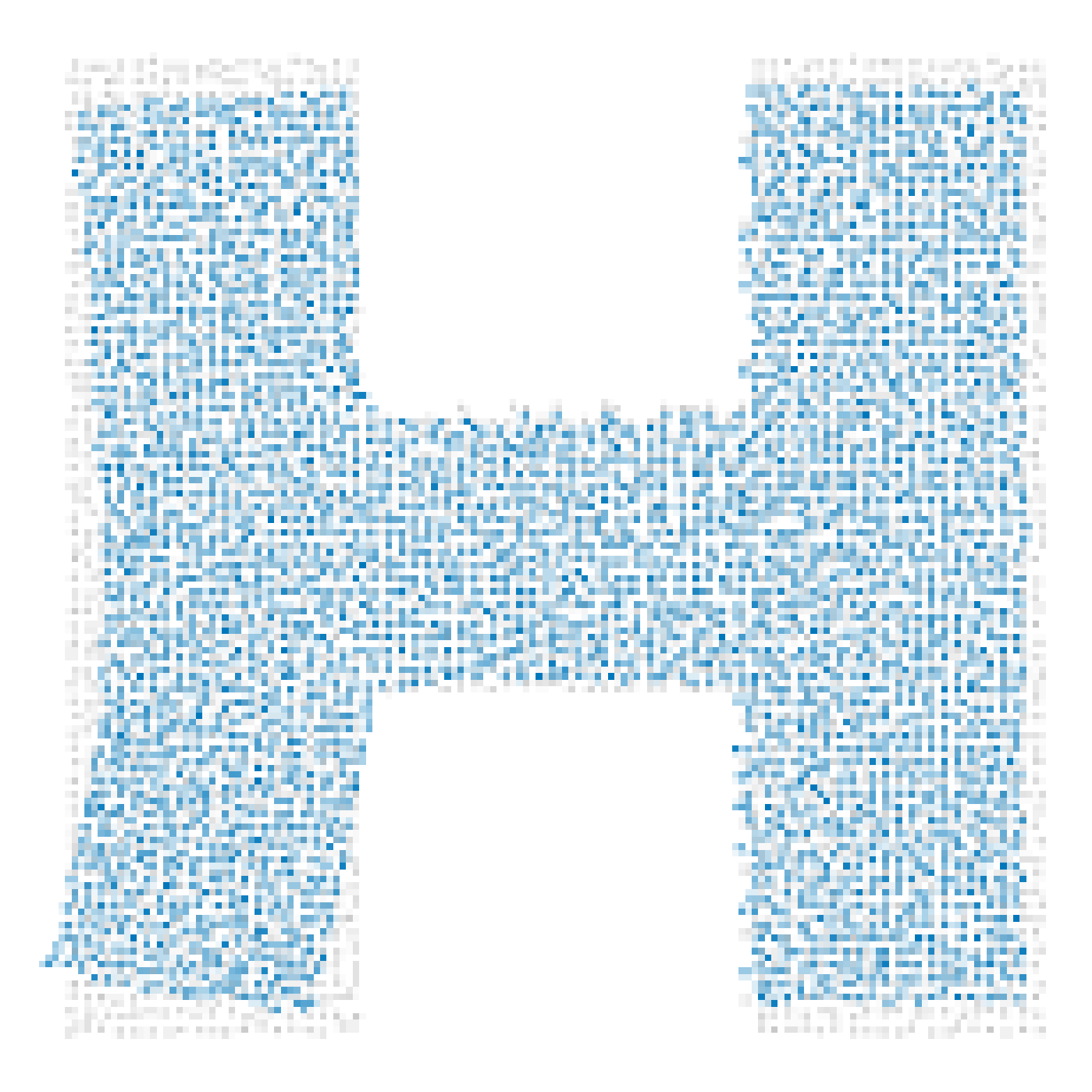} &
\includegraphics[width=.15\linewidth]{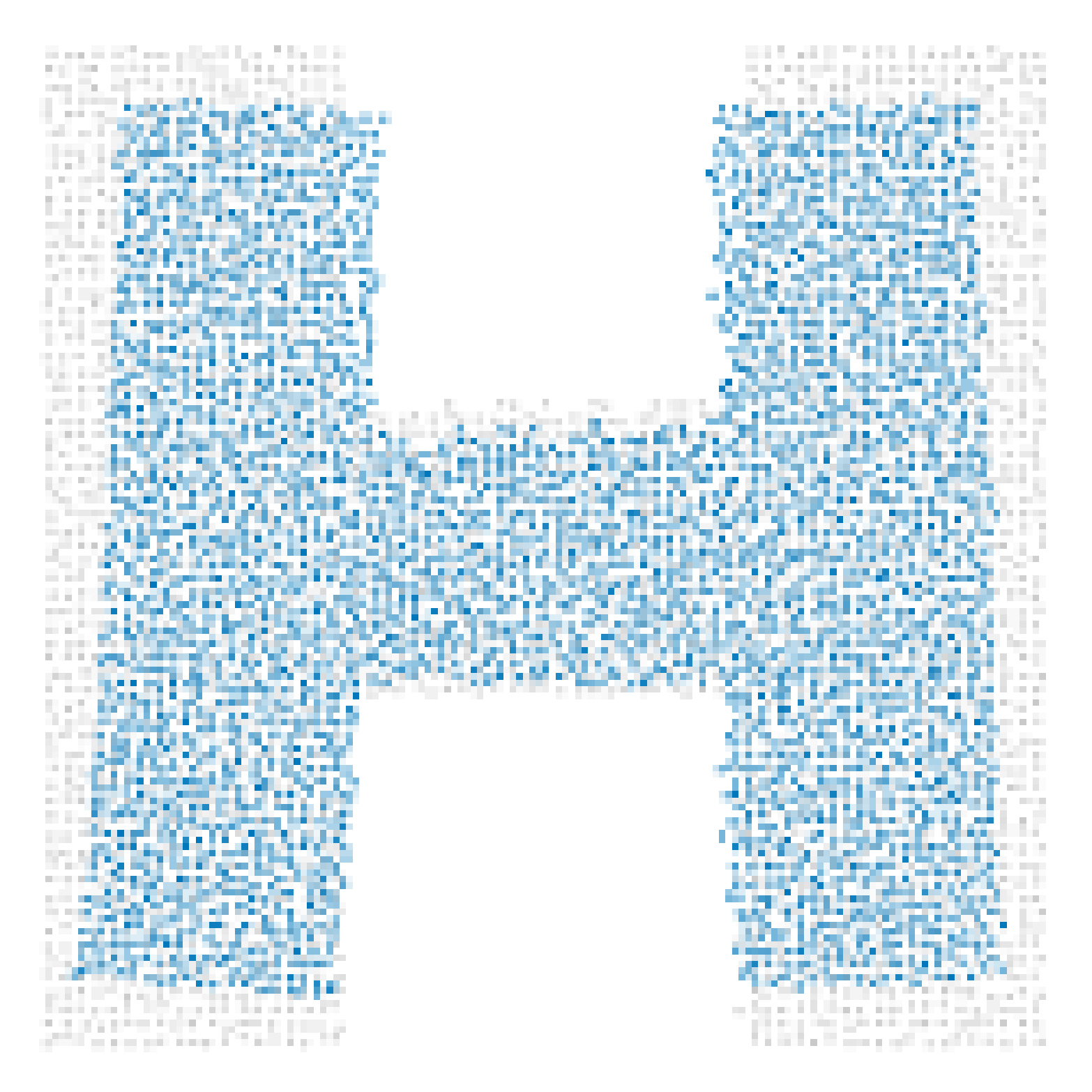} &
\includegraphics[width=.15\linewidth]{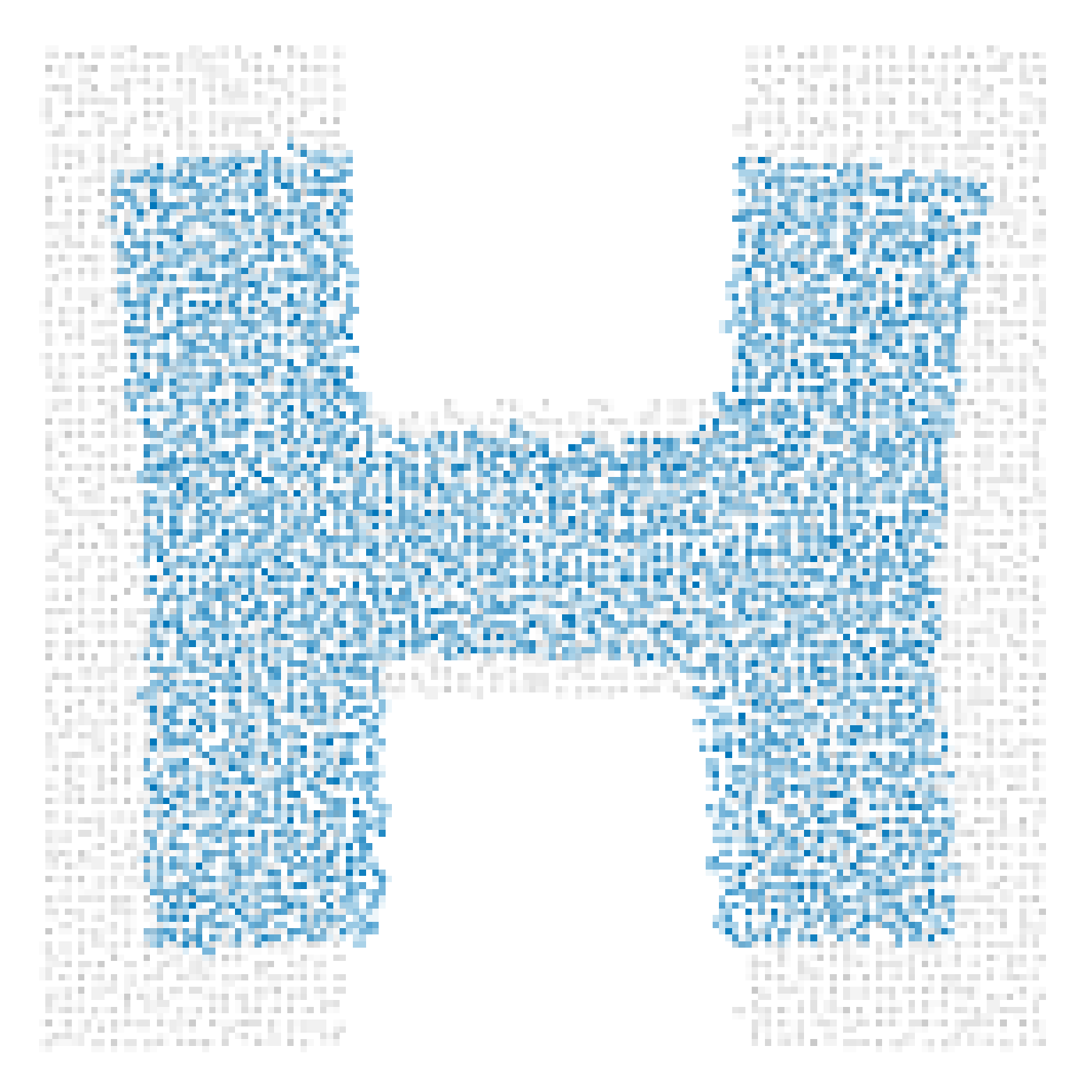} \\
\includegraphics[width=.15\linewidth]{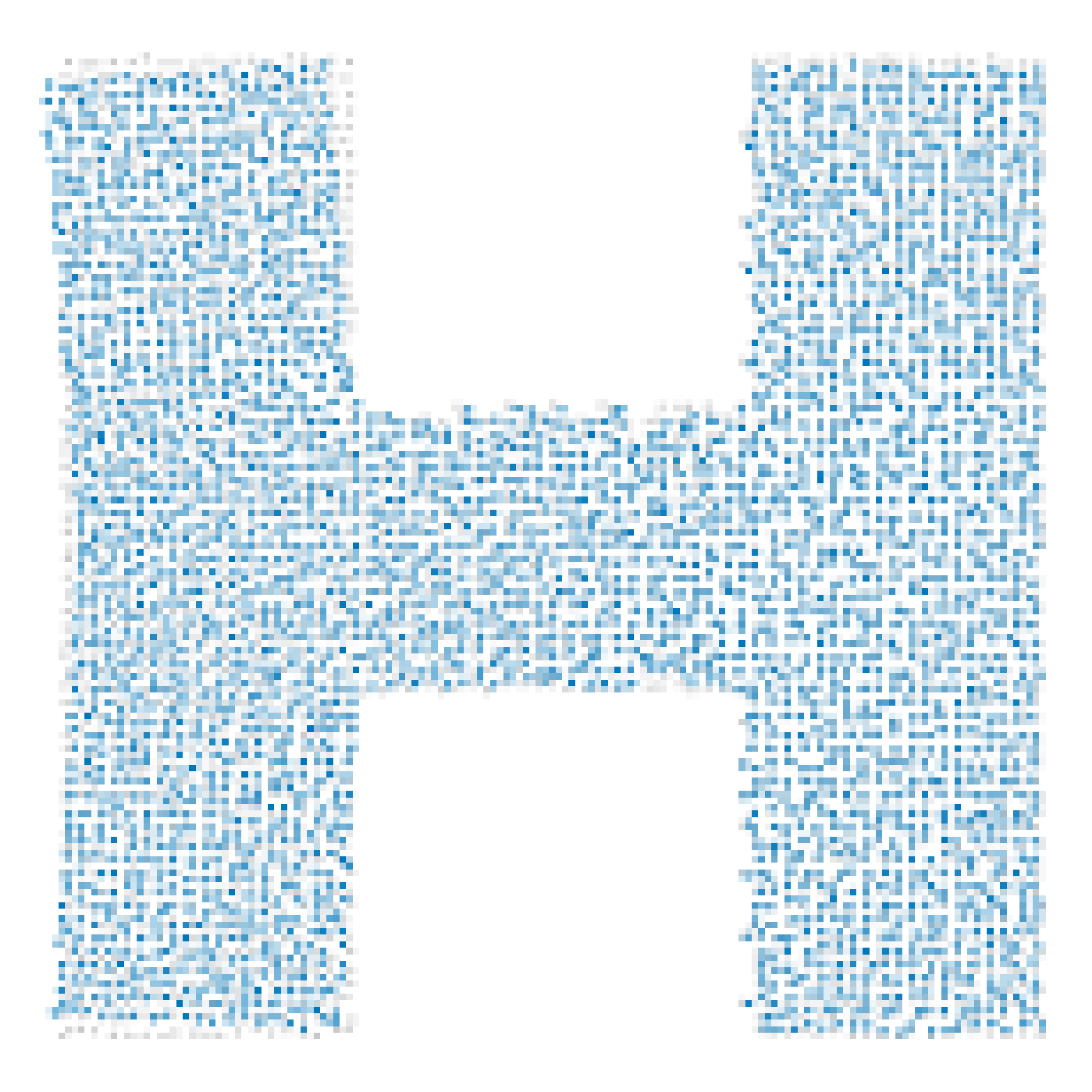} &
\includegraphics[width=.15\linewidth]{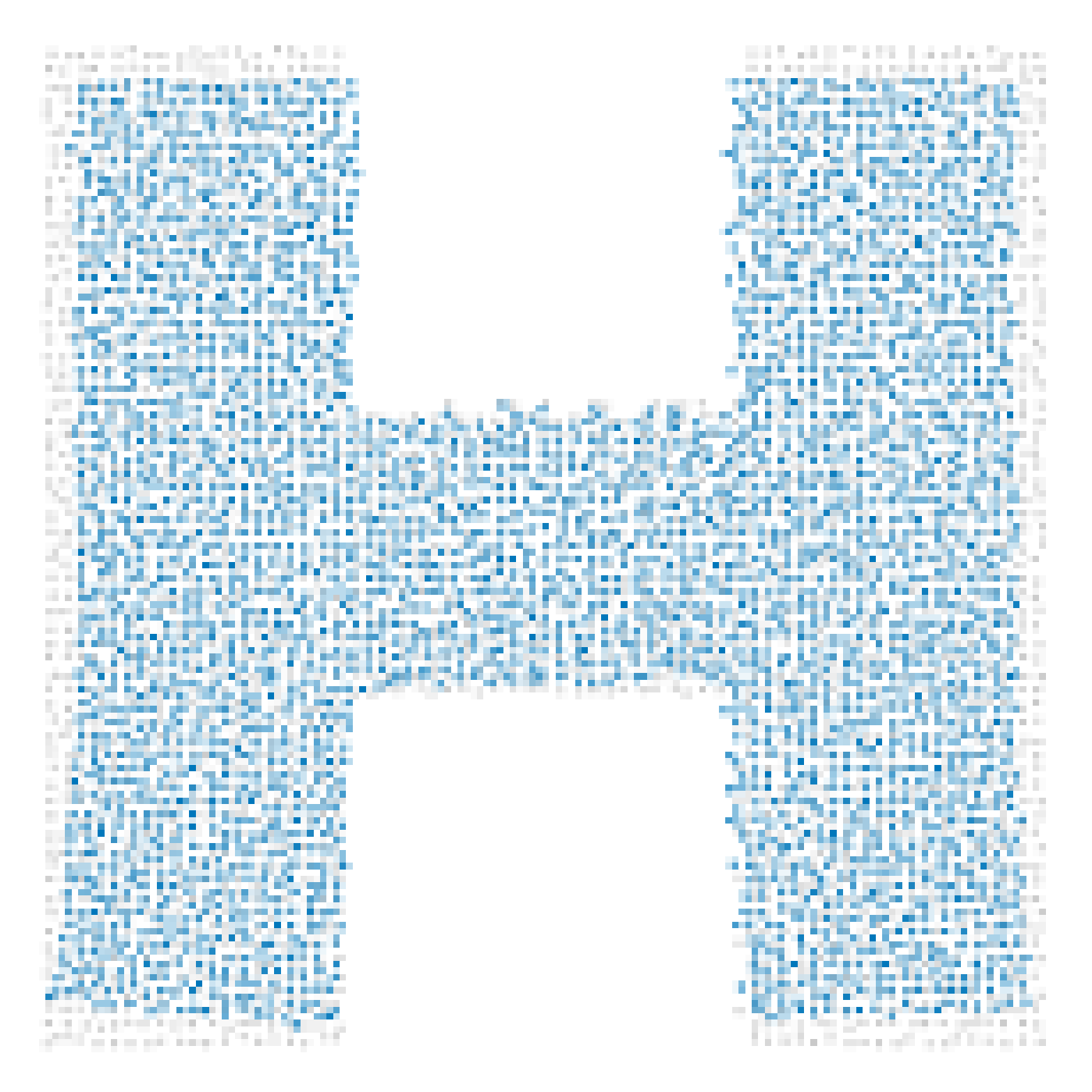} &
\includegraphics[width=.15\linewidth]{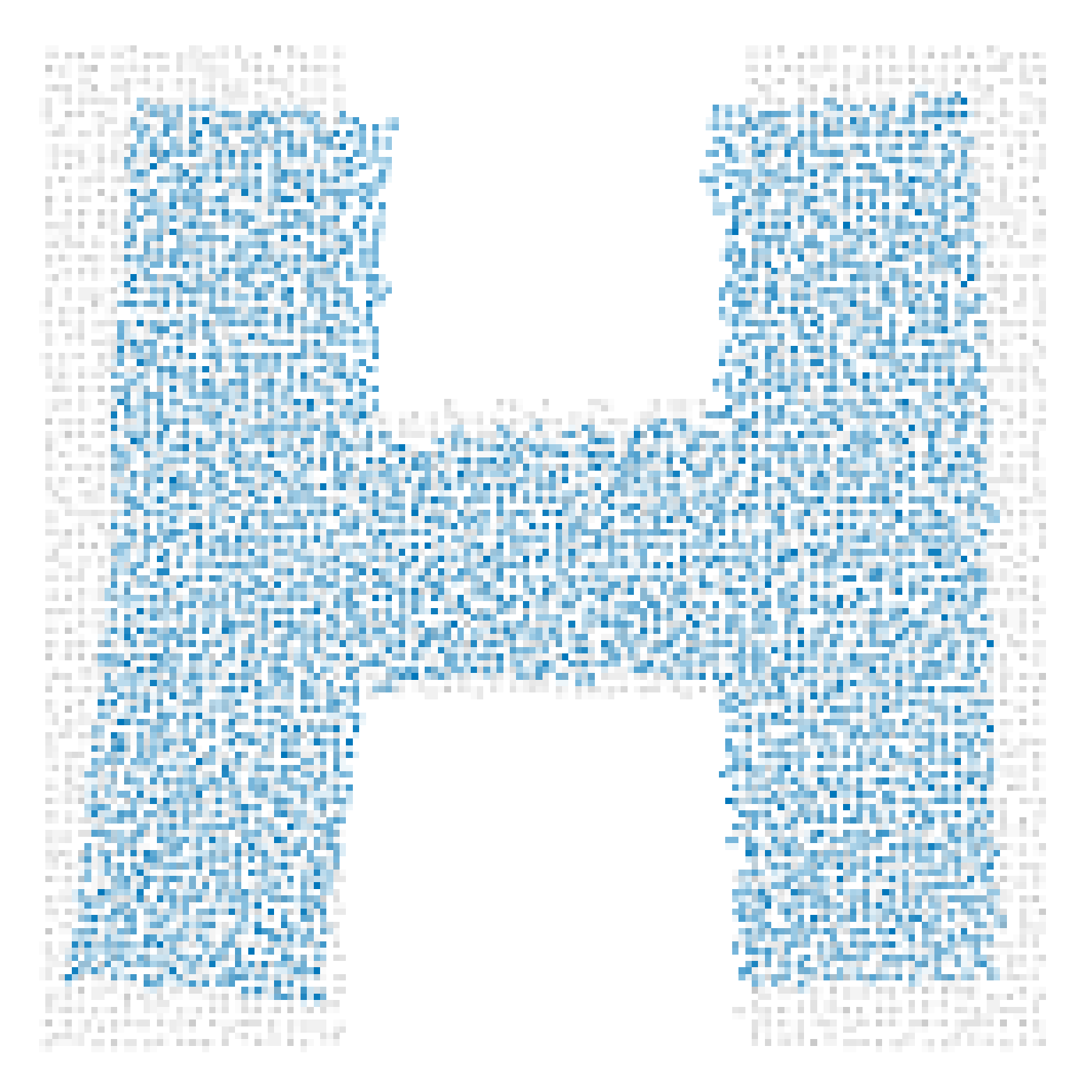} &
\includegraphics[width=.15\linewidth]{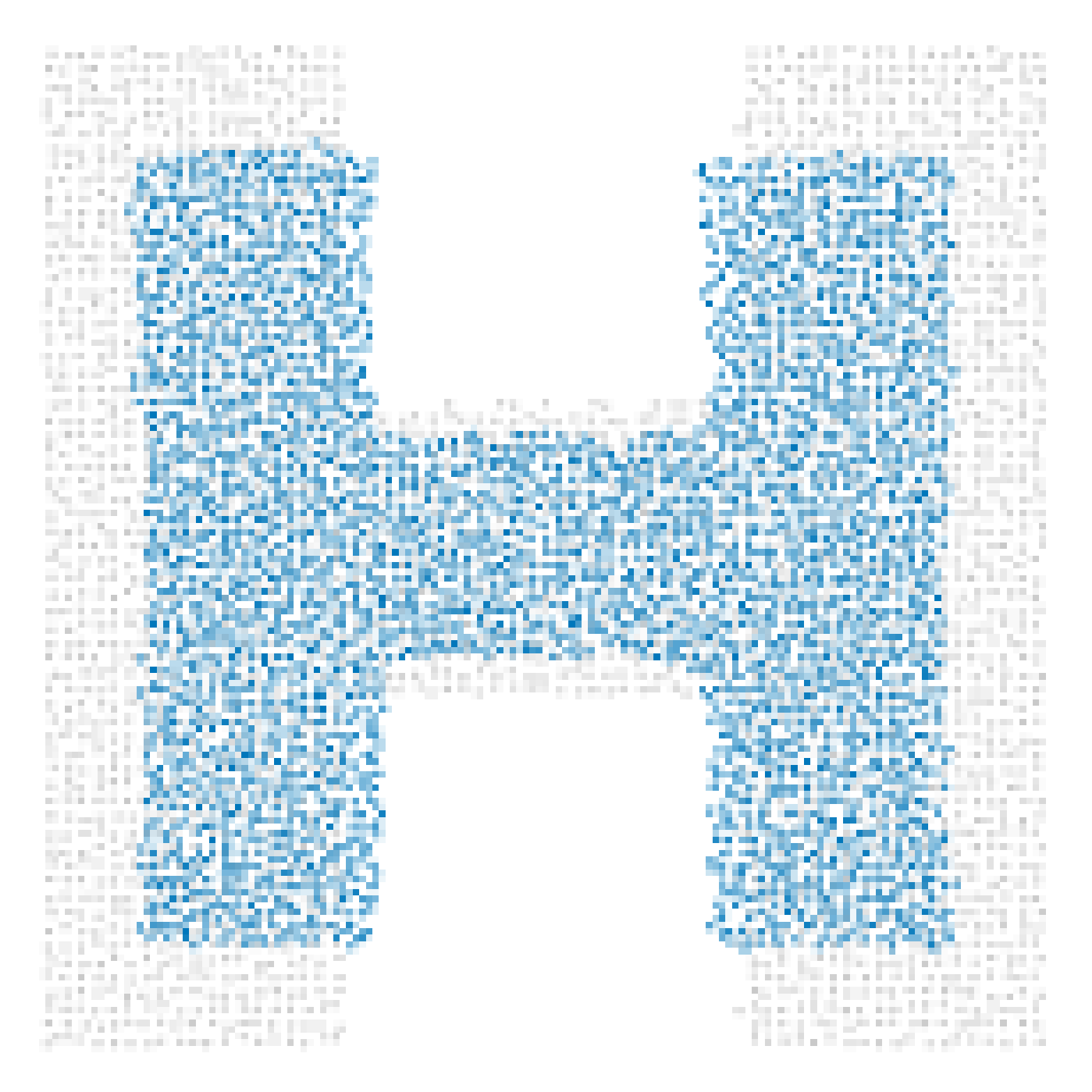} \\
\end{tabular}
\caption{In this experiment, we look at the effect of noise on the optimal parameter choice. The dataset features $n=2866$ points on an H-shaped domain or `dumbbell'. Figures in the $i$th column have noise $\sigma=0.1i$ and the $j$th row uses number of hops $=j$, for $1 \le i \le 4$ and $1 \le j \le 7$ integers. As the noise increases, the embeddings tend to shrink. The shrinkage appears to be caused by the under-estimation of some of the distances by graph distances when applying MDS-D.}
\label{fig:dumbbell_noise_level}
\end{figure}

We work with some emblematic shapes: a rectangle in Figures~\ref{fig:rect_grid_a}--\ref{fig:rect_grid_b}; a hollow rectangle in Figures~\ref{fig:rect_hole_a}--\ref{fig:rect_hole_b}; a C-shaped domain in Figures~\ref{fig:cshape_a}--\ref{fig:cshape_b}; and an H-shaped domain or `dumbbell' in Figures~\ref{fig:dumbbell_a}--\ref{fig:dumbbell_b} and also in \figref{dumbbell_noise_level} for different noise levels.
For each dataset, we apply the variant of MDS-MAP(P) described in \algref{MDS-MAP(P)} with different choices for the number of hops and track the (average) stress and the (average) embedding error. 
We align the output configuration with the true configuration by (orthogonal) Procrustes.
We work with rather sparse graphs to better showcase the result of applying MDS-MAP(P) with different choices for the number of hops. 

% tabular

\subsubsection{Real data: intercity distances}
\begin{figure}[ht!]
\centering
\includegraphics[width=0.47\textwidth]{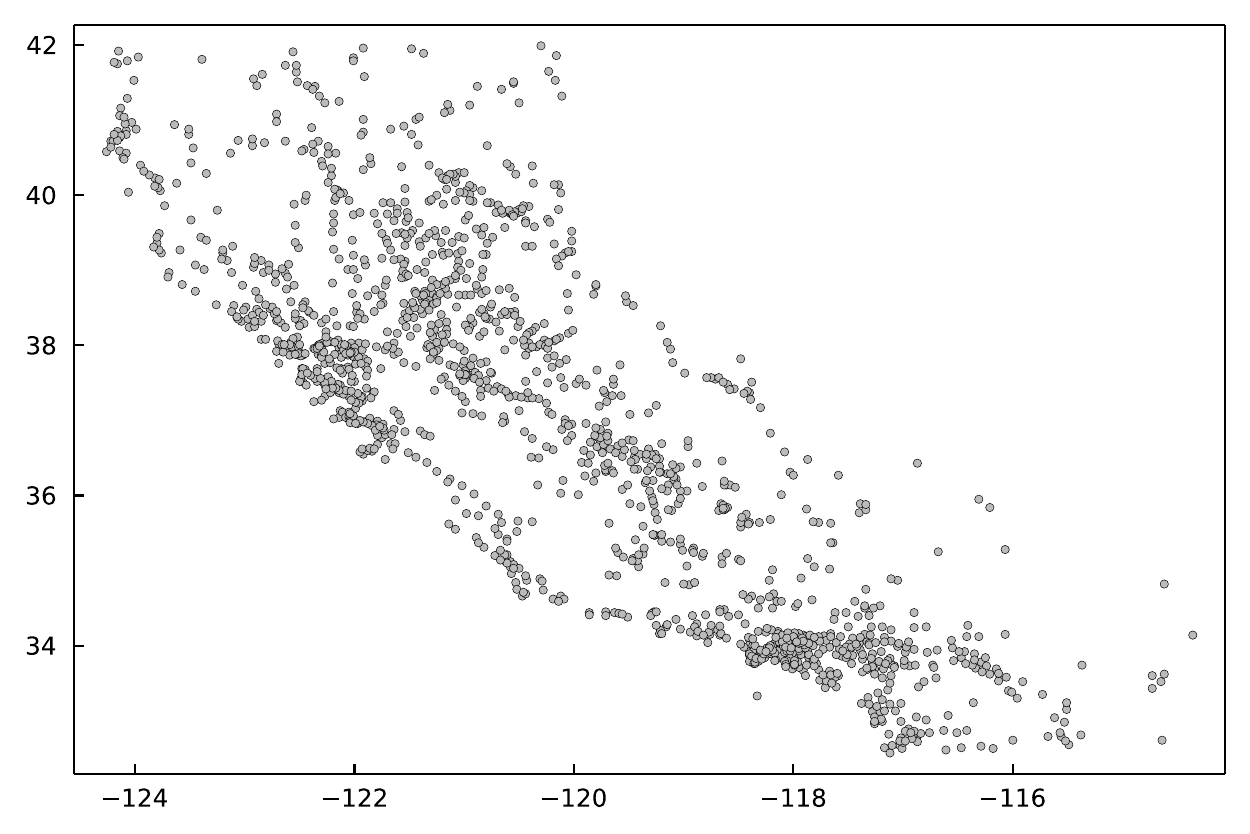}
\includegraphics[width=0.47\textwidth]{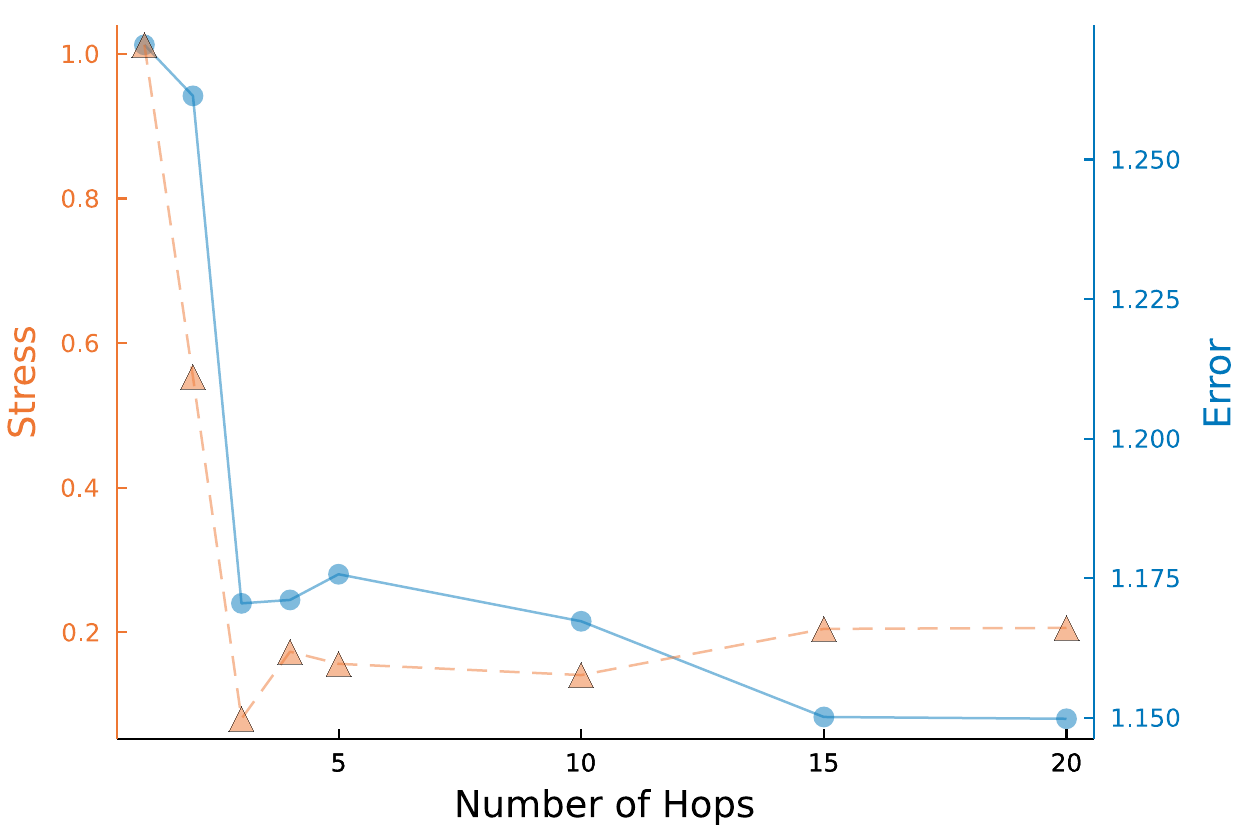}

\includegraphics[width=0.32\textwidth]{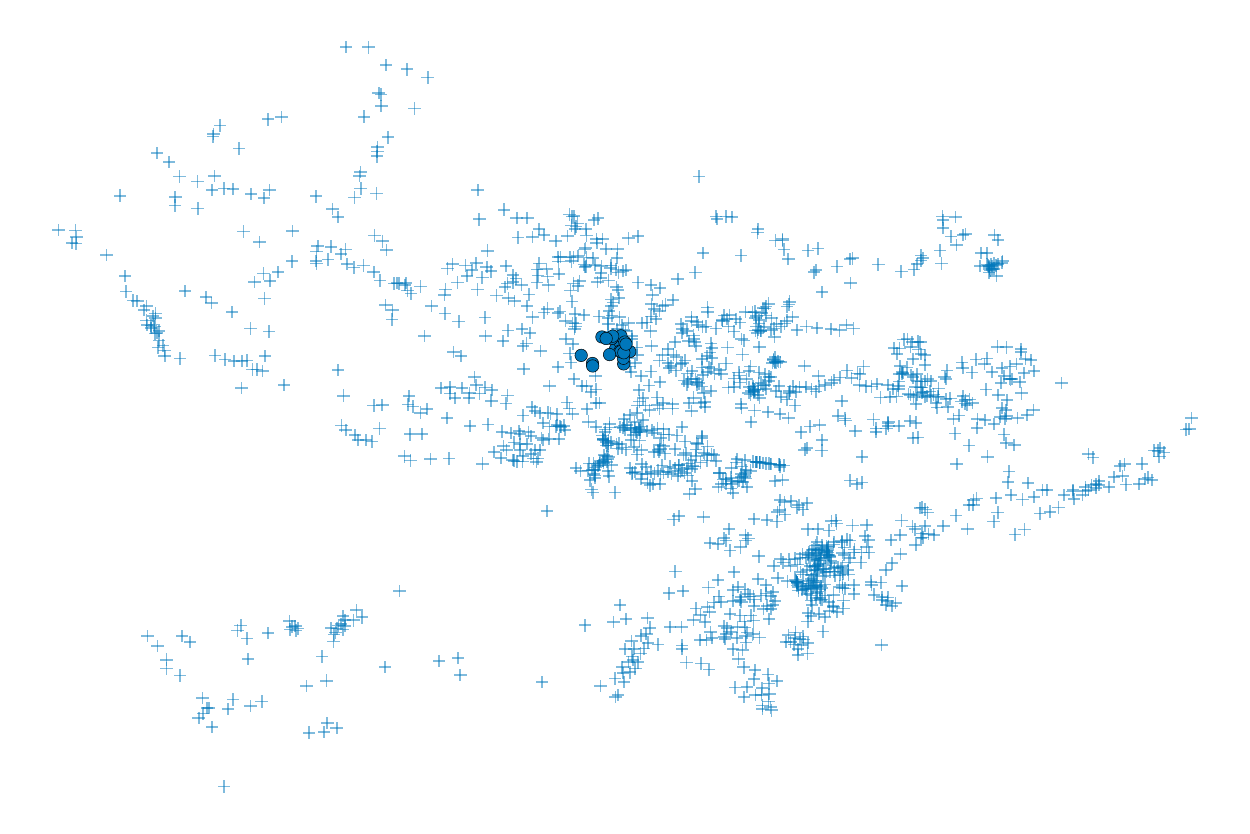}
\includegraphics[width=0.32\textwidth]{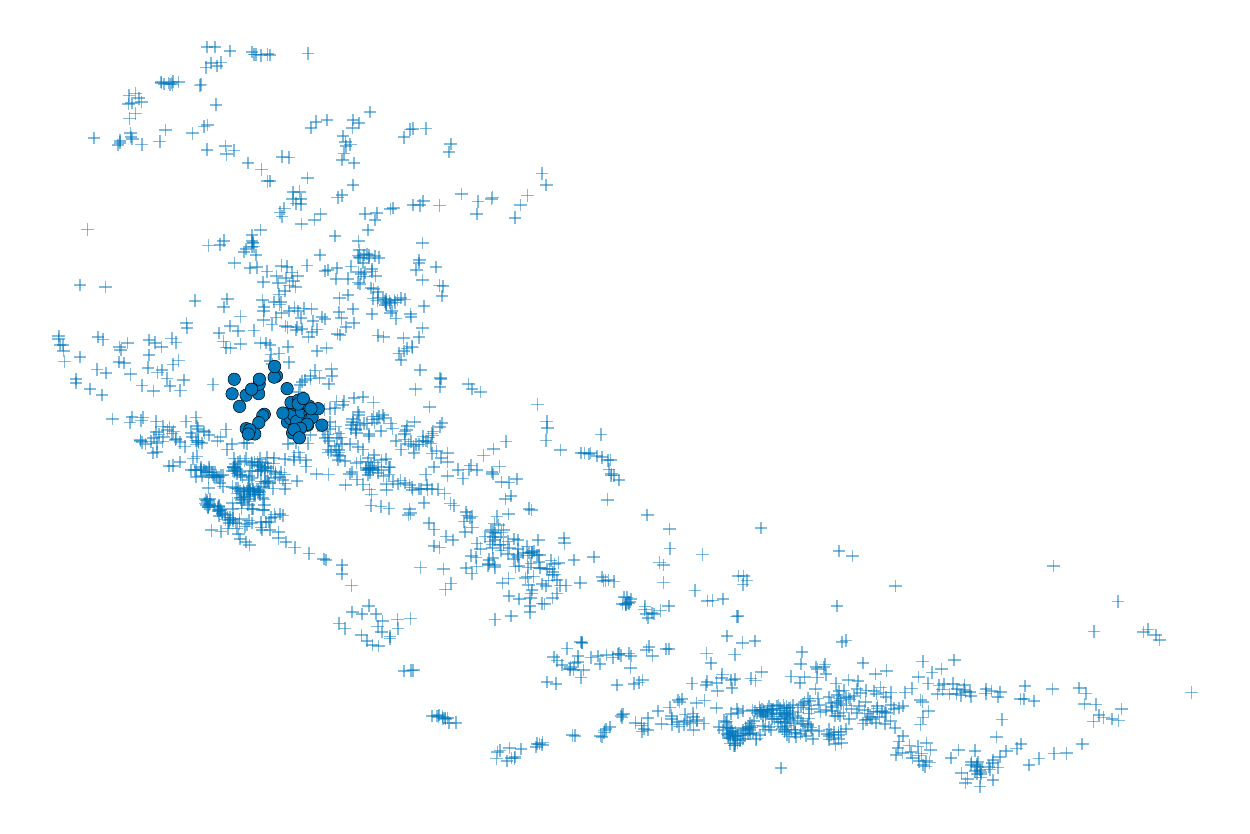}
\includegraphics[width=0.32\textwidth]{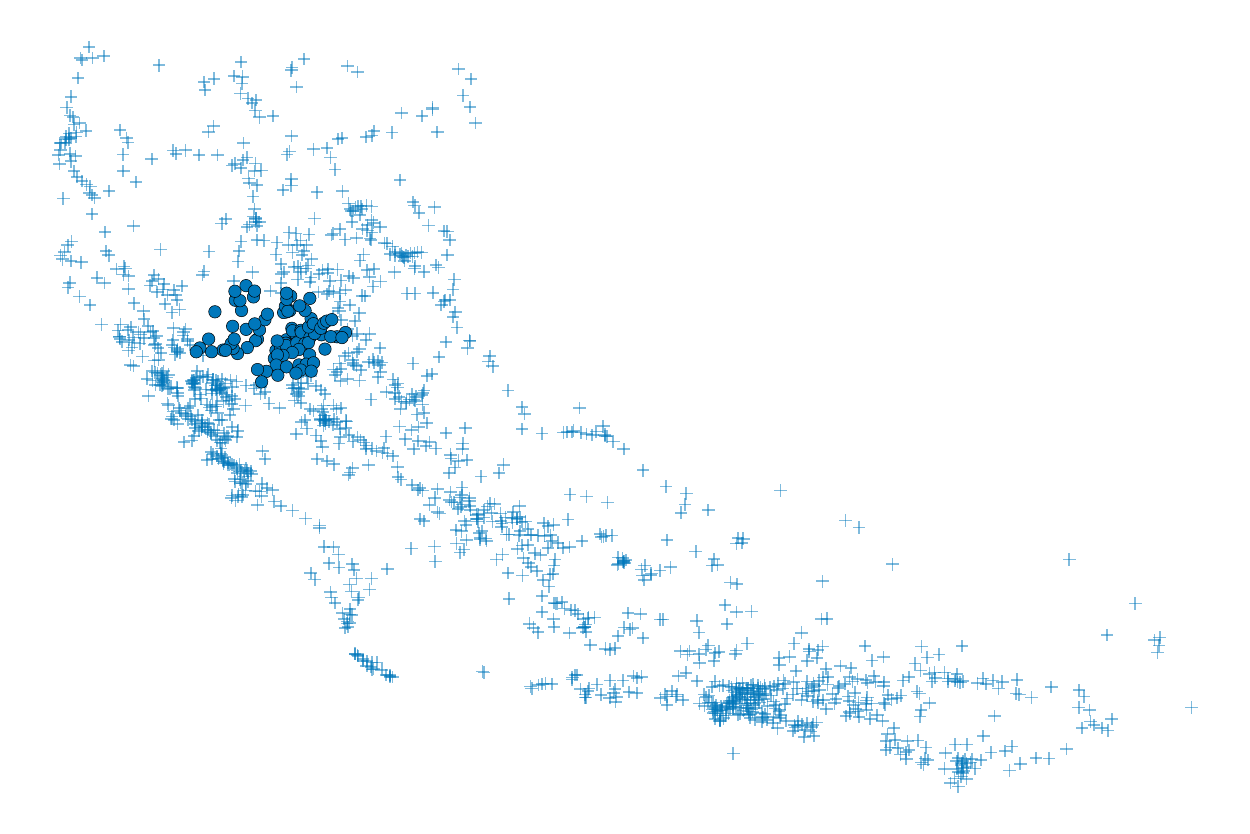}

\includegraphics[width=0.32\textwidth]{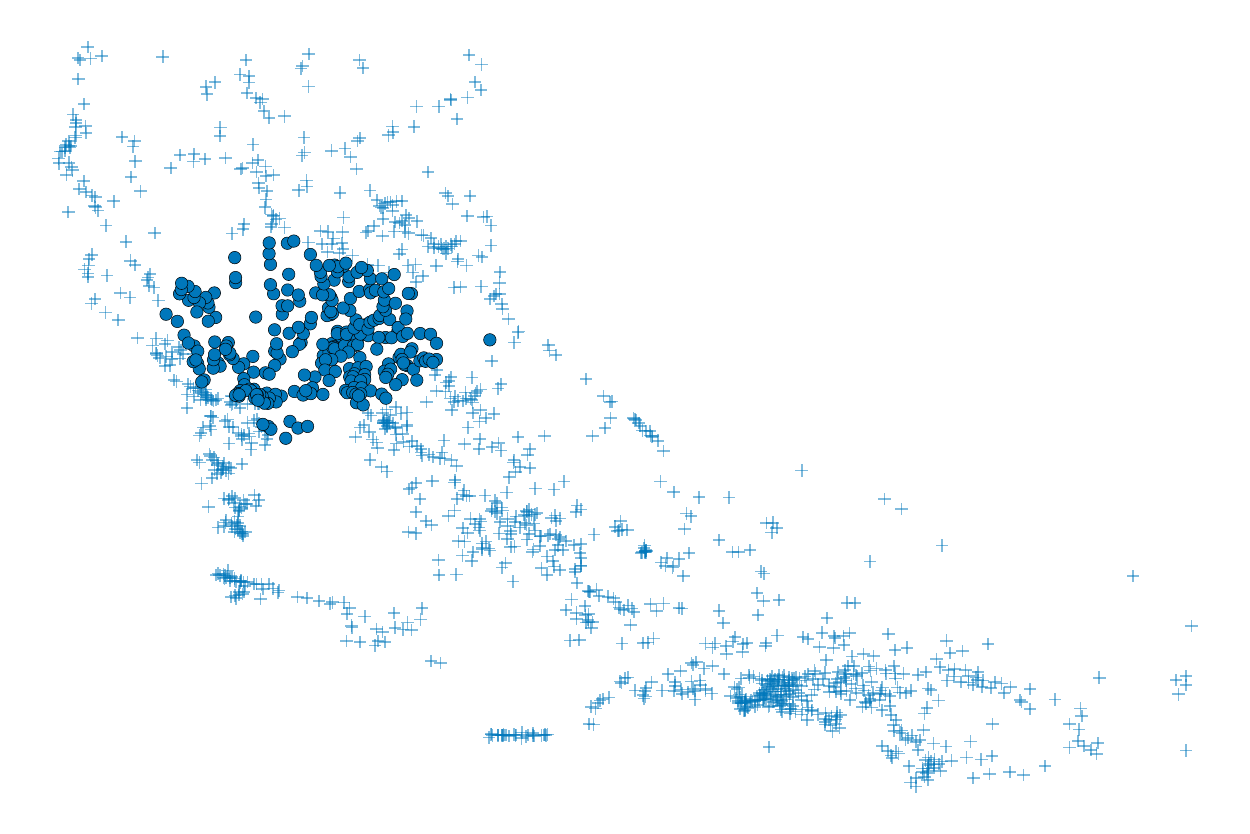}
\includegraphics[width=0.32\textwidth]{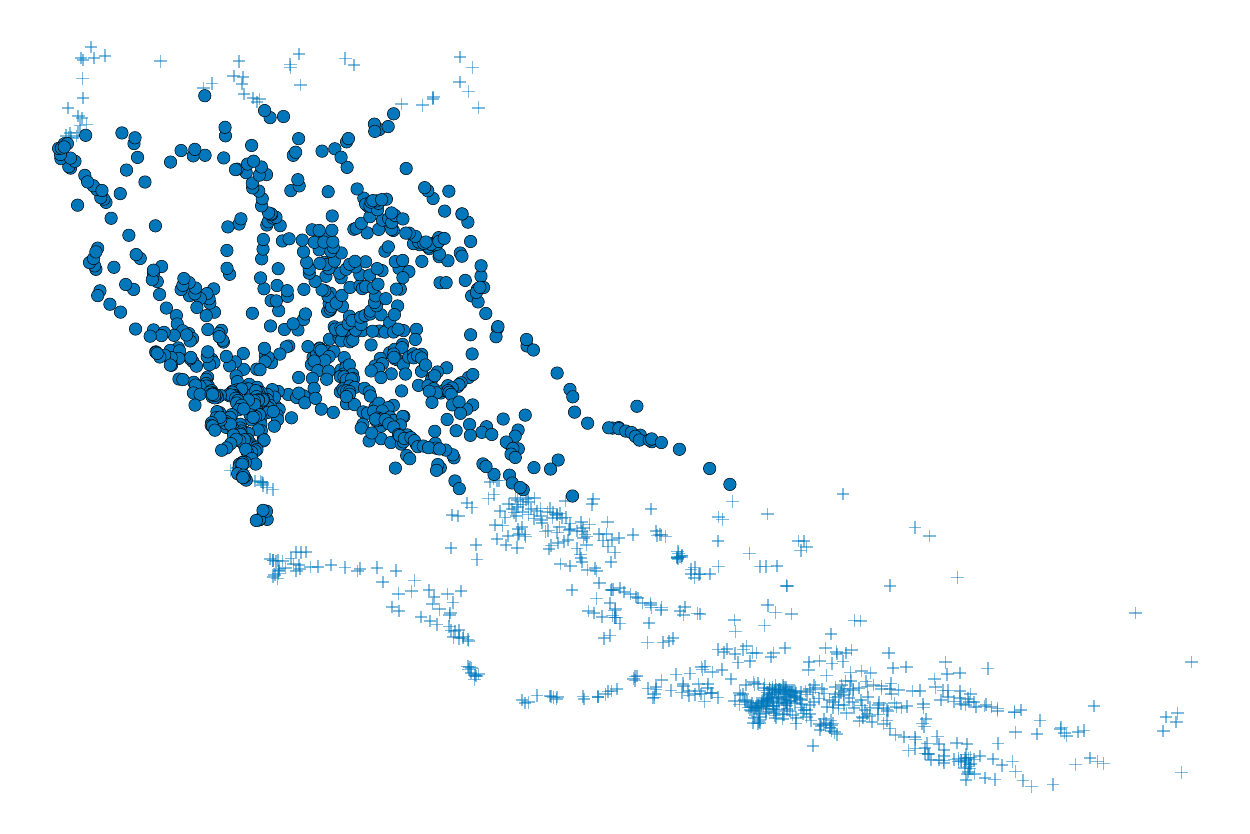}
\includegraphics[width=0.32\textwidth]{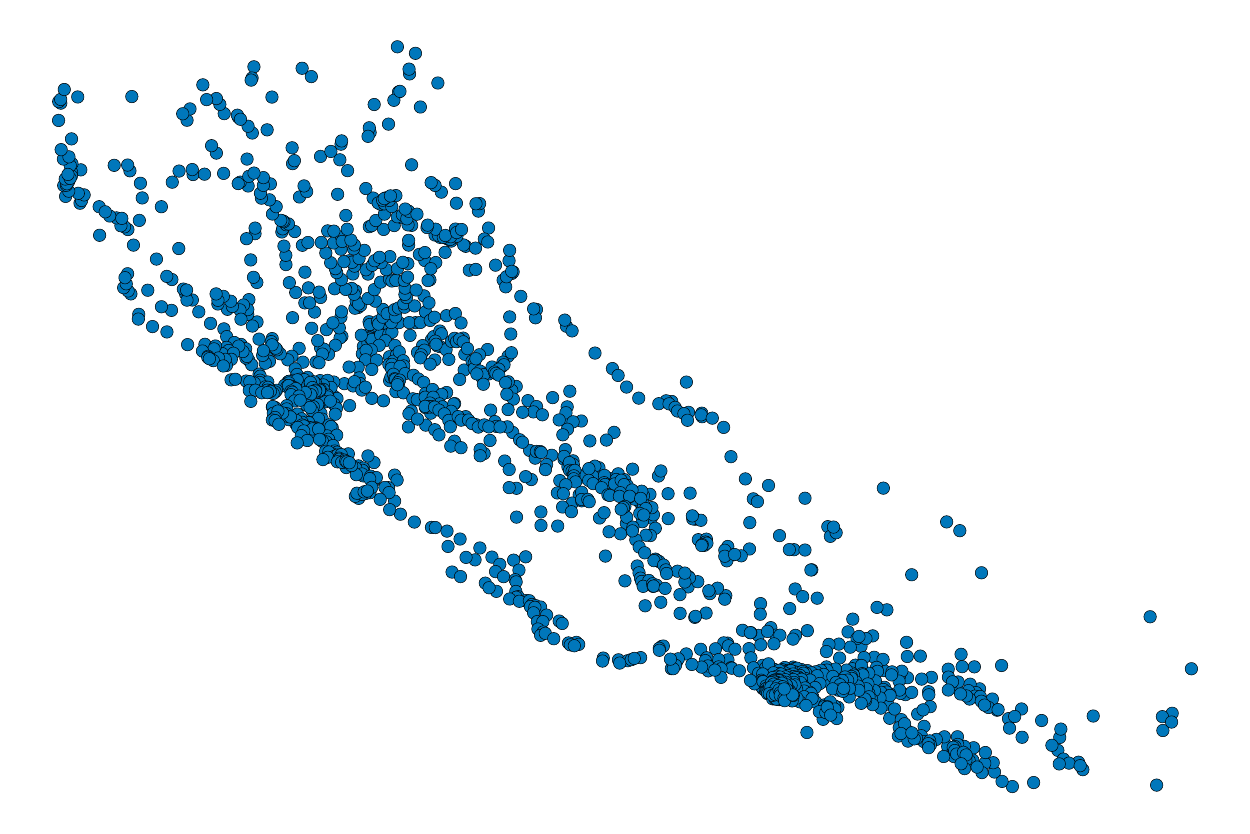}
\caption{MDS-MAP(P) applied to the California intercity dataset. Top left: Plot of cities in California using ground truth latitude and longitude. Top right: Comparison between stress and embedding error. Bottom: In reading order, output of MDS-MAP(P) with number of hops $h = 1, 2, 3, 5, 10, 20$. In each case, the patch originating in Sacramento is highlighted.}
\label{fig:ca}
\end{figure}

\begin{figure}[ht!]
\centering
\includegraphics[width=0.47\textwidth]{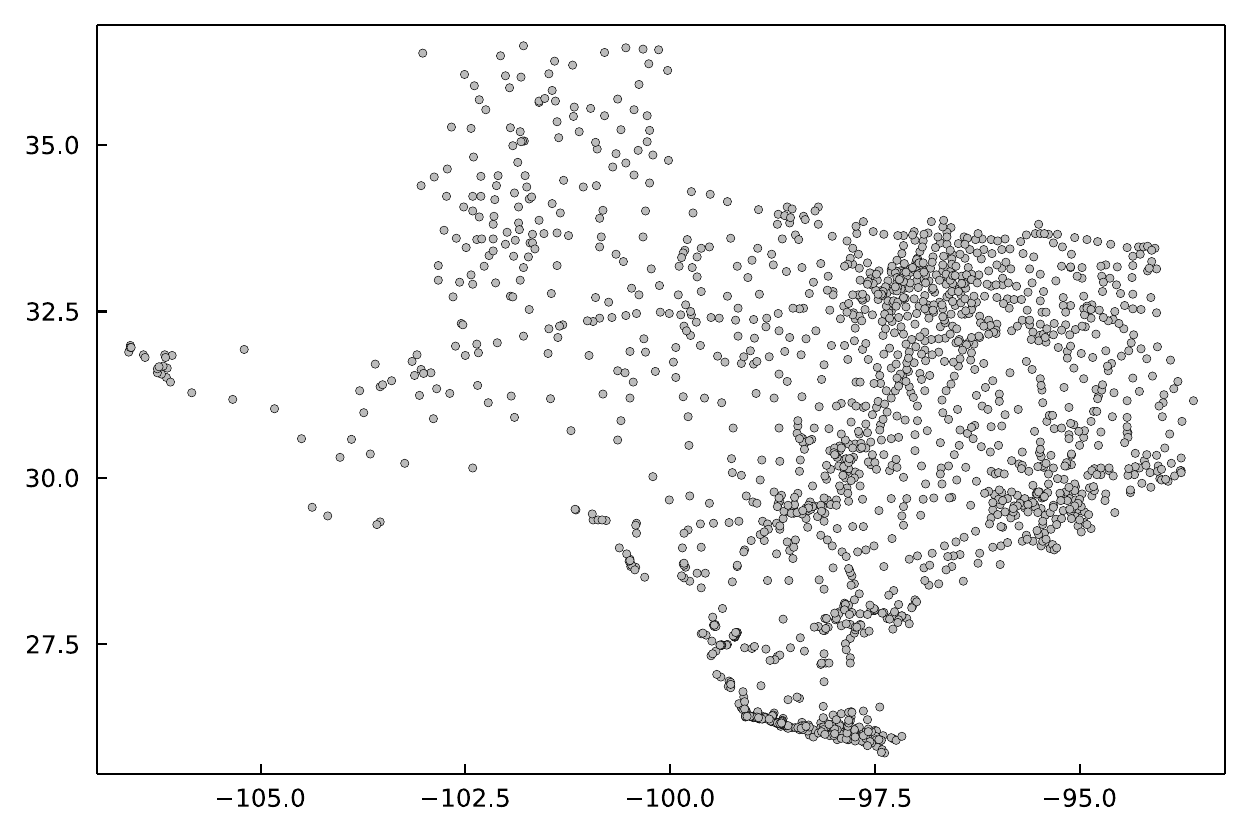}
\includegraphics[width=0.47\textwidth]{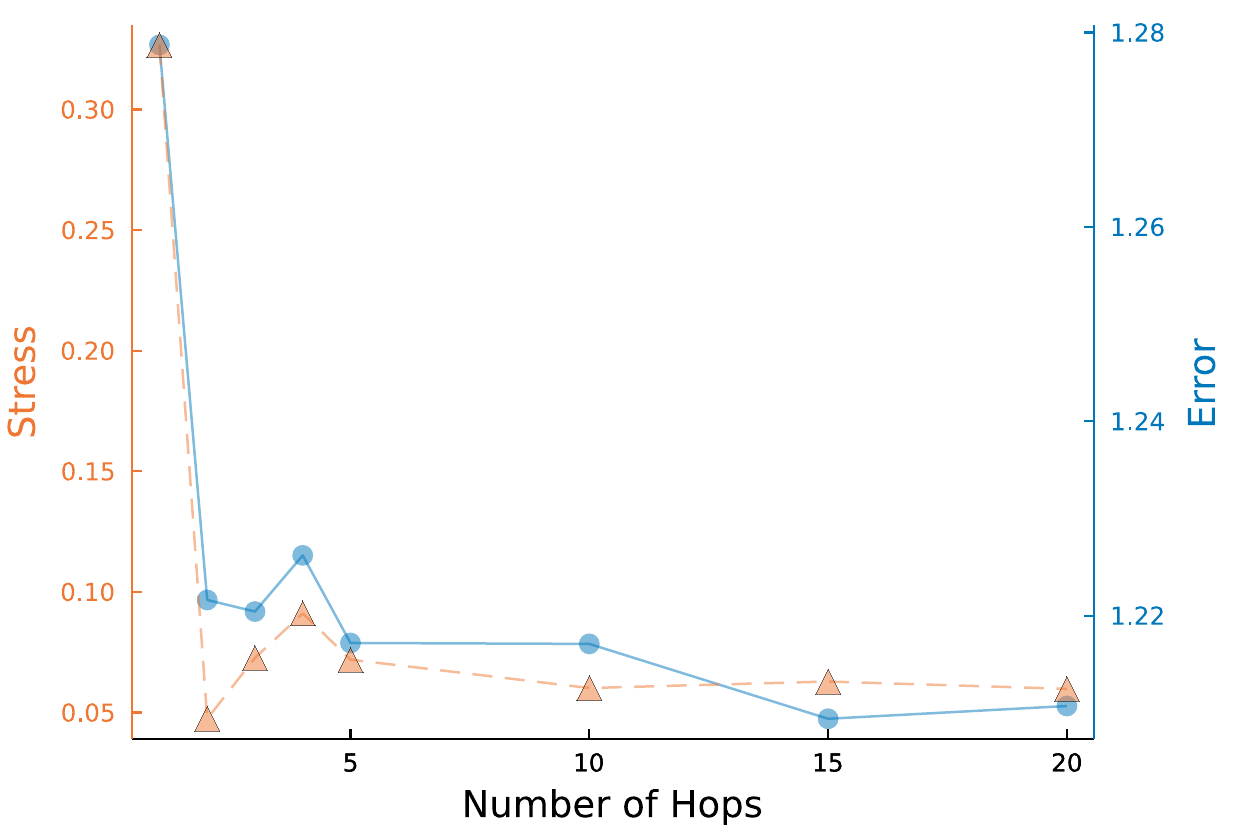}
\includegraphics[width=0.32\textwidth]{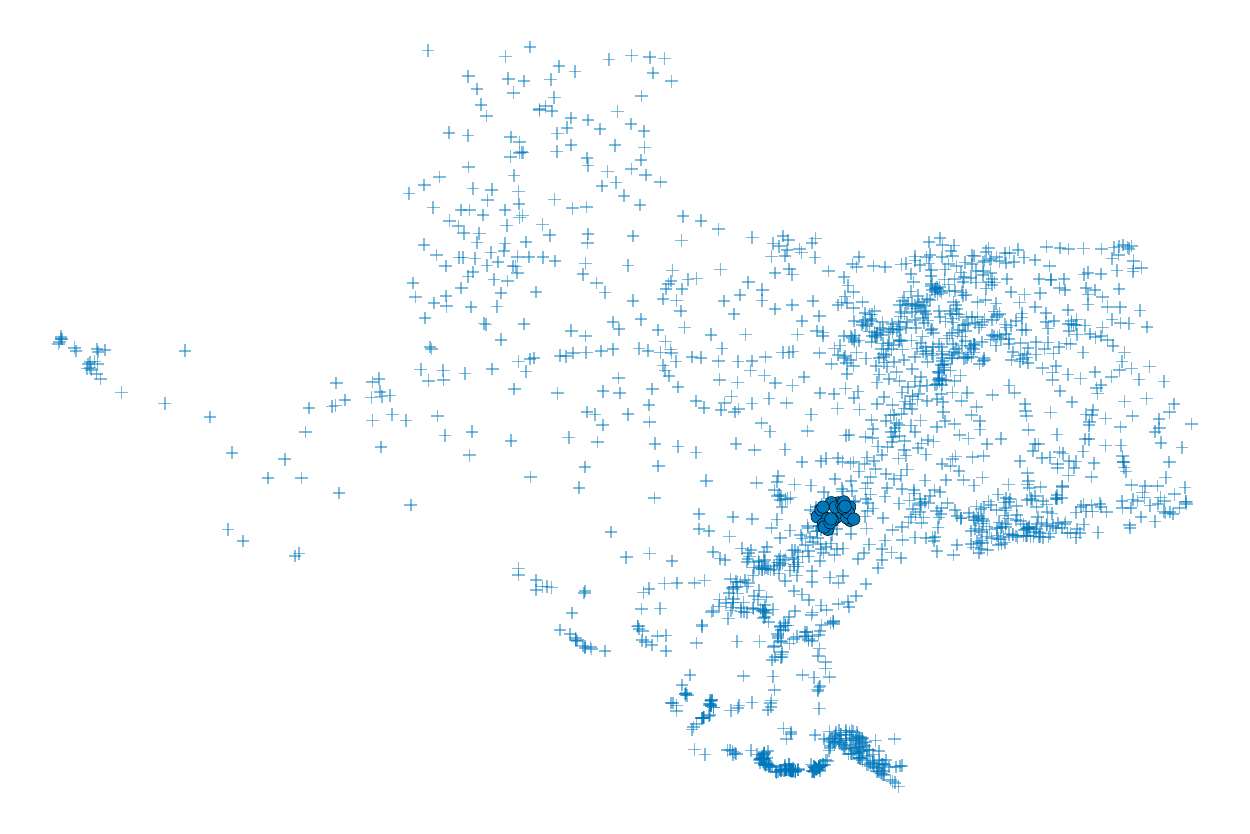}
\includegraphics[width=0.32\textwidth]{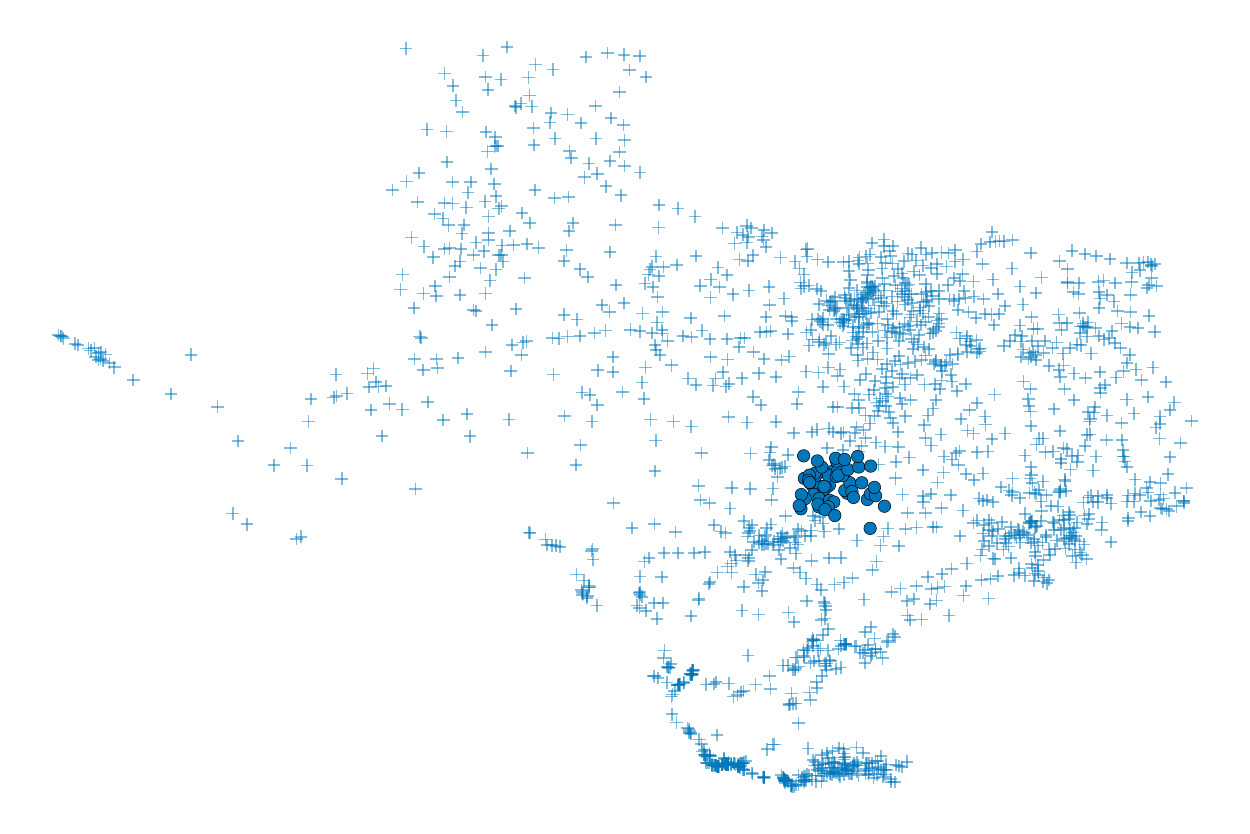}
\includegraphics[width=0.32\textwidth]{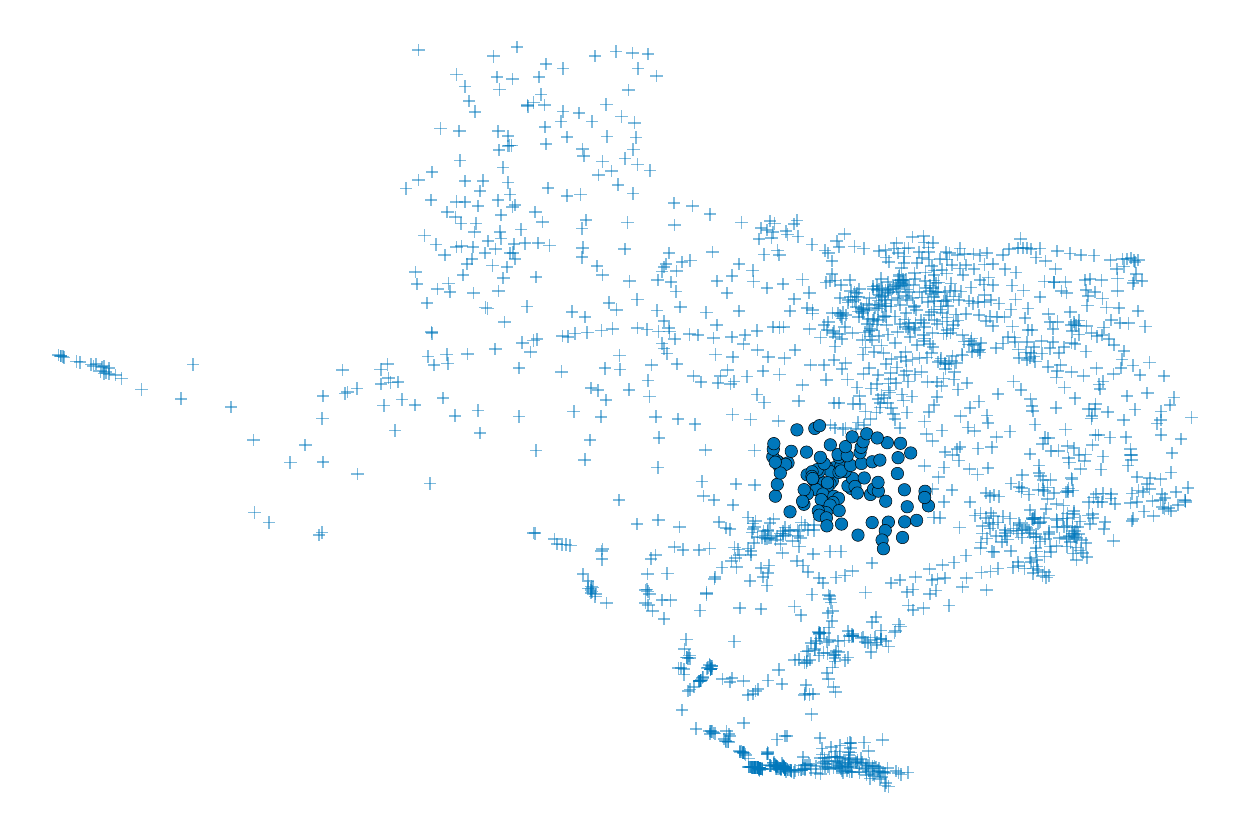}

\includegraphics[width=0.32\textwidth]{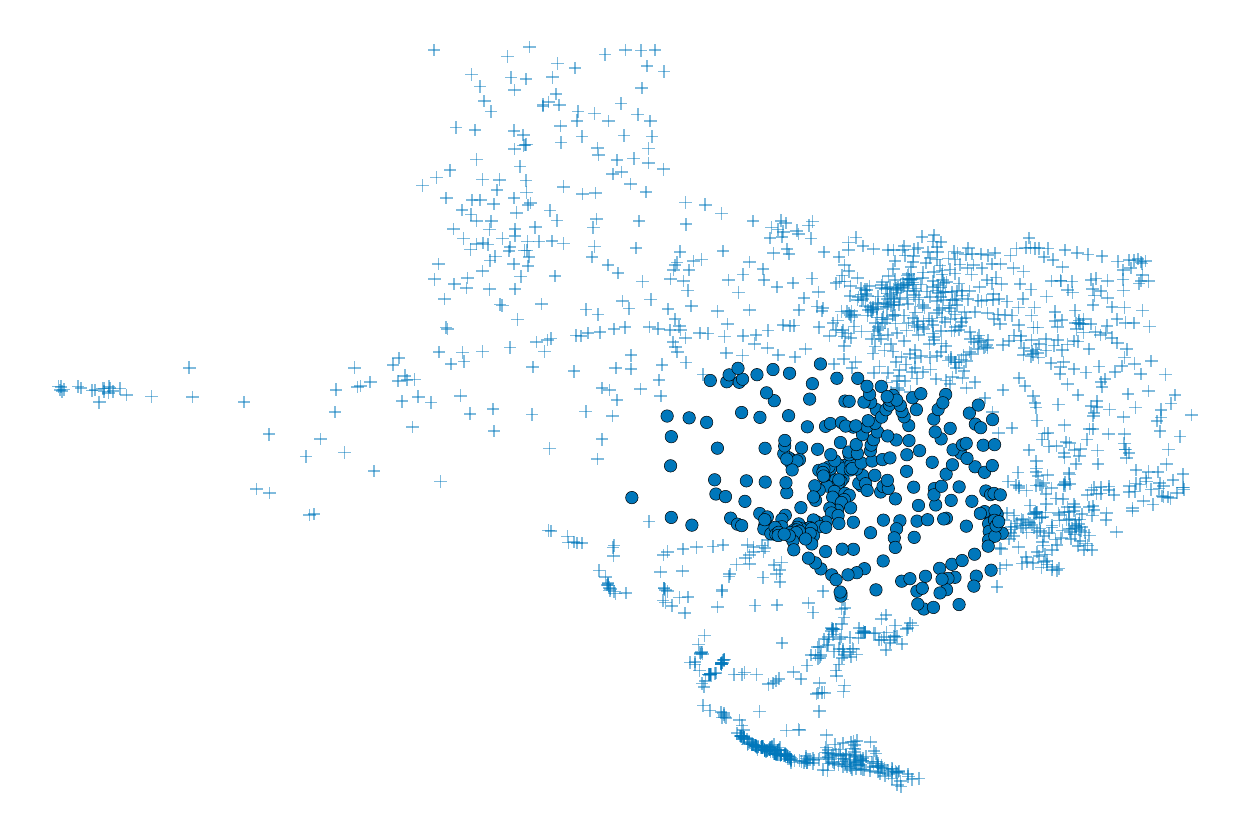}
\includegraphics[width=0.32\textwidth]{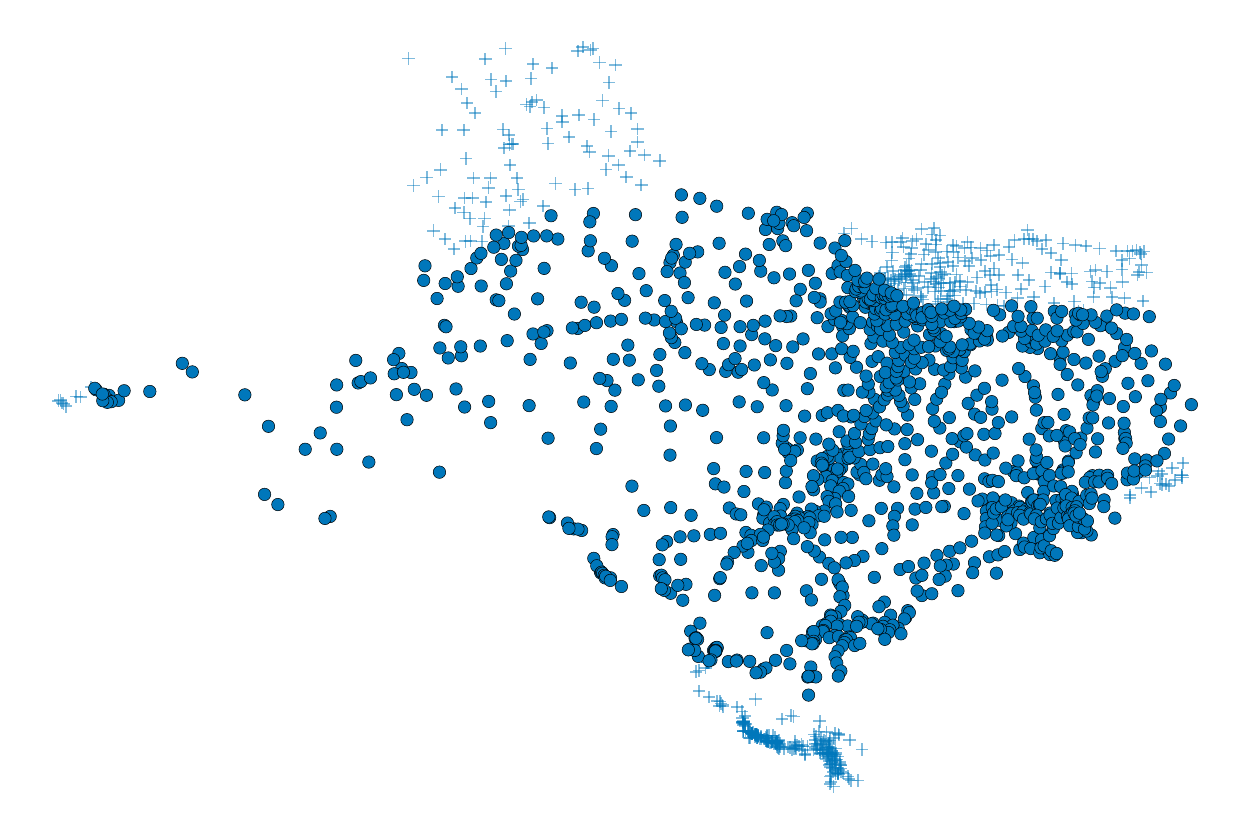}
\includegraphics[width=0.32\textwidth]{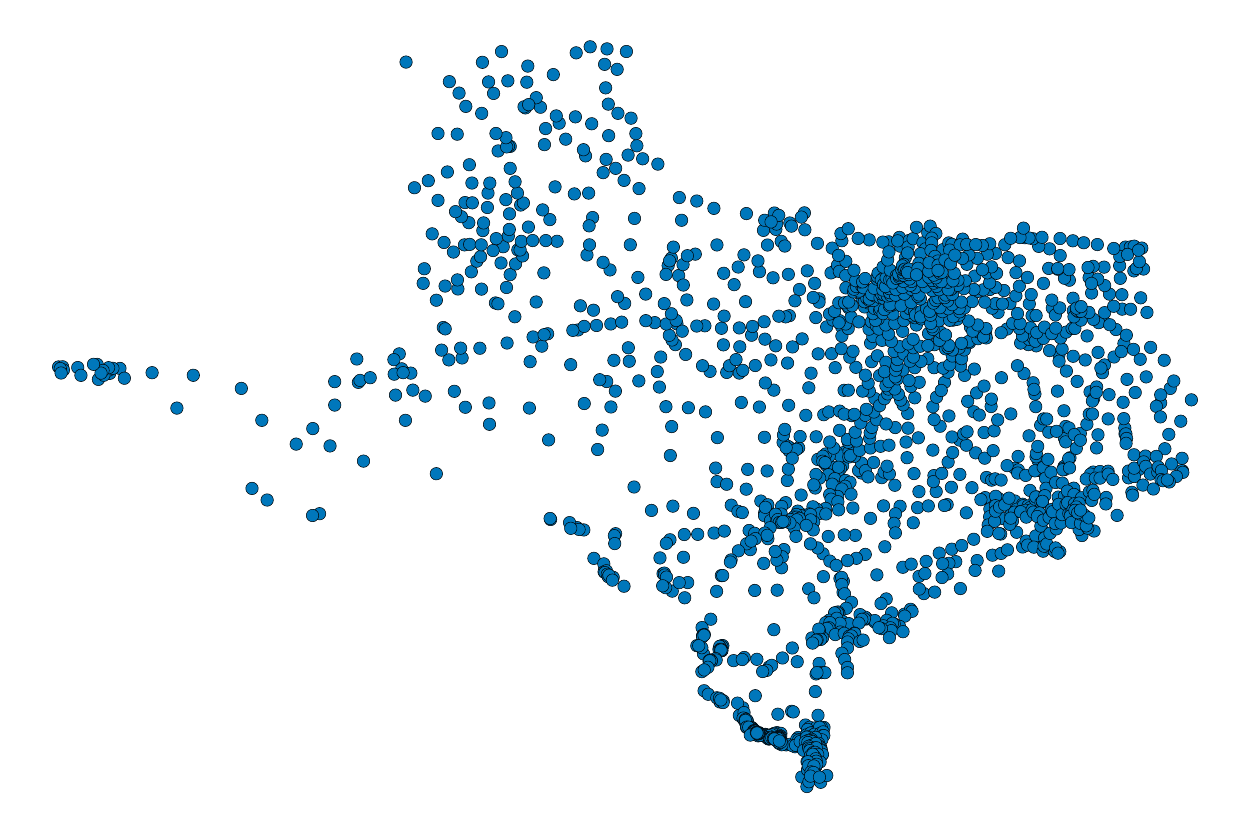}
\caption{MDS-MAP(P) applied to the Texas intercity dataset. Top left: Plot of cities in Texas using ground truth latitude and longitude. Top right: Comparison between stress and embedding error. Bottom: In reading order, output of MDS-MAP(P) with number of hops $h = 1, 2, 3, 5, 10, 20$. In each case, the patch originating in Austin is highlighted.}
\label{fig:tx}
\end{figure}

Besides synthetic datasets, we also examined the application of our approach to the problem of locating cities in a geographical region (California and Texas) using intercity distances.
The latitude and longitude of each city are readily available online\footnote{~For example, at \url{https://simplemaps.com/data/us-cities}}. 
The haversine formula is used to construct the observed dissimilarity matrix of as-the-crow-flies distances from the geographical coordinates.
That is, if $(\lambda_1, \varphi_1), (\lambda_2, \varphi_2)$ denote the latitude and longitude of a pair of cities, then their distance is computed as follows
\[2 \arcsin \sqrt{\sin^2(\tfrac12 (\varphi_2 - \varphi_1)) + \cos(\varphi_1) \cos(\varphi_2) \sin^2(\tfrac12 (\lambda_2 - \lambda_1))}~.
\]
We work with the $k=12$ nearest neighbor graph. Although no noise is added, we note that even without noise an exact realization in the plane is not possible since the points are effectively on a curved surface (the surface of the Earth).
\figref{ca} displays the result of applying MDS-MAP(P) to the intercity distances of California.
To illustrate the size of patches for the different choices of number of hops, the patch originating in the capital Sacramento is highlighted.
\figref{tx} shows the result for Texas with the patch originating in the capital Austin being highlighted.

\section{Dimensionality reduction}
\label{sec:DR}

After discussing multidimensional scaling (MDS) in \secref{MDS}, we now turn to dimensionality reduction (DR). As is well-known to the expert, the two problems are intimately related. In fact, some of the most emblematic methods in DR can be recovered by applying methods for MDS to the pairwise Euclidean distances after discarding the larger distances. This is most famously true of PCA, whose embedding can be obtained by an application of classical scaling; but it is also true of Isomap \cite{tenenbaum2000global}, which can be obtained in this fashion from MDS-D \cite{kruskal1980designing}; of Laplacian eigenmaps \cite{belkin2003laplacian}, which corresponds to applying  \cite{hall1970r}; of maximum variance unfolding \cite{weinberger2006introduction}, which corresponds to semidefinite embedding \cite{weinberger2006graph}; and even recent approaches such as t-SNE \cite{van2008visualizing} and UMAP \cite{mcinnes2018umap} have been shown to be in correspondence with force-directed layouts popular in graph drawing in \cite{bohm2022attraction} and in \cite{damrich2021umap}, respectively. 
Because of this strong parallel, we are able to draw a parallel with the MDS setting discussed in the previous section. The structure of the section is very similar. 

\subsection{Setting}
In DR, the data consist of points $z_1, \dots, z_n \in \bbR^{p_0}$, and given a dimension $p < p_0$, the goal is to embed these points into $\bbR^p$ as faithfully as possible. If by this we mean to preserve the pairwise distances as much as possible, then it can be done by principal component analysis (PCA), which is in fact optimal among linear projections for a particular way of quantifying the accuracy. 
We adopt the manifold learning setting in which the data points are assumed to be on or near a smooth submanifold of given dimension $p$ and the goal is to preserve as much as possible the pairwise distances on the submanifold. 
In that case, PCA will not succeed unless the submanifold is affine or nearly so. Most of the DR methods suggested in recent times have been proposed for this setting and, as already noted, can be seen as 
{\em (i)} computing the Euclidean distances between the data points, i.e., $d_{ij} := \|z_i-z_j\|$ for all $i,j \in [n]$;
{\em (ii)} only keeping the smallest distances, i.e., for the neighborhood graph with edge set $\cE = \{(i,j): d_{ij} \le r\}$ where $r$ is the connectivity radius and a tuning parameter; and then applying a method for MDS to the resulting weighted graph.
The rationale for only keeping or trusting the smallest Euclidean   distances is because, in the limit of an infinitesimally small neighborhood around a point on a smooth submanifold, the Euclidean distances are close to the distances on the submanifold.

\subsection{Methods}
Manifold learning has a substantial literature. 
We already mentioned that Isomap \cite{silva2002global, tenenbaum2000global, tenenbaum1997mapping} is in correspondence with graph distance methods in MDS such as MDS-D \cite{kruskal1980designing} and MDS-MAP \cite{shang2003localization};  
Laplacian eigenmaps \cite{belkin2003laplacian}, and the closely related diffusion maps \cite{coifman2006diffusion}, are in correspondence with spectral methods in MDS \cite{hall1970r};
maximum variance unfolding \cite{weinberger2006introduction} is an SDP method that was in fact simultaneously proposed for DR and MDS by the authors \cite{weinberger2006graph};
and some of the latest methods, such as t-SNE \cite{van2008visualizing} and UMAP \cite{mcinnes2018umap},  are in correspondence with force-directed approaches in graph drawing \cite[Sec 5.7]{klimenta2012extending} as argued in \cite{bohm2022attraction} and \cite{damrich2021umap}. 
Not all methods proposed for DR can be derived from a method originally proposed for MDS.
For example, self-organizing maps \cite{kohonen1982self}, principal surfaces \cite{hastie1989principal}, and Kernel PCA \cite{scholkopf1998nonlinear} approximate the data with a surface of given dimension directly in the ambient space.  

\subsubsection{Patch-stiching methods}
Here too, we work with a class of methods that could also be referred to as patch-stitching methods, and work very much in the same way. To quote \citet{brand2003charting}, the protypical steps are ``to decompose the sample data into locally linear low-dimensional patches, [and] merge these patches into a single low-dimensional coordinate system''.
Local linear embedding \cite{roweis2000nonlinear,saul2003think} and manifold charting \cite{brand2003charting} are clearly of that type, but even more geometrical methods such as Hessian eigenmaps \cite{donoho2003Hessian} and local tangent space alignment \cite{zhang2004principal} operate in a similar fashion. 
This parallel has been known for quite some time, at least by some, including \citet{chen2009local}, who draw inspiration from the extensive literature on graph drawing, and in particular, force-directed methods, to suggest their local MDS algorithm. 

To illustrate the choice of tuning parameter in the context of DR, we simply leverage our variant of MDS-MAP(P) (\algref{MDS-MAP(P)}) into a method for manifold learning obtaining a local variant of isomap (\algref{local isomap}). In light of this close connection between MDS and DR, this is a natural idea, which has already been proposed, including in \cite{schwartz2019intrinsic}.

\begin{algorithm}[!t]
\KwData{Data points $z_1, \dots, z_n \in \bbR^{p_0}$, connectivity radius $r$, embedding dimension $p$}
\KwResult{Configuration $y_1, \dots, y_n \in\bbR^{p}$}
Form the neighborhood graph on $x_1, \dots, x_n$ with connectivity radius $r$\;
Apply MDS-MAP(P) to the resulting weighted graph to obtain an embedding $y_1, \dots, y_n$\;
\caption{Local Isomap via MDS-MAP(P)}
\label{alg:local isomap}
\end{algorithm}

\subsubsection{Tuning by stress minimization}
Assuming, as we have done, that the embedding dimension $p$ is given (see \secref{dimension} for a discussion), local isomap (\algref{local isomap}) relies on two tuning parameters: the connectivity radius $r$ in Step~1 and the number of hops $h$ required by MDS-MAP(P) in Step~2. 

While we continue to advocate that the number of hops be chosen by minimization of a notion of stress such as \eqref{stress}, the choice of connectivity radius $r$ cannot be chosen in the same way for the simple reason that the connectivity radius defines the graph. 
In our experiments, we follow standard practice and choose $r$ as a small multiple of what is needed for the resulting graph to be connected.
We discuss the choice of connectivity radius further in \secref{radius}.  

\subsubsection{Bias--variance tradeoff}
A similar manifestation of bias--variance tradeoff as discussed in \secref{tradeoff MDS} in the MDS setting is at play in the DR setting when tuning the patch size parameter, and so for similar reasons.

\subsection{Experiments}
Although one could anticipate that local isomap behaves in the context of DR in a way that is parallel to how MDS-MAP(P) behaves in the context of MDS, we perform some simple numerical experiments to confirm this. 

\begin{figure}[ht!]
\centering
\includegraphics[width=0.40\textwidth]{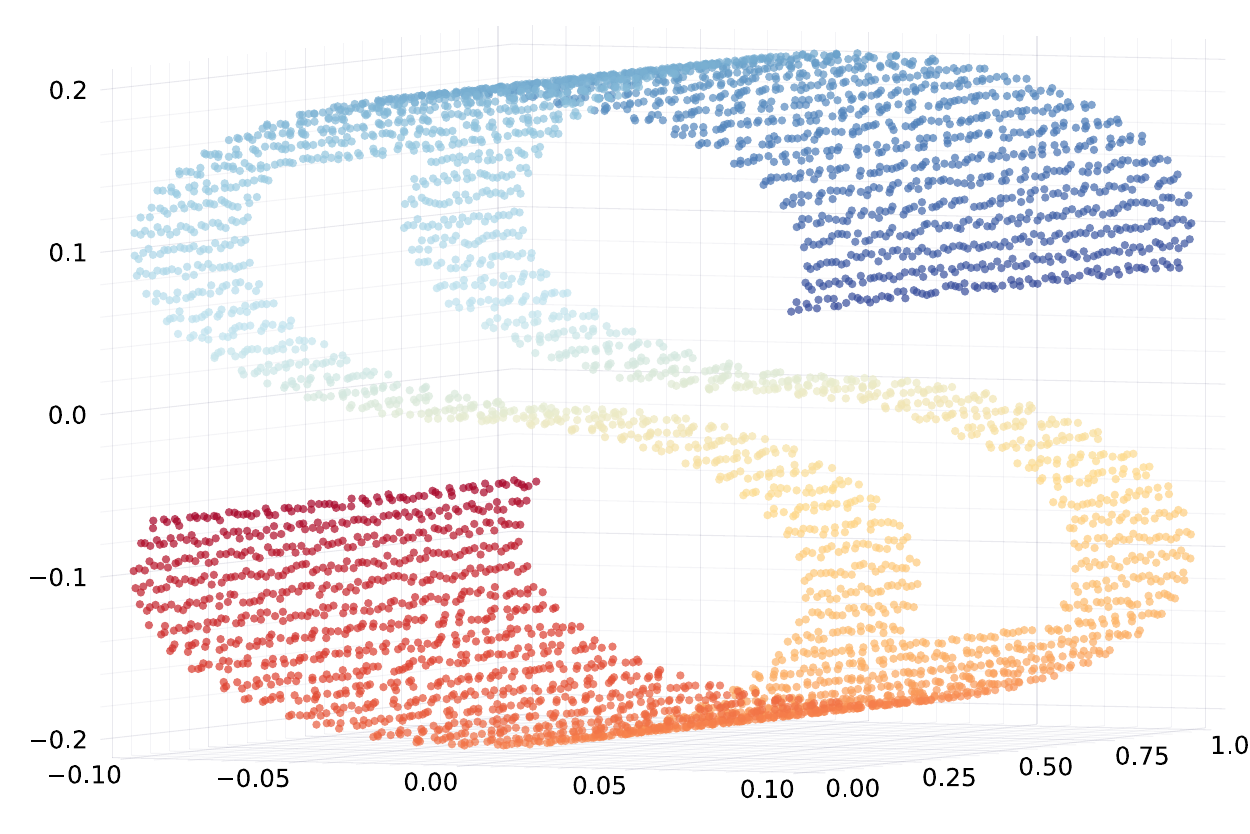} \quad
\includegraphics[width=0.40\textwidth]{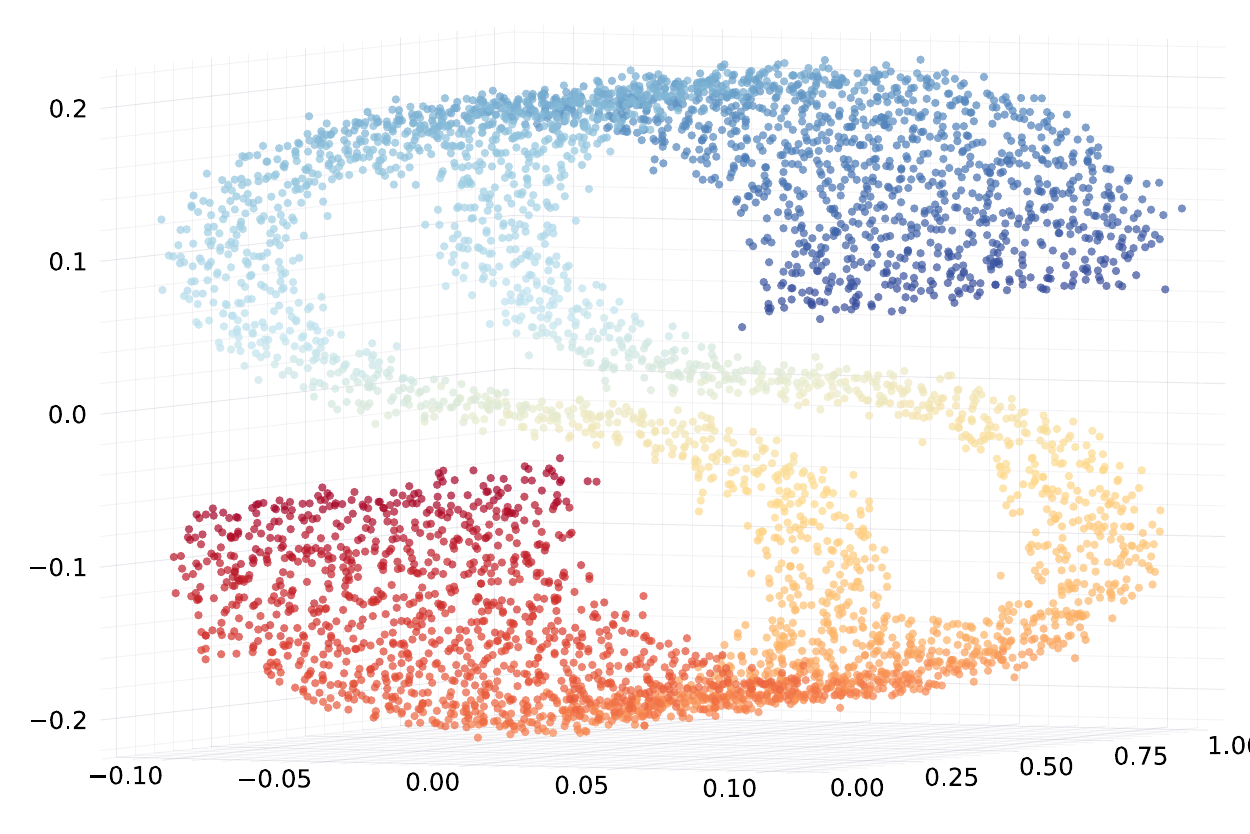}
\caption{Data points (about $n\approx4000$ of them) generated based on embedding the hollow rectangle of \figref{rect_hole_b} as an `S' surface using \eqref{S_varphi}, without (left) and with (right) added noise.}
\label{fig:S}
\end{figure}

\begin{figure}[ht!]
\centering
\includegraphics[width=0.4\textwidth]{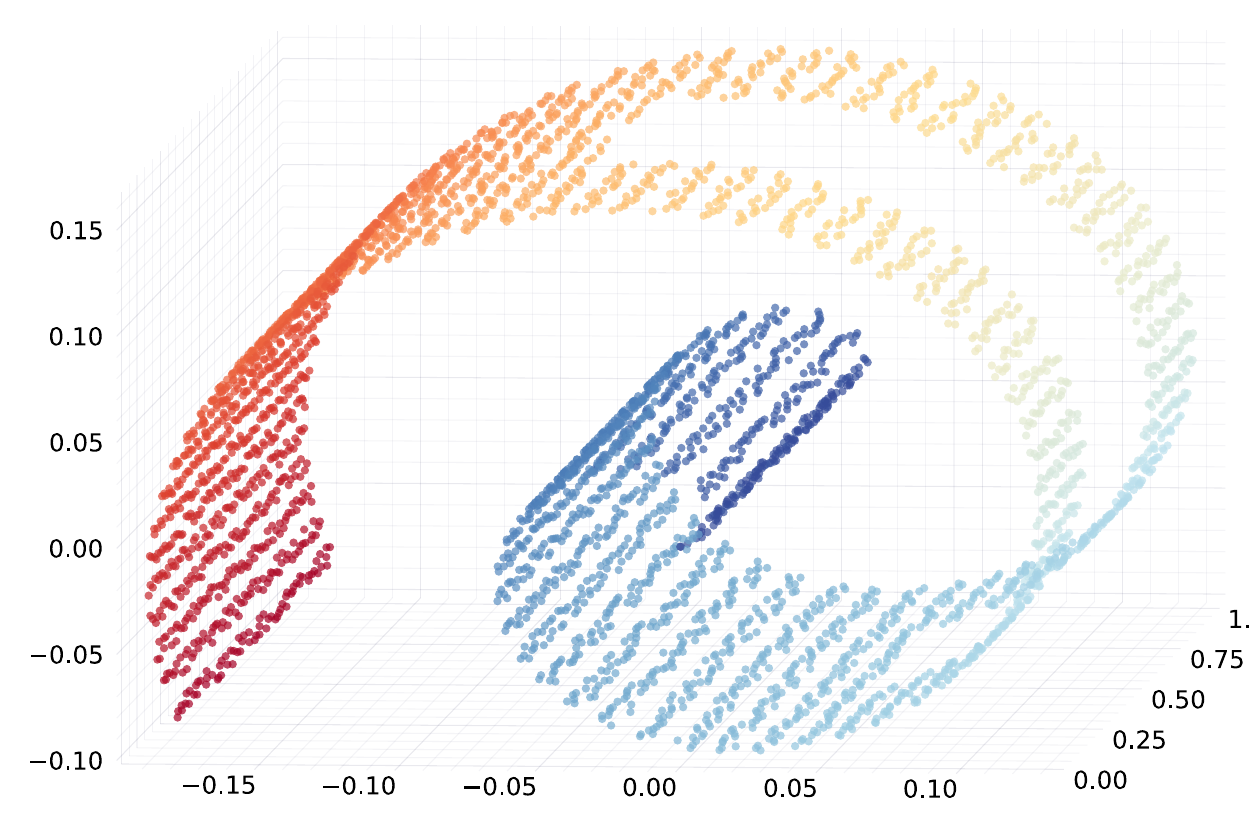} \quad
\includegraphics[width=0.4\textwidth]{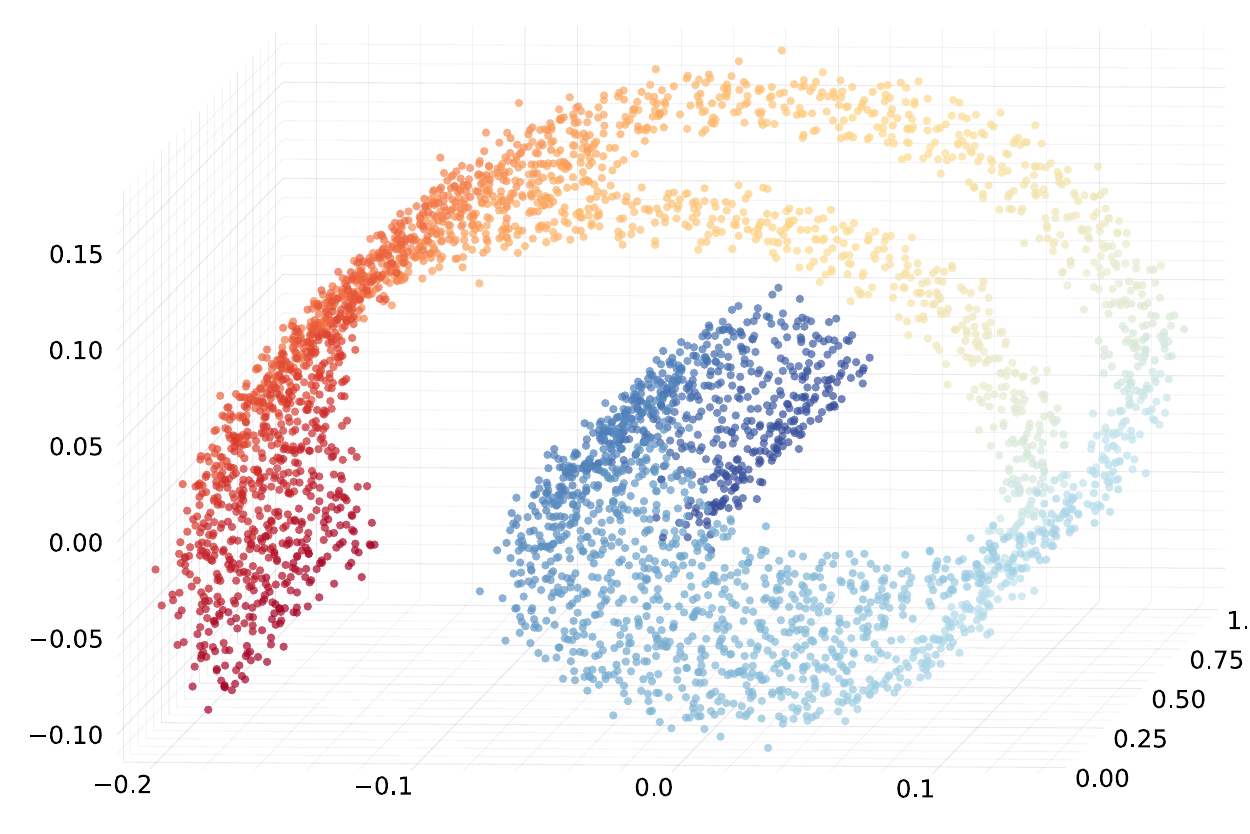}
\caption{Data points (about $n\approx4000$ of them) generated based on embedding the hollow rectangle of \figref{rect_hole_b} as a Swiss roll using \eqref{Swiss_varphi}, without (left) and with (right) added noise.}
\label{fig:Swiss}
\end{figure}

\begin{figure}[ht!]
\centering
\includegraphics[width=0.5\textwidth]{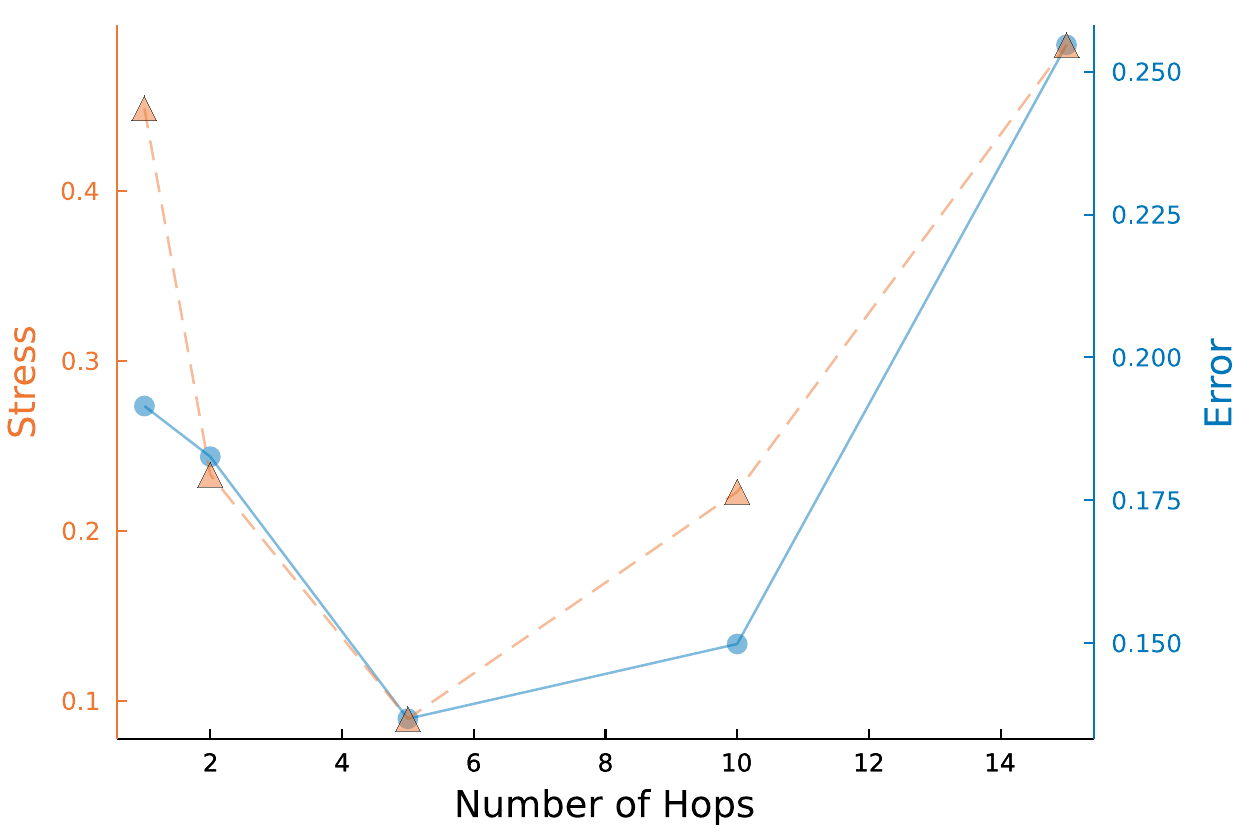}
\caption{Experiment with $n = 5008$ points near an `S' surface.}
\label{fig:S_a}
\end{figure}

\begin{figure}[ht!]
\centering
\includegraphics[width=0.2\textwidth, trim={0 0 2.5in 0}, clip]{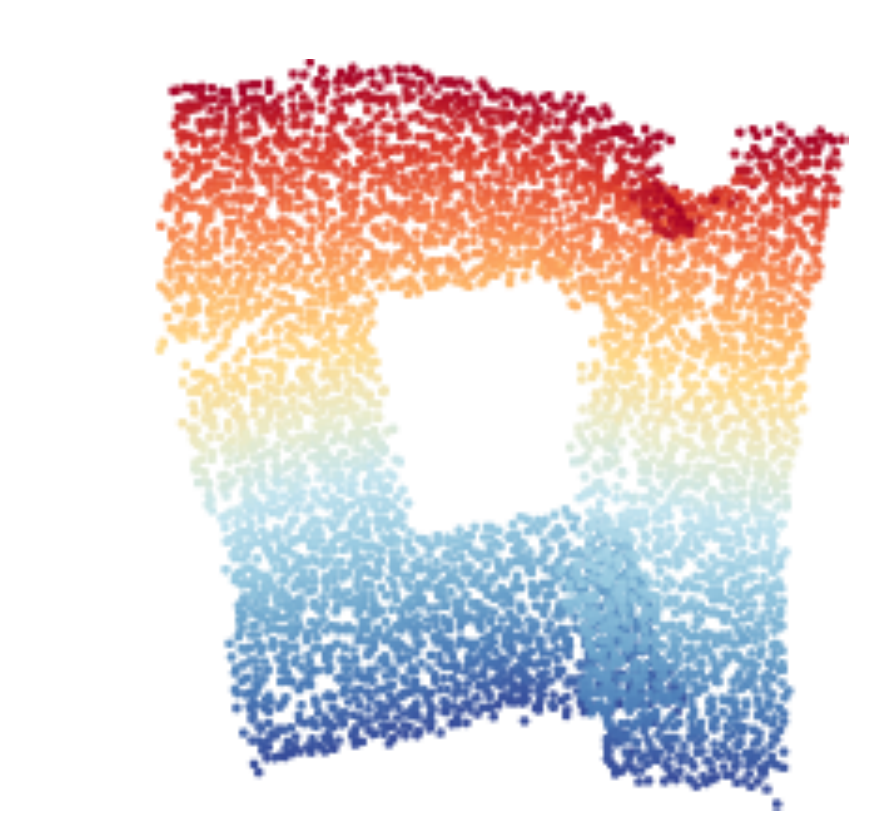}
\includegraphics[width=0.18\textwidth, trim={0 0 2.5in 0}, clip]{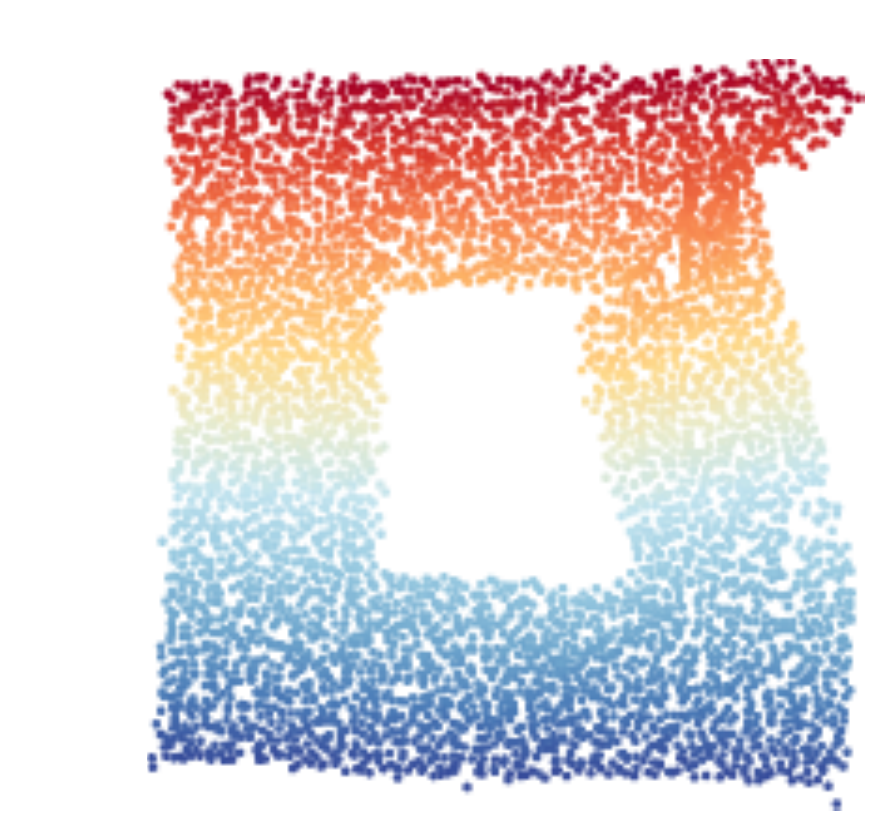}
\includegraphics[width=0.18\textwidth, trim={0 0 2.5in 0}, clip]{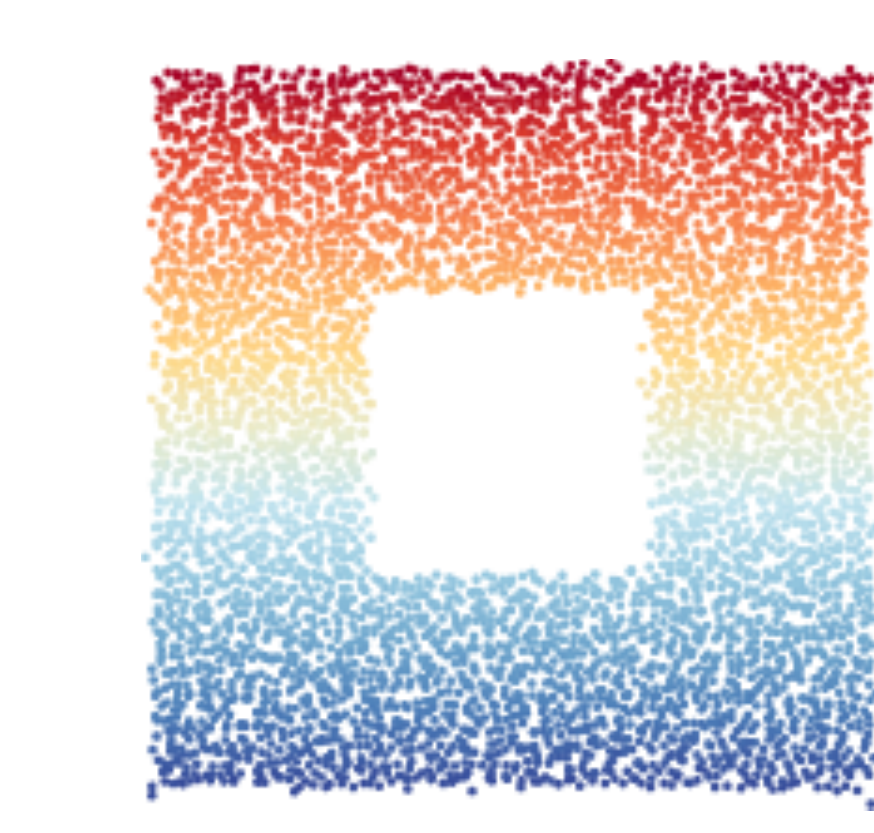}
\includegraphics[width=0.18\textwidth, trim={0 0 2.5in 0}, clip]{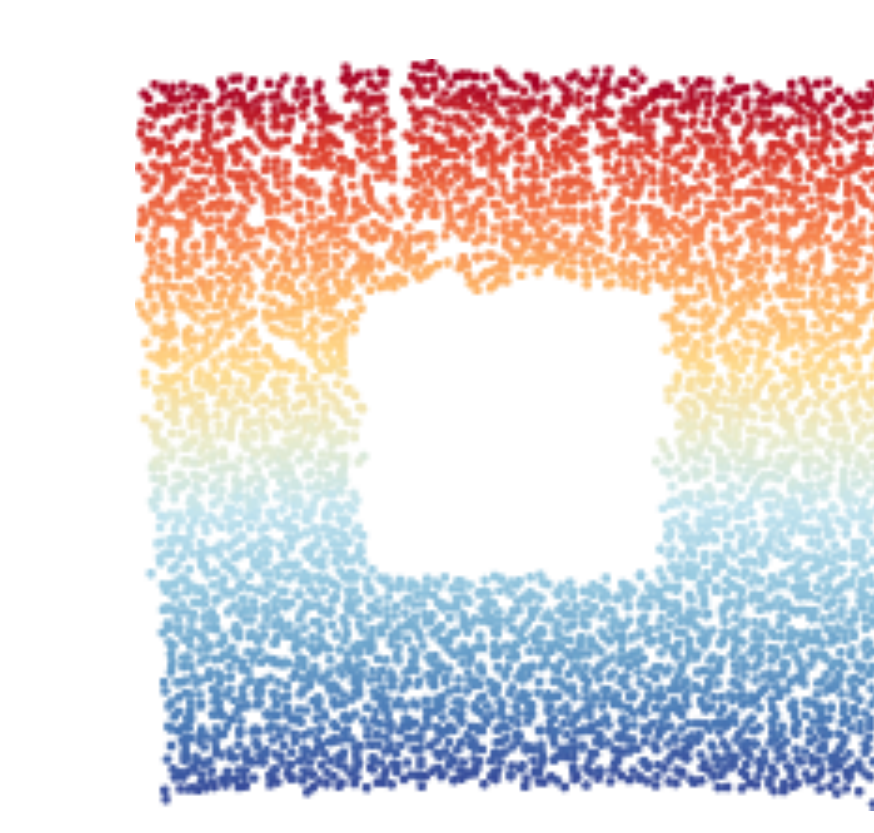}
\includegraphics[width=0.18\textwidth, trim={0 0 2.5in 0}, clip]{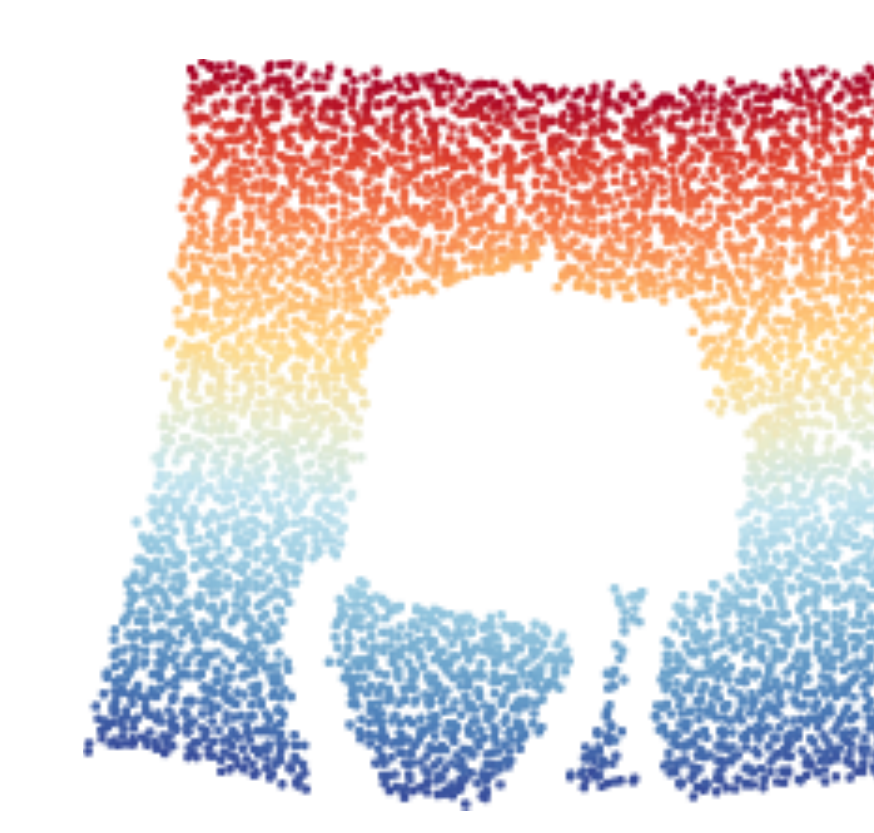}
\caption{Same setting as \figref{S_a}. Examples of embeddings with number of hops $h = 1, 2, 5, 10, 15$.}
\label{fig:S_b}
\end{figure}

\begin{figure}[ht!]
\centering
\includegraphics[width=0.5\textwidth]{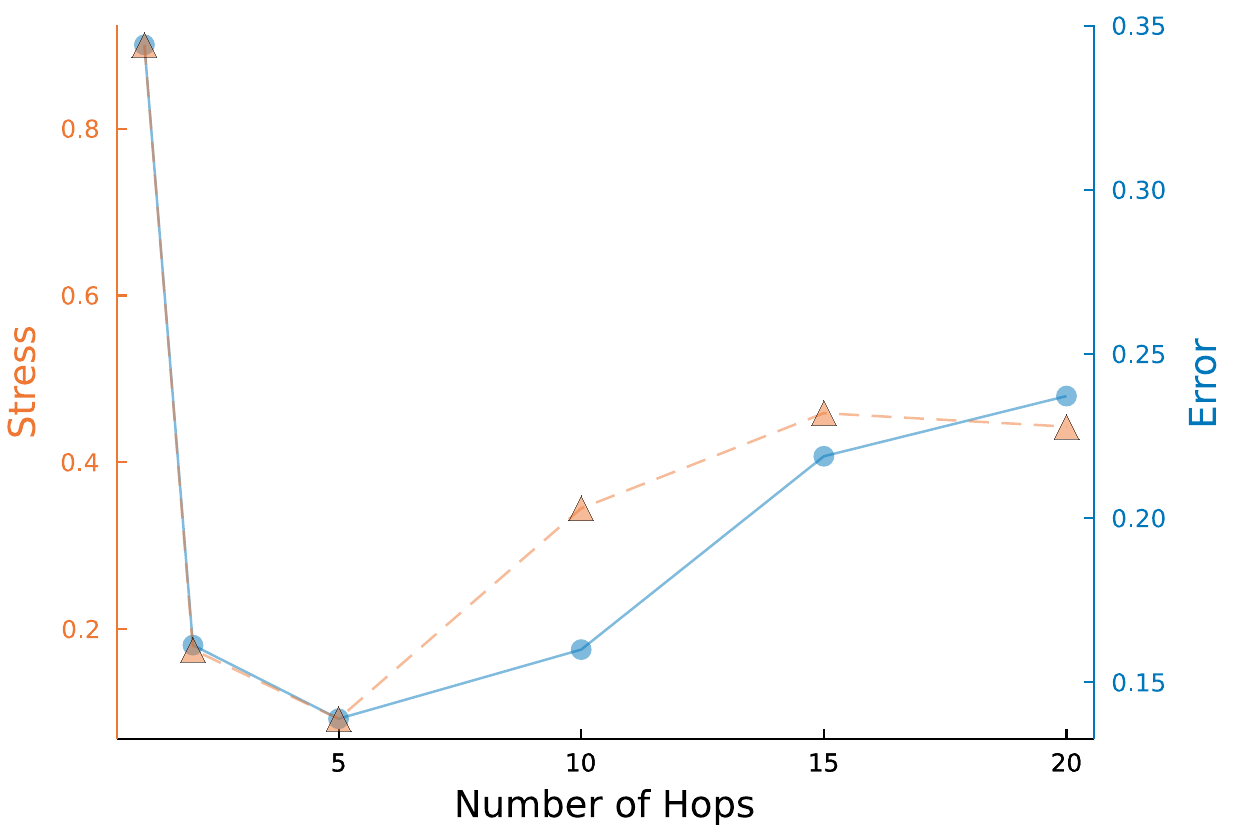}
\caption{Experiment with $n = 4137$ points near an Swiss roll surface.}
\label{fig:Swiss_a}
\end{figure}

\begin{figure}[ht!]
\centering
\includegraphics[width=0.18\textwidth, trim={0 0 2.5in 0}, clip]{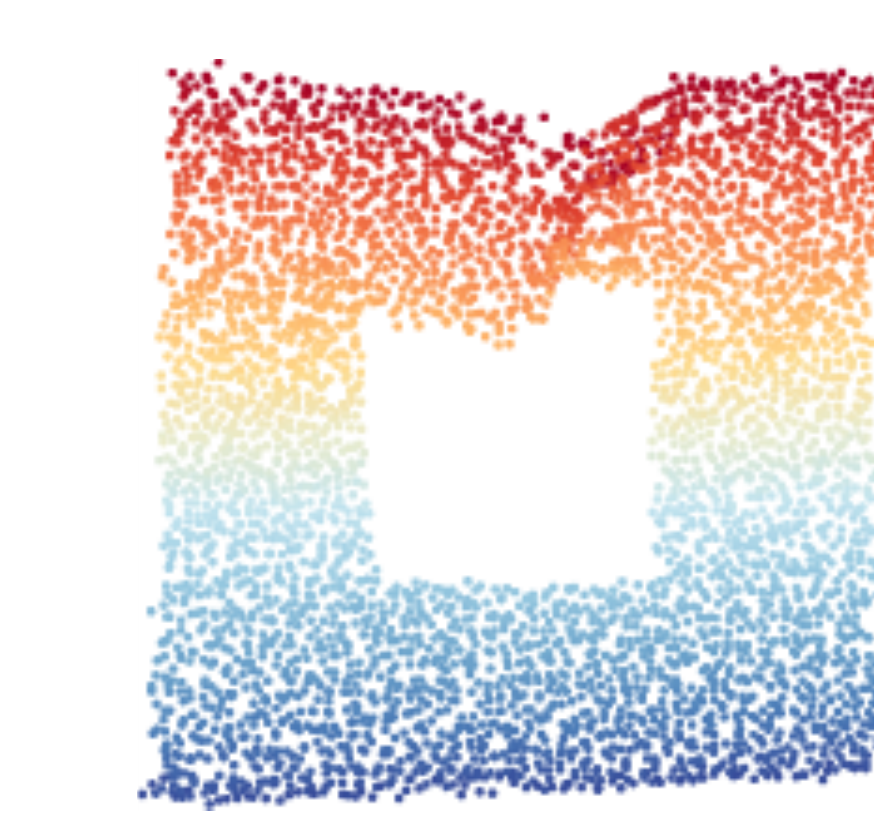}
\includegraphics[width=0.18\textwidth, trim={0 0 2.5in 0}, clip]{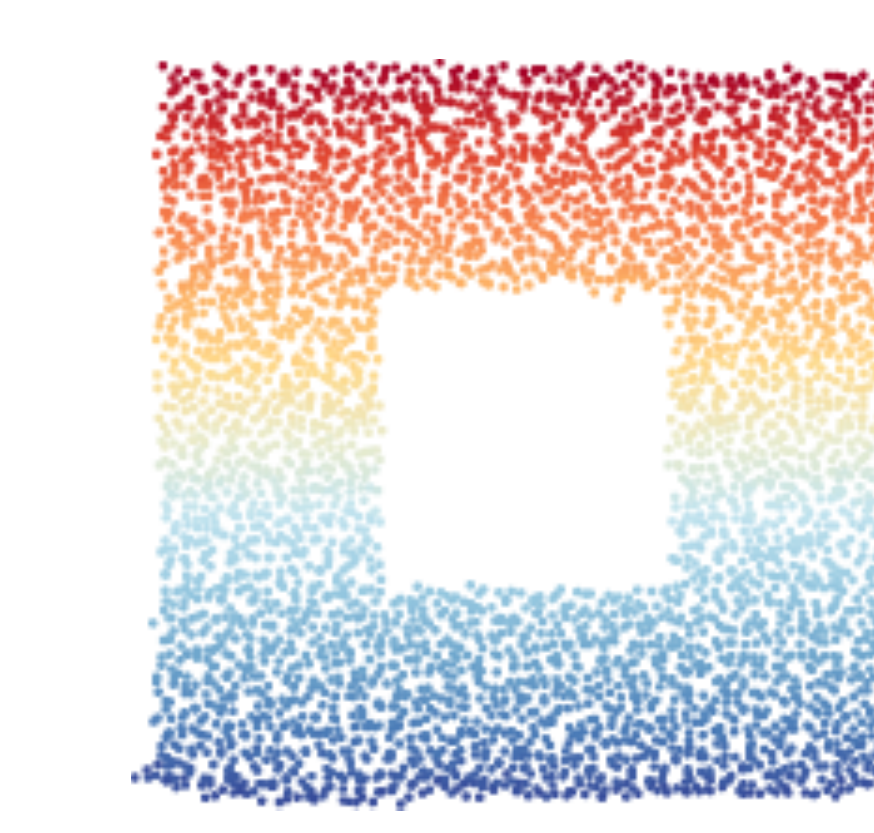}
\includegraphics[width=0.18\textwidth, trim={0 0 2.5in 0}, clip]{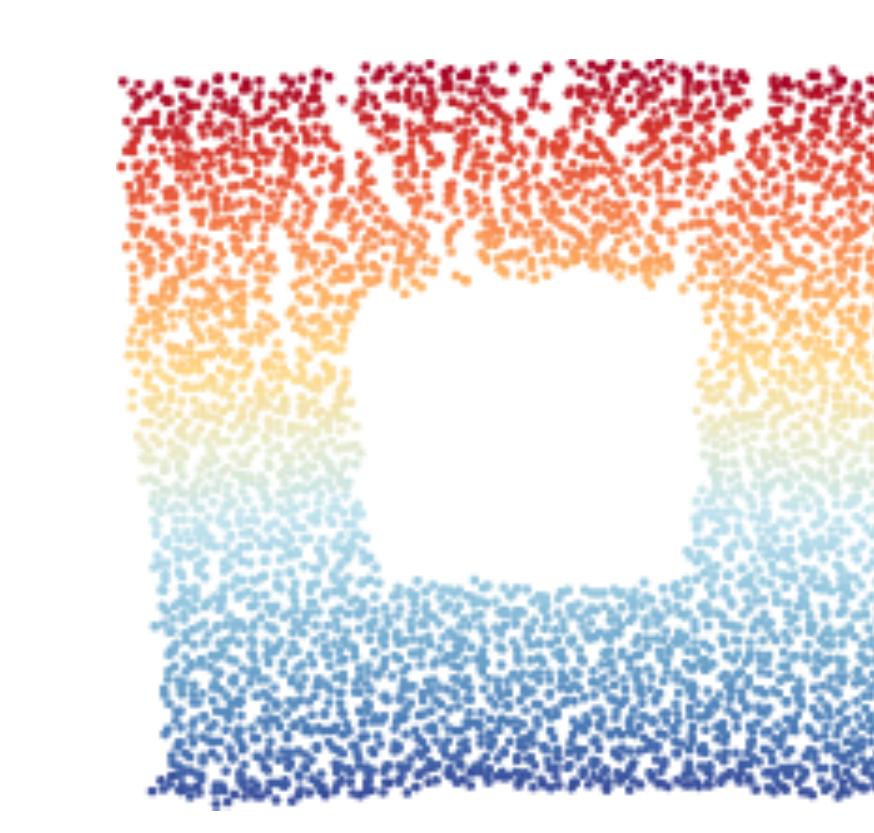}
\includegraphics[width=0.19\textwidth, trim={0 0 2.5in 0}, clip]{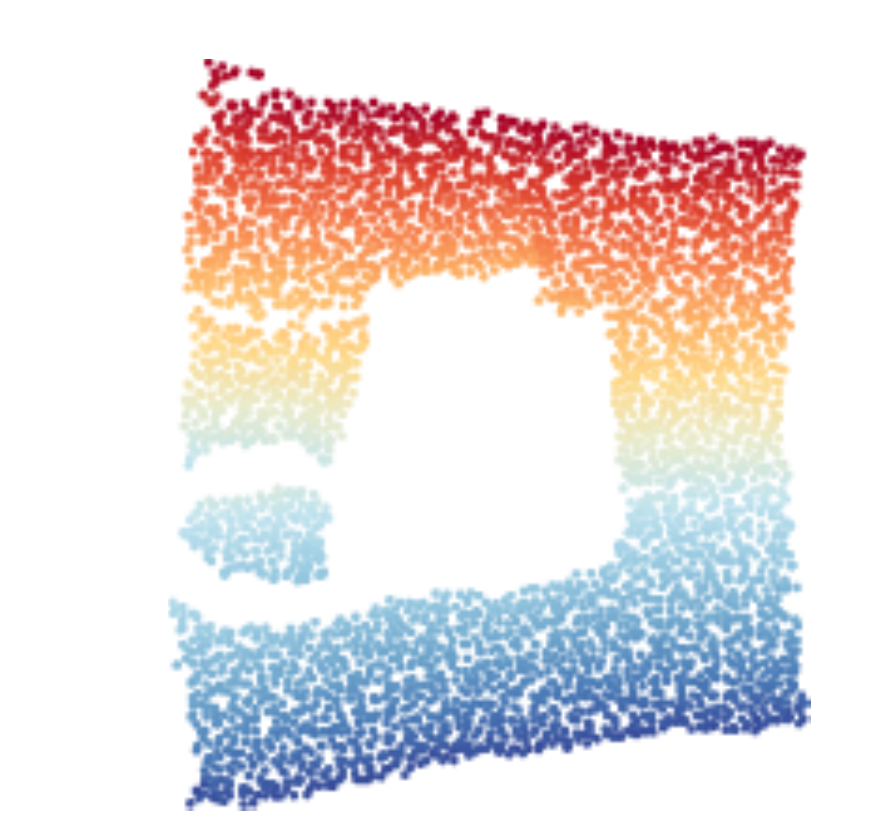}
\includegraphics[width=0.19\textwidth, trim={0 0 2.5in 0}, clip]{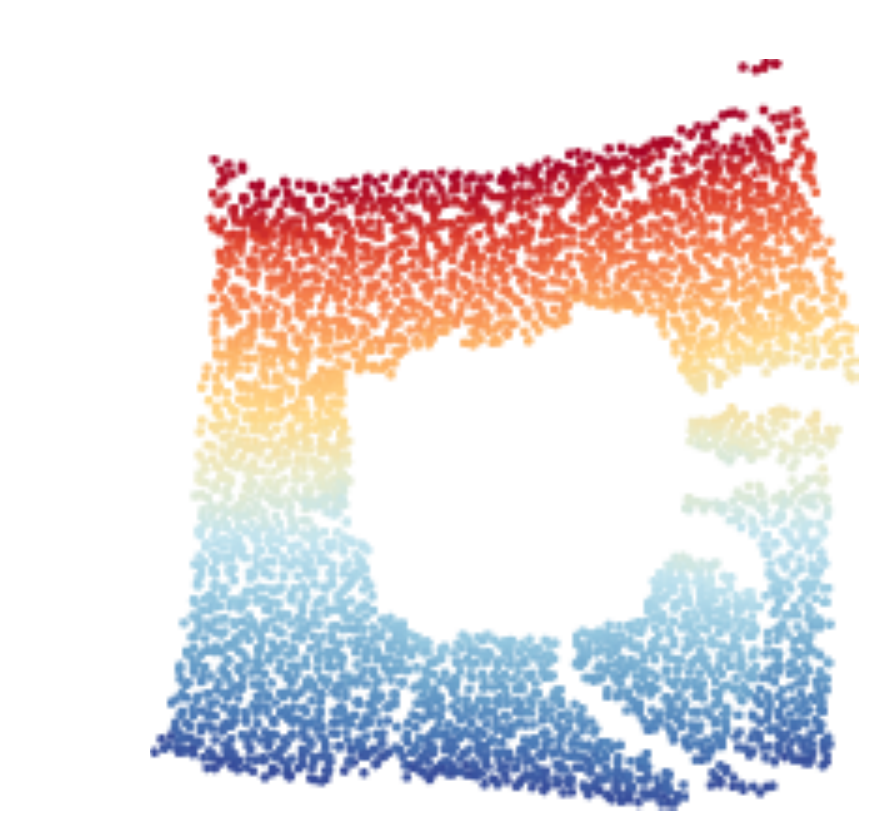}
\caption{Same setting as \figref{Swiss_a}. Examples of embeddings with number of hops $h = 2, 5, 10, 15, 20$.}
\label{fig:Swiss_b}
\end{figure}

To simulate the data in the manifold learning setting, we start with data points in $\bbR^2$ as in \secref{experiments MDS}, and then embed these into $\bbR^3$. We chose to work with the hollow rectangle of \figref{rect_hole_b}. Note that here, unlike in the MDS setting, the graph structure is not given but needs to be chosen. This is done in Step~1 of local isomap.
We used two different embeddings that seem popular in the literature: a cylindrical surface based on an S curve and a cylindrical surface based on a spiral, often referred to as a Swiss roll. 
The `S' surface is obtained via the following embedding of $[0,1]^2$:
\begin{align}
\label{S_varphi}
\varphi_{S, \alpha} (u, v) = 
\big(\alpha^{-1} \sin(\alpha v) ,\ u,\ \alpha^{-1}(\cos (\alpha v) - 1)\big);
\end{align}
the Swiss roll is obtained via the following embedding of $[0,1]^2$:
\begin{align}
\label{Swiss_varphi}
\varphi_{\rm Swiss, \alpha} (u, v) 
= \big(s(v) \cos(\alpha s(v)),\ u,\ s(v) \sin (\alpha s(v))\big),
\end{align}
where $s(v)$ is the solution to $\int_0^s \sqrt{1 + (\alpha t)^2} \, \d t = v$.
The parameter $\alpha$ allows us to increase the curvature of the resulting surface. 
In our experiments, $\alpha = 10$ for the `S' surface and $\alpha = 50$ for the Swiss roll.
Although the estimation of the local intrinsic distances by the ambient Euclidean distances implies a bias which already plays the role of noise, we add a small amount of Gaussian noise to obtain the data points: see \figref{S} for the `S' surface and \figref{Swiss} for the Swiss roll. 
The result of applying local isomap for various choice of the number of hops is displayed in Figures~\ref{fig:S_a}--\ref{fig:S_b} for the `S' surface and Figures~\ref{fig:Swiss_a}--\ref{fig:Swiss_b} for the Swiss roll. 

\section{Discussion}
\label{sec:discussion}

\subsection{Choice of embedding dimension}
\label{sec:dimension}
Some applications, as in sensor network localization where the items are known to be in a 2D physical space, the embedding dimension comes with the problem itself. In other situations, it needs to be chosen by the analyst. In the context of MDS, this is discussed in \cite[Sec~3.5]{borg2005modern}, where the suggested approach consists, for a particular method under consideration, in plotting the stress for the output embedding as a function of the embedding dimension, and look for an `elbow' in the resulting plot indicating that the gains in stress from increasing the dimension have started to dampen. 
Similar strategies have been suggested in DR, for example, \cite{grassberger1983measuring}, although a number of competing methods have also been proposed --- \cite[Sec 2.1]{mordohai2010dimensionality} provides a partial review.

While this ad hoc approach can be formalized for particular methods (e.g., the one proposed in \cite{grassberger1983measuring} is shown to be consistent in \cite{higher-order}), we can already see that the situation is very different as compared to choosing a tuning parameter such as the number of hops for MDS-MAP(P) because the stress is decreasing as a function of the embedding dimension. Consequently, using the stress to choose the embedding dimension is useless as it would result in choosing the largest possible dimension, meaning $p = n-1$. 
($n$ points, even in an infinite-dimensional linear space, are always contained in the affine subspace that they span, which is of dimension at most $n-1$.)
Thus, stress minimization is not a good strategy for choosing the embedding dimension.

\subsection{Choice of connectivity radius}
\label{sec:radius}
Most modern methods in manifold learning rely on a construction of a neighborhood graph, and this necessitates the choice of a connectivity radius.\footnote{~While we speak of ball graphs, nearest neighbor graphs are sometimes preferred in practice. But the core issue remains, and simply becomes the choice of the number of nearest neighbors.} In the literature, the choice of connectivity radius appears to be ad hoc. One of them is a small multiple of what is needed for the resulting graph to be connected, which was our choice in our numerical experiments.

It turns out that the connection with MDS can inform that choice in a more principled manner. Indeed, considerations of rigidity --- in that we want the result to be well-defined up to a rigid transformation --- would prompt us to choose the connectivity radius a little larger than what is needed for the resulting graph to be generically globally rigid. Actionable, sufficient conditions for that to be true exist. For example, in the important case of $p = 2$, it is enough that the graph be 6-connected \cite[Th 7.2]{jackson2005connected}, and this can be checked using a variety of algorithms \cite{esfahanian1984computing}. 

\subsection*{Acknowledgements} 
This work was partially supported by the US National Science Foundation (DMS 1916071).

\small
\bibliographystyle{chicago}
\bibliography{ref}

\end{document}